\documentclass[11pt,draftclsnofoot,onecolumn] {IEEEtran}
\usepackage{bbm}
\usepackage{psfrag}
\usepackage{color}
\usepackage[compress]{cite}
\usepackage[pdftex]{graphicx}
\usepackage{amssymb}
\usepackage[cmex10]{amsmath}
\usepackage{amsthm}
\usepackage{algorithm, algorithmic}
\usepackage{array,enumerate}
\usepackage{multirow}
\usepackage[tight, footnotesize]{subfigure}
\usepackage{breqn}

\graphicspath{{./Figures/}}

\newtheorem{definition}{Definition}
\newtheorem{assumption}{Assumption}
\newtheorem{theorem}{Theorem}
\newtheorem{proposition}{Proposition}

\newtheorem{lemma}{Lemma}
\newcounter{dummy} \numberwithin{dummy}{section}
\newtheorem{lem}[dummy]{Lemma}

\newcommand{\bX}{\mathbf{X}}
\newcommand{\tX}{\tilde{X}}

\renewcommand{\P}{\mathbb{P}}

\newcommand{\Sus}{\mathbf{s}}
\newcommand{\Inf}{\mathbf{i}}
\newcommand{\Rec}{\mathbf{r}}

\newcommand{\Non}{\mathbf{n}}
\newcommand{\cT}{\mathcal{T}}
\newcommand{\cX}{\mathcal{X}}

\newcommand{\pInf}[1]{p_{\Inf}({#1})}
\newcommand{\pSus}[1]{p_{\Sus}({#1})}
\newcommand{\pRec}[1]{p_{\Rec}({#1})}

\newcommand{\cA}{\mathcal{A}}

\newcommand{\Real}{\mathbb{R}}

\newcommand{\sng}{{\tilde{G}(S,X^t)}}

\newcommand{\parent}[1]{\mathrm{pa}(#1)}
\newcommand{\children}[1]{\mathrm{ch}(#1)}
\newcommand{\sn}[1]{\textrm{Supernode}(#1)}
\newcommand{\minimum}[1]{\min \left\{#1\right\}}
\newcommand{\maximum}[1]{\max \left\{#1\right\}}

\newcommand{\psmin}{\alpha}
\newcommand{\psmax}{\beta}

\newcommand{\blue}[1]{{{\color{blue} #1}}}

\usepackage[normalem]{ulem}

\begin{document}
\title{On the Universality of Jordan Centers for Estimating Infection Sources in Tree Networks}

\author{
    Wuqiong~Luo,~\IEEEmembership{Member,~IEEE}, Wee~Peng~Tay,~\IEEEmembership{Senior Member,~IEEE} and Mei~Leng,~\IEEEmembership{Member,~IEEE}
    \thanks{
        Part of this work was presented at the 1st IEEE Global Conference on Signal and Information Processing, Austin, TX, December 2013. This work was supported in part by the Singapore Ministry of Education Academic Research Fund Tier 2 grants MOE2013-T2-2-006 and MOE2014-T2-1-028.
    }
    \thanks{W. Luo was with the Nanyang Technological University, Singapore, and is currently with Micron Semiconductor Asia. W.~P. Tay is with the Nanyang Technological University, Singapore. M. Leng is with the Temasek Laboratories@NTU, Singapore. E-mail: \texttt{wluo1@e.ntu.
edu.sg, wptay@ntu.edu.sg, lengmei@ntu.edu.sg}.}
}

\maketitle

\begin{abstract}
Finding the infection sources in a network when we only know the network topology and infected nodes, but not the rates of infection, is a challenging combinatorial problem, and it is even more difficult in practice where the underlying infection spreading model is usually unknown a priori. In this paper, we are interested in finding a source estimator that is applicable to various spreading models, including the Susceptible-Infected (SI), Susceptible-Infected-Recovered (SIR), Susceptible-Infected-Recovered-Infected (SIRI), and Susceptible-Infected-Susceptible (SIS) models. We show that under the SI, SIR and SIRI spreading models and with mild technical assumptions, the Jordan center is the infection source associated with the most likely infection path in a tree network with a single infection source. This conclusion applies for a wide range of spreading parameters, while it holds for regular trees under the SIS model with homogeneous infection and recovery rates. Since the Jordan center does not depend on the infection, recovery and reinfection rates, it can be regarded as a \emph{universal} source estimator. We also consider the case where there are $k>1$ infection sources, generalize the Jordan center definition to a $k$-Jordan center set, and show that this is an optimal infection source set estimator in a tree network for the SI model. Simulation results on various general synthetic networks and real world networks suggest that Jordan center-based estimators consistently outperform the betweenness, closeness, distance, degree, eigenvector, and pagerank centrality based heuristics, even if the network is not a tree.
\end{abstract}


\section{Introduction}\label{sec:Introduction}

We define an infection to be a property that can be spread probabilistically from one node to another in a network. Examples of infection spreading include a rumor or a piece of news spreading in a social network, a contagious disease spreading in a community, and a computer virus spreading on the Internet. Various models have been developed to describe the spreading process of an infection. In this paper, we consider only discrete time stochastic spreading models. The two simplest models are the Susceptible-Infected (SI) model \cite{Bai2007,Wu2012,Shang2013,Chou2013} and the Susceptible-Infected-Recovered (SIR) model \cite{Bailey1975, Easley2010}. In the SI model, a susceptible node becomes infected probabilistically at each time step, while an infected node retains the infection forever once it is infected. In the SIR model, an infected node can recover from an infection with a given probability at each time step, upon which it gains immunity from further infections.

With increasing interconnectedness of the world, both physically and online, prompt identification and isolation of infection sources is crucial in many practical applications in limiting the damage caused by the infection, and dealing with the aftermath effectively. Therefore, the problem of infection sources estimation has attracted immense interest from the research community after the pioneering work of \cite{Shah2011}, which investigates the problem of identifying a single infection source in the SI model. The reference \cite{Dong2013} considers single source estimation with a priori knowledge of the set of suspect nodes, while \cite{Wang2014} investigates the use of multiple infection spreading instances to identify a source. These methods are based on variants of the distance or rumor centrality of the network graph. We have also developed procedures to identify a source with limited observations of the set of infected nodes \cite{Luo2014}, and to identify multiple infection sources in \cite{Luo2013}. All these works adopt the SI model. Identification of a single infection source in the SIR model was considered in \cite{Zhu2012,Zhu2013}, which showed that the Jordan center\footnote{The Jordan center of the infected node set is the node in the network with the smallest maximum distance to any observed infected node.} gives the optimal estimator associated with a most likely infection path. Infection source estimators using a dynamic message passing (DMP) approach \cite{Lokho2013}, and the belief propagation (BP) approach \cite{Altarelli2013} have also been developed for the SIR model. These two approaches however require significant a priori knowledge of the infection spreading process like the infection and recovery rates of each node in the network.

The SI and SIR models have been widely adopted in the literature due to their simplicity, but these models do not adequately reflect many practical situations in which an infected node recovers and becomes infected again at some future time through either a relapse or reinfection. If an individual recovers from a disease such as bovine tuberculosis or human herpes virus, he may later experience a relapse and exhibit infection symptoms again \cite{Blower1998,Driessche2007, Cruz2013, Georgescu2013 }. The spread of such diseases are often modeled using a Susceptible-Infected-Recovered-Infected (SIRI) model \cite{Driessche2007, Cruz2013, Georgescu2013 }. On the other hand, if an individual recovers from a disease such as gonorrhea \cite{Hethcote1984}, he does not acquire any immunity from his previous infection and may later become reinfected with the same disease. These types of diseases are often modeled using a Susceptible-Infected-Susceptible (SIS) model \cite{Hethcote1976,Allen1994,Newman2003}. A further example of SIRI and SIS type of infection spreading is rumor spreading in an online social network, as monitored by an external agency that does not have access to the full database of the social network. An individual in the network may post a rumor, remove it, and repost the rumor subsequently. If the external agency only has access to a limited set of the most recent postings of each user (for example, due to storage constraints), then trying to identify the source of the rumor based purely on the time-stamps of the rumor posts will lead to an erroneous result.

To the best of our knowledge, finding infection sources under the SIRI and SIS models have not been investigated. Moreover, all the existing works assume that the underlying infection spreading model is known, and in most cases, the infection and recovery rates of each node are also known. This knowledge may be difficult to obtain in practice. For example, when a new type of infectious disease breaks out, the spreading characteristics of the disease is usually unclear before its epidemiology is determined. Therefore, it would be highly desirable if a source estimator can be shown to be robust, under a reasonable non-trivial statistical criterion, to the underlying spreading mechanism, and \emph{universal} to a wide range of parameters governing the spreading process. Indeed, it is unclear that such an estimator even exists for the SI, SIR, SIRI and SIS models.

In this paper, we adopt the most likely infection path (MLIP) criterion of \cite{Zhu2012,Luo2014} to find the optimal infection source estimator. Finding optimal source estimators is in general NP-hard, and proving the optimality of an estimator is also in general very challenging, with similar results in the current literature restricted to tree networks and the SI or SIR spreading models \cite{Shah2011,Zhu2012,Zhu2013,Luo2014}. Therefore, any hope of obtaining theoretical optimality guarantees is restricted to special classes of networks. Our work is a small step towards finding optimal source estimators for the more general SIRI and SIS models. Our main contributions are the following:
\begin{enumerate}[(i)]
\item\label{con:SIRI} For an infection spreading from a single source under the SI, SIR, and SIRI models,\footnote{By setting the recovery probability and relapse probability in the SIRI model to zero, we obtain the SI and SIR models, respectively. However, in this paper, for clarity and due to some differences in the assumptions we make under each of these models, we explicitly differentiate the SIRI model from the SI and SIR models.} and over an infinite tree network in which nodes may have different infection and recovery probabilities, we show that the Jordan center of the observed infected node set is an optimal infection source estimator under the MLIP criterion and under some mild technical assumptions. \blue{Our result corroborates that in \cite{Zhu2012,Zhu2013}, which shows that the Jordan center is the optimal source estimator for the SIR model under assumptions slightly different from ours (cf.\ Section \ref{sec:problem_formulation} for a detailed discussion), and that in \cite{Luo2014}, which gives the same result for the case where the infection spreading follows the SI model, but only a limited set of infected nodes are observed.}

\item\label{con:SIS} We show that if an infection spreads according to the SIS model over an infinite regular tree in which all nodes have the same infection and recovery probabilities, then the Jordan center is again the optimal infection source estimator under the MLIP criterion.

\item We introduce the concept of a $k$-Jordan center set, and show that if an infection spreads from $k > 1$ sources in an infinite tree network where nodes may have different infection probabilities, and in accordance to the SI model, then the $k$-Jordan center set is an optimal estimator of the infection source set under the MLIP criterion. A heuristic procedure was proposed in \cite{Luo2013} to determine multiple infection sources in the SI model based on the single source maximum likelihood (ML) estimator for regular trees, but not shown to be optimal. Simulation results suggest that our estimator outperforms that in \cite{Luo2013} in terms of the average error distance.

\item  We extend the Jordan center-based estimators above heuristically to general graph networks, and perform extensive simulations to verify the performance of our estimators. We perform infection spreading simulations on random trees, part of the Facebook network, and the western states power grid network of the United States. In our simulation results, the Jordan center-based estimators consistently achieve the lowest average error distance compared to the betweenness, closeness, distance, degree, eigenvector, and pagerank centrality based heuristics.
\end{enumerate}

Finding the Jordan center does not require knowledge of each node's infection and recovery probabilities. Therefore, our result in item \eqref{con:SIRI} shows that the Jordan center is a universal source estimator for the SI, SIR and SIRI models, under a wide range of spreading parameters. In contribution \eqref{con:SIS}, we show that the Jordan center is also optimal for the SIS model in regular tree networks. Although we are not able to show that this is true for general graphs and for multiple infection sources, our simulation results suggest that Jordan center-based source estimators outperform many other source estimators, which similarly do not require knowledge of the underlying infection spreading parameters, regardless of which of the four considered infection spreading models is used. This is somewhat surprising since the SI, SIR, SIRI, and SIS spreading mechanisms are quite different from each other. Note that although \cite{Lokho2013} and \cite{Altarelli2013} have reported better source detection rates in numerical experiments using the DMP and BP approaches respectively, these methods require the knowledge of the underlying infection spreading parameters, and are applicable only to the SIR model. There is also a lack of theoretical results on the optimality of the DMP and BP approaches, and extending them to the SIRI and SIS models is highly non-trivial \cite{Hu2014}. We hope that the insights derived from our current work will inform future design of better source estimators in the case where the exact values of infection parameters are unknown.

The rest of this paper is organized as follows. In Section \ref{sec:problem_formulation}, we present our system model, assumptions and problem formulation. In Section \ref{sec:single_source_estimation}, we show, under some technical conditions, that the Jordan center is an optimal source estimator for tree networks when there is a single infection source. In Section \ref{sec:multiple_sources_estimation}, we derive an estimator for tree networks when there is an infection spreading from multiple sources under the SI model. In Section \ref{sec:sources_estimation_for_general_graphs}, we heuristically extend the proposed estimators to general graphs and propose heuristic algorithms to find them. We present simulation results in Section \ref{sec:simulation_results} to verify the effectiveness of the proposed estimators. Finally we conclude and summarize in Section \ref{sec:conclusion}.

\section{Problem Formulation}\label{sec:problem_formulation}
In this section, we present our system model, assumptions, and various notations used throughout this paper. We also describe the most likely infection path criterion. A table summarizing the most commonly used notations is provided at the end of this section.

\subsection{Infection Spreading Model}
We model the underlying network over which an infection spreads as an undirected graph $G=(V,E)$, where $V$ is the set of nodes and $E$ is the set of edges. Two nodes connected by an edge are called neighbors or neighboring nodes. Suppose that an infection starts spreading from one or more source nodes. In most of this paper, we will assume a single source node, and extend this to multiple source nodes for the SI model described below. We adopt a discrete time spreading model in which time is divided into discrete slots, and the states of the nodes in the graph $G$ follow a Markov process with probability measure $\P$. Our goal is to infer the infection sources from observations of the infected nodes at a particular point in time. We consider the following four discrete time infection spreading models.
\begin{enumerate}
	\item \emph{SI model}:
In the SI model, each node takes on one of 3 possible states: \emph{susceptible} ($\Sus$),  \emph{infected} ($\Inf$) and \emph{non-susceptible} ($\Non$). At any time slot, if a node is infected, we say that it is in state $\Inf$. The set of uninfected nodes that have infected neighbors are in state $\Sus$, and are called susceptible nodes. In the SI model, an infected node remains infected forever, and a susceptible node becomes infected probabilistically in the next time slot. All other nodes are in state $\Non$, and are called non-susceptible nodes. A non-susceptible node has probability zero of becoming infected in the next time slot.
\item \emph{SIR model}:
In the SIR model, the possible node states are \emph{susceptible} ($\Sus$),  \emph{infected} ($\Inf$), \emph{non-susceptible} ($\Non$), and \emph{recovered} ($\Rec$). The only difference with the SI model is that an infected node in state $\Inf$ in a time slot may recover to state $\Rec$ in the next time slot with a positive probability. A recovered node then stays in the recovered state $\Rec$ forever. In other words, a recovered node will never become infected again.

\item \emph{SIRI model}: The possible nodes states in the SIRI model are the same as for the SIR model. The difference from the SIR model is that a recovered node (in state $\Rec$) may become infected again at a future time slot with a positive probability. This infection relapse is spontaneous, and can take place even if the node does not have any infected neighbors. Here, we reserve the state $\Sus$ for those nodes that have infected neighbors and have never been infected before.

\item \emph{SIS model}:
In the SIS model, the possible node states are \emph{susceptible} ($\Sus$),  \emph{infected} ($\Inf$) and \emph{non-susceptible} ($\Non$). This model describes a more complicated spreading process where once an infected node recovers from the infection (with a positive probability), it immediately becomes a susceptible node (if it has at least one infected neighbor) or non-susceptible node (if it does not have any infected neighbor). There is therefore no \emph{recovered} state in this model.
\end{enumerate}

For any node $v \in V$, we let $\pSus{v}$, $\pInf{v}$ and $\pRec{v}$ be the probability for $v$ to be in state $\Inf$ in the next time slot conditioned on $v$ being susceptible, infected, or recovered in the current time slot, respectively. These probabilities characterize different infection spreading models, and we assume that they satisfy the following Assumptions \ref{assump:infection_prob_SI}--\ref{assump:infection_prob_SIS}. Let $\psmin = \min_{u \in V} \pSus{u}$ and $\psmax = \max_{u \in V} \pSus{u}$. For simplicity, we assume that $\pSus{v}$, $\pInf{v}$ and $\pRec{v}$ do not change over time slots for each $v$, although all our results and proofs (with slight modifications) are still valid if these probabilities are time-varying as long as Assumptions \ref{assump:infection_prob_SI}--\ref{assump:infection_prob_SIS} hold over all time slots.

\begin{assumption}
\label{assump:infection_prob_SI}
Under the SI model, for every $v \in V$, we have
\begin{align}
\psmax & \le \blue{\frac{\psmin}{(1-\psmin)^2}}. \label{ineq:SI_ps}
\end{align}
\end{assumption}

\begin{figure}[!ht] 
  \centering
  \includegraphics[width=0.4\textwidth]{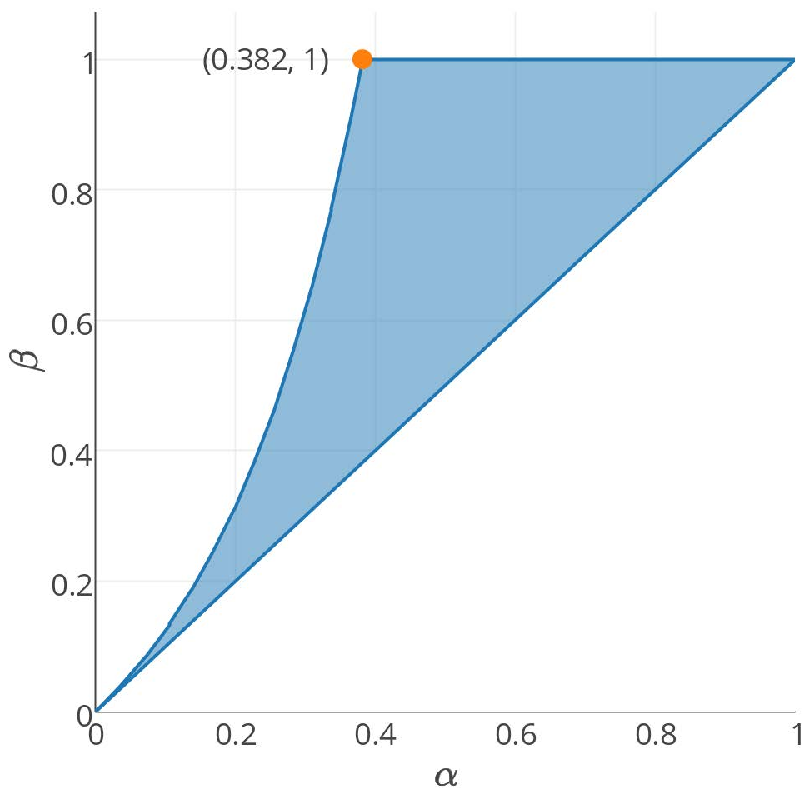}
  \caption{Illustration of the region where $(\psmin, \psmax)$ satisfies \eqref{ineq:SI_ps}.}
  \label{fig:assumption1}
\end{figure}

\blue{See Fig. \ref{fig:assumption1} for the region where $(\psmin, \psmax)$ satisfies the inequality \eqref{ineq:SI_ps}. For example, if $\alpha \geq 0.382$, then \eqref{ineq:SI_ps} holds since its right hand side is greater than 1.} In the inequality \eqref{ineq:SI_ps}, we assume that the infection probabilities at each node in the network does not differ drastically for the SI model.  This is required because in this work, we do not assume knowledge of the exact infection rates at each node. Therefore, if part of the network has nodes that are much easier to infect than other nodes, then any estimator with no knowledge of the infection rates will result in a highly biased result, which may not do better on average than making random choices for the infection sources. \blue{We provide an example in Fig. \ref{fig:example_for_assump_infection_prob_SI} to show that Assumption \ref{assump:infection_prob_SI} is a necessary condition for Theorem \ref{theorem:single_source_estimate_Jordan_infection_center} to hold for the SI model.}

\begin{assumption}
\label{assump:infection_prob_SIR}
Under the SIR model, for every $v \in V$, we have
\begin{align}
0 & \le \pInf{v} \le \sqrt{\frac{\psmin}{\psmax}}. \label{ineq:SIR_pi}
\end{align}
\end{assumption}
\blue{The reference \cite{Zhu2012} assumes that $\pSus{v}$ is the same for every $v \in V$, which implies that $\psmin = \psmax$, and \eqref{ineq:SIR_pi} then reduces to the trivial condition $0 \le \pInf{v} \le 1$. It also assumes that $\pInf{v}$ is the same for every $v \in V$. Therefore, the setup in \cite{Zhu2012} is a special case of the problem studied in this paper. On the other hand, the reference \cite{Zhu2013} considers the SIR model under a heterogeneous setting, where an infection is transmitted across each edge $(u,v)$ with probability $p(u,v)$ so that $\pSus{v} = 1 - \prod_{u\in N_v} (1-p(u,v))$, where $N_v$ is the set of infected neighbors of node $v$ at the beginning of the current time slot. However, since \cite{Zhu2013} considers undirected graphs with $p(u,v) = p(v,u)$ for all edges $(u,v)$ (see Fig.~\ref{fig:example_for_assump_infection_prob_SI} for a counterexample if edge infection probabilities are not symmetric), no additional assumptions are required to show that the Jordan center is an optimal estimator under the MLIP criterion for an infinite tree, where each node has degree at least 2. In a social network, the strength of influence might not be symmetric between each pair of friends. Therefore, we do not make this assumption.}

\begin{assumption}
\label{assump:infection_prob_SIRI}
Under the SIRI model, for every $v \in V$, we have
\begin{align}
\frac{\psmax -\psmin}{1 - \psmin} &\le \pInf{v} \le \sqrt{\frac{\psmin}{\psmax}}, \label{ineq:SIRI_pi} \\
1-\sqrt{\frac{\psmin}{\psmax}} &\le \pRec{v} \le \minimum{1, \sqrt{\frac{\psmin}{\psmax}}  \frac{\pInf{v}}{1- \pInf{v}}}.  \label{ineq:SIRI_pr}
\end{align}
\end{assumption}

\begin{figure}[!ht] 
  \centering
  \includegraphics[width=0.4\textwidth]{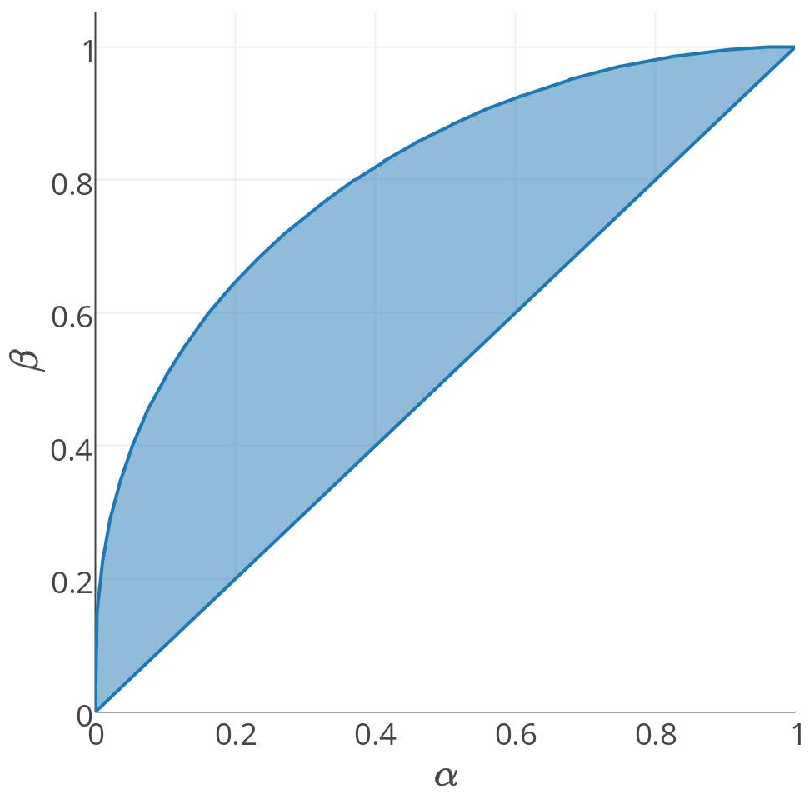}
  \caption{Illustration of the region where $(\psmin, \psmax)$ satisfies \eqref{ineq:SIRI_pi}.}
  \label{fig:assumption3_eqn3}
\end{figure}

\blue{See Fig.~\ref{fig:assumption3_eqn3} for the region of $(\psmin, \psmax)$ that makes \eqref{ineq:SIRI_pi} feasible.} Note that if $\psmin = \psmax$, \eqref{ineq:SIRI_pi} reduces to $0 \le \pInf{v} \le 1$, and \eqref{ineq:SIRI_pr} reduces to $0 \le \pRec{v} \le \minimum{1, \frac{ \pInf{v}}{1- \pInf{v}}}$. Inequality \eqref{ineq:SIRI_pr} implies that a node does not easily relapse into an infected state (i.e., small $p_{\Rec}$) if it recovers quickly (i.e., small $p_{\Inf}$). This is intuitively appealing as it corresponds to the case where if an infected node has a low probability of staying infected in the next time slot, then it is unlikely for the node to relapse into the infection once it has recovered. A practical example is: it is hard to re-convince someone to believe a rumor if he already has a reason to reject the rumor.

\begin{assumption}
\label{assump:infection_prob_SIS}
Under the SIS model, for every $v \in V$, we have
\begin{align}
\pSus{v} &= p_\Sus, \nonumber \\
\pInf{v} &= p_\Inf, \nonumber \\
0 \le p_{\Sus} &\le p_{\Inf} \le 1. \label{ineq:SIS_ps}
\end{align}
\end{assumption}

Inequality \eqref{ineq:SIS_ps} helps us to avoid the case where an infection spreads very fast (i.e., large $p_\Sus$) and infected nodes also recover relatively quickly (i.e., $p_\Inf < p_\Sus$) from happening. In such cases, infected nodes close to sources are likely to have recovered by the time we observe the state of the network, while there may be a significant set of infected nodes at a distance away from the source. Therefore, trying to estimate the source nodes will result in a large bias.

In Assumptions \ref{assump:infection_prob_SI}-\ref{assump:infection_prob_SIRI} for the SI, SIR and SIRI models, the infection, recovery, and relapse probabilities can vary between different nodes. We call such networks \emph{heterogeneous}. On the other hand, in Assumption \ref{assump:infection_prob_SIS}, the infection and recovery probabilities are the same for all nodes in the network. We call such networks \emph{homogeneous}.


\subsection{Most Likely Infection Path Sources Estimator}

In this subsection, we present the MLIP statistical criterion that we adopt to find the optimal infection sources in this paper. We focus on the single source formulation in the following description, and extend to the multiple sources case in Section \ref{sec:multiple_sources_estimation}. The following exposition and definitions follow mainly from \cite{Luo2014}, and is repeated here for completeness.

Let $\bX(u,t)$ be a random variable denoting the state of a node $u$ in time slot $t$. At time 0, suppose that there is a single infected node $s^* \in V$, which we call the infection source. Let $\bX^t = \{ \bX(u,\tau): u \in V, 1\leq \tau \leq t\}$ be the collection of the states of all nodes in $V$ from time $1$ to $t$. A realization $X^t= \{X(u,\tau): u \in V, 1\leq \tau \leq t\}$ of $\bX^t$ is an \emph{infection path}. At time $t$, we observe the set of nodes that are currently infected. The observed set of infected nodes is denoted $V_\Inf$ and is assumed to be non-empty. We assume that the elapsed time $t$ is unknown. We say that an infection path $X^t$ is \emph{consistent} with $V_\Inf$ if we have $X(u,t) = \Inf$ for all $u \in V_\Inf$ and no other nodes in $V$ is infected in $X^t$. Conditioned on $s$ being the infection source, we let $\cX_s$ be the set of all possible infection paths consistent with $V_\Inf$, and $\cT_s$ be the set of the corresponding feasible elapsed times.

We want to estimate the infection source based only on knowledge of $V_\Inf$ and the underlying graph $G$. Finding the ML estimator for a single infection source in the SI model for a general graph network is a \#P-complete problem \cite{Shah2011}. \blue{(Note that \cite{Shah2011} considers a spreading model in which the propagation time of the infection across an edge has exponential distribution with rate 1. Due to the memoryless property of the exponential distribution, the problem of estimating the source in \cite{Shah2011}'s model can be reduced to the problem of estimating the source in a discrete time spreading model where time is discretized into unit intervals, and the probability of an infection spreading across an edge in each time slot is $1-e^{-1}$. Therefore, under the discrete time spreading model, finding the ML estimator is also a \#P-complete problem.)} We consider instead an alternative statistical criterion first proposed by \cite{Zhu2012}, and given by
\begin{align}
\hat{s} \in \arg \max_{\substack{s \in V \\ t \in \cT_s, X^t \in \cX_s}} \P(\bX^t = X^t \mid s^* = s).
\label{equ:proposed_single_source_estimator}
\end{align}
The basic idea behind \eqref{equ:proposed_single_source_estimator} is to estimate the source as the node associated with a \emph{most likely infection path} out of all possible infection paths that are consistent with $V_\Inf$. The search of a most likely infection path depends not only on the elapsed time but also on the structure of the underlying graph. Even at a given elapsed time, the number of consistent paths cannot be calculated easily, and the most likely infection paths are not unique. Solving \eqref{equ:proposed_single_source_estimator} directly involves searching over both $\cT_s$ and $\cX_s$, whose size increases exponentially fast with the number of nodes. In order to derive insights into an optimal source estimator for \eqref{equ:proposed_single_source_estimator}, we first consider the network with a single source in Section \ref{sec:single_source_estimation}. With the utilization of some properties of the elapsed times, we reduce the objective function to a simpler formulation and derive an estimator for all four considered infection spreading models. In Section \ref{sec:multiple_sources_estimation}, we generalize the idea to a tree network with multiple sources for the SI model.

\subsection{Some Notations and Definitions} \label{sec:notation}

\begin{table}[!t]
    \caption{Summary of notations}\label{table:notation}
    \centering
    \begin{tabular}{|c||l|}
    \hline
    $s^*$ & the true infection source \\ \hline
    $G = (V,E)$ & the underlying graph network \\ \hline
    $\psmin$ & $\min_{u \in V} \pSus{u}$ \\ \hline
    $\psmax$ & $\max_{u \in V} \pSus{u}$  \\ \hline
    $H_v$ & the minimum connected subgraph of $G$ that contains $V_\Inf$ and the node $v$\\ \hline
    $|A|$ & the number of elements in $A$ if $A$ is a set, or the number of nodes in $A$ if $A$ is a graph \\ \hline
    $V(u,i)$ & the set of nodes $i$ hops away from node $u$ \\ \hline
    $T_u(v;A)$ & the subtree rooted at node $u$ of the tree $A$, with the first link of the path from $u$ to $v$ in $A$ removed \\ \hline
    $d(s,u)$ & the length of the shortest path between $s$ and $u$ in the graph $G$ (i.e., the distance between them) \\ \hline
    $t_s$ & a most likely elapsed time conditioned on $s$ being the infection source \\ \hline
    $\cX_s$ & the set of all possible infection paths consistent with $V_\Inf$ conditioned on $s$ being the source \\ \hline
    $\cT_s$ & the set of the feasible elapsed time corresponding to $\cX_s$ \\ \hline
    \end{tabular}
\end{table}

In this subsection, we list some notations and definitions that we use throughout this paper. We refer the reader to a summary of basic notations given in Table \ref{table:notation}.

For a given tree network $A$ with $v$ being the root, we assign directions to each edge of $A$ so that all edges point towards $v$. For any $u \in A$, let $\parent{u}$ be the parent node of node $u$ (i.e., the node with an incoming edge from $u$), and $\children{u}$ be the set of child nodes of $u$ in $A$ (i.e., the set of nodes with outgoing edges to $u$).

For any infection path $X^t$, a subset $J \subset V$, and $0 \leq i \leq j \leq t$, let $X^t(J,[i,j])$ be the states of nodes in $J$ from time slots $i$ to $j$ in the infection path $X^t$. To avoid cluttered expressions, we abuse notations and let	
    	\begin{align*}
    	P_s\left(X^t(J,[i,j]) \right) \triangleq \P\Big(\bX^t(J,[i,j])=X^t(J,[i,j]) \mid s^* =s\Big).
    	\end{align*}
Therefore, $P_s(X^t)$ represents the probability of $X^t$ conditioned on $s$ being the source and $t$ being the elapsed time. Moreover, when we want to remind the reader of the state of a node $u$ at a specific time in the conditional probability $P_s(X^t)$, we use the notation $P_s(X(u,i) = a)$, where $a \in \{\Inf,\Sus,\Rec,\Non\}$ is the state of $u$ at time $i$.

\begin{definition}[Most likely infection paths]
For any $s\in V$ and any feasible elapsed time $t \in \cT_s$, we say that an infection path $X^t$ is most likely for $(s,t)$ if $X^t \in \arg\max_{\tX^t \in \cX_s} P_s(\tX^t)$. Moreover, an infection path $X^t$ is called a most likely infection path if there exists some $s \in V$, and $t \in \cT_s$ such that
\begin{align*}
P_s(X^t) = \max_{u \in V, r\in \cT_u, Y^{r} \in \cX_u} P_u(Y^{r}).
\end{align*}
\end{definition}

\begin{definition}[Jordan center]\label{def:jordan_infection_center}
For any node $s\in V$, let its infection range be
\begin{align*}
\bar{d}(s,V_\Inf) \triangleq \max_{u \in V_\Inf} d(s,u).
\end{align*}
Any node in $G$ with minimum infection range is called the Jordan center of $V_\Inf$.
\end{definition}

Finally, in several of our proofs, we need to differentiate between subtrees that have infected nodes or not.

\begin{definition}[Uninfected subtree and infected subtree]
Suppose that $v$ is the infection source. For any node $u$, we say that $T_u(v;G)$ is an \emph{uninfected subtree} if \footnote{See Table \ref{table:notation} for the definition of $T_u(v;G)$.}
	\begin{align*}
	T_u(v;G) \bigcap V_\Inf = \emptyset;
	\end{align*}
 and we say that $T_u(v;G)$ is an \emph{infected subtree} if
    \begin{align*}
	T_u(v;G) \bigcap V_\Inf \ne \emptyset.
	\end{align*}
\end{definition}

\section{Single Source Estimation for Trees}\label{sec:single_source_estimation}
In this section, we show that a Jordan center of the infected node set $V_\Inf$ is an optimal infection source estimator universally applicable for infection spreading under the SI, SIR, and SIRI models for trees, and the SIS model for regular trees. The Jordan center has previously been shown to be optimal estimators for SI infection spreading \cite{Luo2014} and for the SIR model \cite{Zhu2012,Zhu2013}, but under different technical assumptions.

As noted in Section \ref{sec:Introduction}, proving optimality results for infection source estimators is in general challenging. In most of this paper, we restrict ourselves to the following specific graph networks depending on the infection spreading model. We say that a tree is an \emph{infinite tree} if every node in it has degree at least two.

\begin{assumption}
\label{assump:topology_infinite_regular}
For an infection spreading according to the SI, SIR or SIRI models, the underlying graph $G$ is an infinite tree. For an infection spreading according to the SIS model, the underlying graph $G$ is a regular infinite tree, i.e., every node has the same degree.
\end{assumption}

For the SI, SIR, and SIRI models, Assumption \ref{assump:topology_infinite_regular} is adopted to avoid boundary effects. Consider the extreme case where a source node has only one neighbor.  Then, the infection can spread away from the source in only one direction. In this case, any estimator based only on the graph topology is expected to perform badly. In the SIS model, a recovered node is the same as a susceptible node, which leads to more complex evolution of the node states in the network as compared to the SIRI model in which the state evolution of a recovered node becomes independent from the rest of the network. To simplify the problem, we restrict to regular trees for the SIS model in Assumption \ref{assump:topology_infinite_regular}. The problem of finding optimal source estimators for the SIS model in more general network topologies remains open.

\subsection{Most Likely Elapsed Time}
We assume no knowledge of the elapsed time when the set of nodes $V_\Inf$ is observed. Suppose that $v \in V$ is the source, then the feasible set of all elapsed times is given by $\cT_v = [\bar{d}(v, V_\Inf), +\infty)$, where the lower bound is the minimum amount of time required for the infection to spread from $v$ to all the nodes in $V_\Inf$. It is obviously computationally inefficient to search over all elapsed times. In Proposition \ref{lemma:optimal_t}, we show how to find a \textit{most likely elapsed time} $t_v$ that maximizes the probability of observing $V_\Inf$.

\begin{proposition}\label{lemma:optimal_t}
Suppose that Assumptions \ref{assump:infection_prob_SI}--\ref{assump:infection_prob_SIS} hold, $v \in V$ is the infection source, and a non-empty set of infected nodes $V_\Inf$ is observed.  For an infection under the SI, SIR, SIRI or SIS model in a network satisfying Assumption \ref{assump:topology_infinite_regular}, we have for any $t \in \cT_v$, and any two most likely infection paths $X^t$ for $(v,t)$ and $Y^{t+1}$ for $(v,t+1)$,
\begin{enumerate}[(a)]
      \item \label{lemma:optimal_t_monotonically_decreasing} $P_v(Y^{t+1}) \le \delta P_v(X^t)$, where $\delta = (1-\psmin)^2, \sqrt{\frac{\psmin}{\psmax}}, \sqrt{\frac{\psmin}{\psmax}}$ and 1 for the SI, SIR, SIRI and SIS model, respectively; and
      \item \label{lemma:optimal_t_optimal_t} conditioned on $v$ being the infection source, a most likely elapsed time is given by
      \begin{align*}
      t_v=\bar{d}(v,V_\Inf).
      \end{align*}
\end{enumerate}
\end{proposition}

The proof of Proposition \ref{lemma:optimal_t} is provided in Appendix \ref{appendix:lemma:optimal_t}. Proposition \ref{lemma:optimal_t}\eqref{lemma:optimal_t_optimal_t} shows a universal property that is robust to the underlying infection spreading models: a most likely elapsed time $t_v$ is the infection range of $v$ (cf. Definition \ref{def:jordan_infection_center}). Moreover, Proposition \ref{lemma:optimal_t}\eqref{lemma:optimal_t_monotonically_decreasing} shows that a most likely elapsed time should be as small as possible. This result is intuitive. Consider the conditional probability
\begin{align*}
P_v(X^t) = \prod_{u \in V, \tau \in [1, t]} P_v(u,\tau),
\end{align*}
where the value of each term in the product on the right hand side is at most 1. When $t$ decreases, there are less terms in the product, which in turn increases the value of $P_v(X^t)$.

Following Proposition \ref{lemma:optimal_t}, the problem in \eqref{equ:proposed_single_source_estimator} is now reduced to
\begin{align*}
\hat{s} \in \arg \max_{\substack{v \in V, t_v = \bar{d}(v,V_\Inf) \\ X^{t_v} \in \cX_v}} P_v(X^{t_v}).
\end{align*}
After the most likely elapsed time has been identified, we can now proceed to find the source node associated with the most likely infection path.

\subsection{Source Associated With the Most Likely Infection Path}

In this subsection, we derive the source estimator associated with a most likely infection path for all four considered infection spreading models, under specific graph networks. Although Proposition \ref{lemma:optimal_t} gives a most likely elapsed time $t_v$ conditioned on a node $v \in V$ being the infection source, it is still difficult to count the number of infection paths that are consistent with $V_\Inf$, not to mention finding the most likely infection path for $(v, t_v)$. Therefore, instead of directly looking for the most likely infection path, we first consider the conditional probabilities $P_v(X^{t_v})$ and $P_u(Y^{t_u})$ of two infection paths, where $v$ and $u$ are a pair of neighboring nodes, $X^{t_v}$ is a most likely infection path for $(v, t_v)$, and $Y^{t_u}$ is a most likely infection path for $(u, t_u)$. We then show that if $v$ has a smaller infection range, $P_v(X^{t_v})$ is not less than $P_u(Y^{t_u})$. Upon establishing this neighboring node relationship, we can find a path on which the infection range of each node is decreasing, and the conditional probability of the most likely infection path is non-decreasing. This in turn implies that the Jordan center of $V_\Inf$ is the source estimator we are looking for. The neighboring node relationship is summarized in Proposition \ref{lemma:better_neighbor}, the proof of which is provided in Appendix \ref{appendix:lemma:better_neighbor}.

\begin{proposition}\label{lemma:better_neighbor}
Suppose that $V_\Inf$ is non-empty. For an infection process under the SI, SIR, SIRI or SIS model satisfying Assumptions \ref{assump:infection_prob_SI}-\ref{assump:topology_infinite_regular}, and for any pair of neighboring nodes $u$ and $v$, we have
\begin{align*}
P_v(X^{t_v}) \ge P_u(Y^{t_u}), \text{ if } t_v < t_u,
\end{align*}
where $X^{t_v}$ and $Y^{t_u}$ are most likely infection paths for $(v,t_v)$ and $(u,t_u)$ respectively.
\end{proposition}

We note that Proposition \ref{lemma:optimal_t} and Proposition \ref{lemma:better_neighbor} match Proposition 2 and Lemma 4 in \cite{Luo2014}, respectively. Then following the same proof as Theorem 1 in \cite{Luo2014}, we have the following result.

\begin{theorem}\label{theorem:single_source_estimate_Jordan_infection_center}
Suppose that $V_\Inf$ is non-empty. For an infection process under the SI, SIR, SIRI or SIS model satisfying Assumptions \ref{assump:infection_prob_SI}--\ref{assump:infection_prob_SIS}, respectively, and Assumption \ref{assump:topology_infinite_regular} holds, a Jordan center of $V_\Inf$ is an optimal source estimator for \eqref{equ:proposed_single_source_estimator}.
\end{theorem}

Theorem \ref{theorem:single_source_estimate_Jordan_infection_center} shows that for regular infinite trees, a Jordan center is an optimal source estimator, regardless of which of the four considered infection spreading model the infection is following. This is a somewhat surprising result since the four infection spreading models are fundamentally different. The ``universality'' of the Jordan center makes it highly desirable in practice, where the underlying infection spreading model is usually unknown a priori. A distributed linear time complexity algorithm has been proposed in \cite{Luo2014} to find the Jordan center in a tree, which makes timely estimation of the infection source possible.

\begin{figure}[!ht] 
  \centering
  \psfrag{a}[][][0.8][0]{$a$}
  \psfrag{c}[][][0.8][0]{$b$}
  \psfrag{e}[][][0.8][0]{$c$}
  \psfrag{m}[][][0.8][0]{$d$}
  \psfrag{n}[][][0.8][0]{$e$}
  \psfrag{o}[][][0.8][0]{$f$}
  \psfrag{v}[][][0.8][0]{$g$}
  \psfrag{r}[][][0.8][0]{$\pSus{i}$:}
  \psfrag{s}[][][0.8][0]{$\psmax$}
  \psfrag{x}[][][0.8][0]{$\psmin$}
  \includegraphics[width=0.6\textwidth]{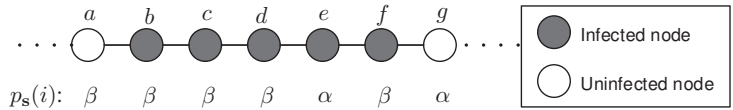}
  \caption{\blue{An example that shows Assumption \ref{assump:infection_prob_SI} is a necessary condition for Theorem \ref{theorem:single_source_estimate_Jordan_infection_center} to hold for the SI infection process. Suppose the infection process follows the SI model. If node $d$ (Jordan center of $V_\Inf$) is the infection source, for an infection path $X^2$ with elapsed time $2$, we have $P_d(X^2) = \psmax^3 \psmin$. On the other hand, if node $e$ is the infection source, for an infection path $Y^3$ with elapsed time $3$ and in which node $f$ is infected in the first time slot, we have $P_e(Y^3) = \psmax^4 (1-\psmin)^2$. For $d$ to be the optimal MLIP estimator, we require $P_d(X^2) \ge P_e(Y^3)$, which in turn requires $\psmax \le \psmin / (1-\psmin)^2$.}}
  \label{fig:example_for_assump_infection_prob_SI}
\end{figure}

\blue{In Fig. \ref{fig:example_for_assump_infection_prob_SI}, we provide an example to show that Assumption \ref{assump:infection_prob_SI} is a necessary condition for Theorem \ref{theorem:single_source_estimate_Jordan_infection_center} to hold for the SI infection process. Similar examples can be used to show that Assumptions \ref{assump:infection_prob_SIR}, \ref{assump:infection_prob_SIRI} and \ref{assump:infection_prob_SIS} are necessary for Theorem \ref{theorem:single_source_estimate_Jordan_infection_center} to hold for the SIR, SIRI and SIS infection process, respectively.}

\section{Multiple Sources Estimation for SI Infection Spreading in Trees} \label{sec:multiple_sources_estimation}

In this section, we restrict our discussion to an infection spreading under the SI model. Since the optimal single infection source estimator has been shown to be the Jordan center of $V_\Inf$ in Section \ref{sec:single_source_estimation}, we consider here the case where there are $k > 1$ infection sources, i.e., $S^* = \{s^*_1, s^*_2, \cdots, s^*_k\}$. Then the most likely infection path based sources estimation problem becomes
\begin{align}
\hat{S} \in \arg \max_{\substack{S \subset V, |S|=k \\ t \in \cT_S, X^t \in \cX_S}} \P(\bX^t = X^t \mid S^* = S).
\label{equ:proposed_multiple_sources_estimator}
\end{align}

The definitions in Section \ref{sec:notation} are similarly generalized to the case of $k$ infection sources by replacing $s$ with $S$ in the definitions. In particular, we generalize the Jordan center definition to $k$-Jordan infection center set.

\begin{definition}[$k$-Jordan center set]\label{def:jordan_infection_centers}
The infection range of a set of source nodes $S=\{s_1, s_2, \cdots, s_k\}$ is defined as
\begin{align*}
\bar{d}(S,V_\Inf) = \max_{u \in V_\Inf} \min_{s_i \in S} d(s_i, u).
\end{align*}
The set of $k$ nodes in $G$ with minimum infection range is called the $k$-Jordan center set of $V_\Inf$.
\end{definition}

Without loss of generality, we assume that the minimum subgraph $B$ of $G$ that contains $V_\Inf$ is connected, otherwise the same estimation procedure can be applied to each component of $B$. We first show a similar result as that in Proposition \ref{lemma:optimal_t}. The proof of Proposition \ref{lemma:optimal_t_k_sources} is provided in Appendix \ref{appendix:lemma:optimal_t_k_sources}.

\begin{proposition}\label{lemma:optimal_t_k_sources}
Suppose that the underlying network $G$ is an infinite tree, the infection sources are $S= \{s_1, s_2, \cdots, s_k\}$, and the set of observed infected nodes $V_\Inf$ is non-empty. For an infection spreading under the SI model, any most likely infection path $X^t$ for $(S,t)$ has the following properties:
\begin{enumerate}[(a)]
  \item \label{lemma:optimal_t_k_sources_monotonically_decreasing}$P_{S}(X^t)$ is non-increasing in $t \in \cT_{S}$; and
  \item \label{lemma:optimal_t_k_sources_optimal_t} conditioned on $S$ being the infection sources, a most likely elapsed time for $X^t$ is given by
  \begin{align*}
  t_{S}=\bar{d}(S,V_\Inf).
  \end{align*}
\end{enumerate}
\end{proposition}

\begin{figure}[!ht] 
  \centering
  \psfrag{a}[][][0.8][0]{$s_1$}
  \psfrag{b}[][][0.8][0]{$s_2$}
  \psfrag{c}[][][0.8][0]{$m_1$}
  \psfrag{d}[][][0.8][0]{$m_{i-1}$}
  \psfrag{u}[][][0.8][0]{$m_i$}
  \psfrag{v}[][][0.8][0]{$m_{i+1}$}
  \psfrag{e}[][][0.8][0]{$m_{i+2}$}
  \psfrag{h}[][][0.8][0]{$m_p$}
  \psfrag{i}[][][0.8][0]{$n_1$}
  \psfrag{k}[][][0.8][0]{$n_{k_1}$}
  \psfrag{m}[][][0.8][0]{$n_{k_1+1}$}
  \psfrag{n}[][][0.8][0]{$n_{k_1+k_2}$}
  \psfrag{o}[][][0.8][0]{$B$}
  \psfrag{r}[][][0.8][0]{$C$}
  \psfrag{s}[l][][0.8][0]{$\sn{S}$}
  \psfrag{1}[][][1.2][0]{$A_1$}
  \psfrag{2}[][][1.2][0]{$A_2$}
  \subfigure[Given any infection path $X^t$, suppose $m_i$ becomes susceptible at time $\tau_1$ in $X^t$ for the first time. Suppose $m_{i+1}$ becomes susceptible at time $\tau_2 > \tau_1$ in $X^t$ for the first time when $m_{i+2}$ becomes infected, while $m_i$ stays susceptible at time $\tau_2$. Then nodes in $B$ belong to component $A_1$ and nodes in $C$ belong to component $A_2$.]{
    \label{subfig:two_source_network}
    \includegraphics[width=0.6\textwidth]{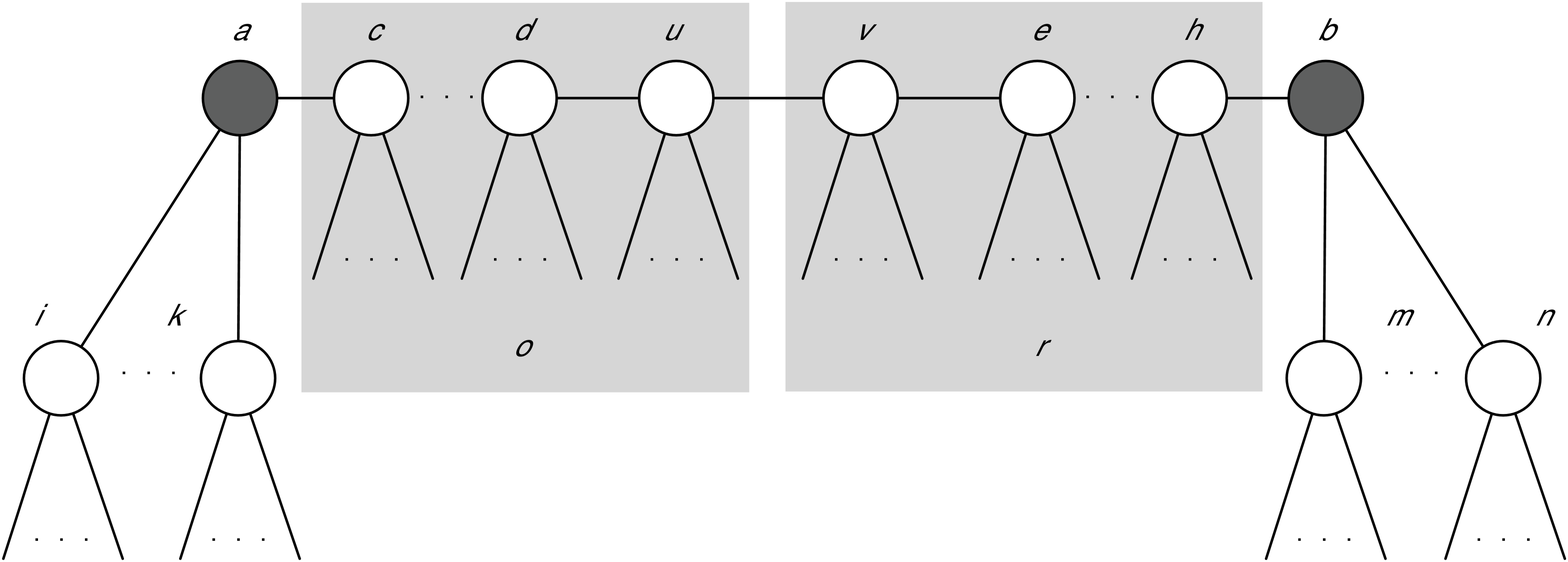}}
  \hspace{0.1in}
  \subfigure[Partition $G$ into $\cA = A_1 \bigcup A_2$.]{
    \label{subfig:two_source_network_divide}
    \includegraphics[width=0.6\textwidth]{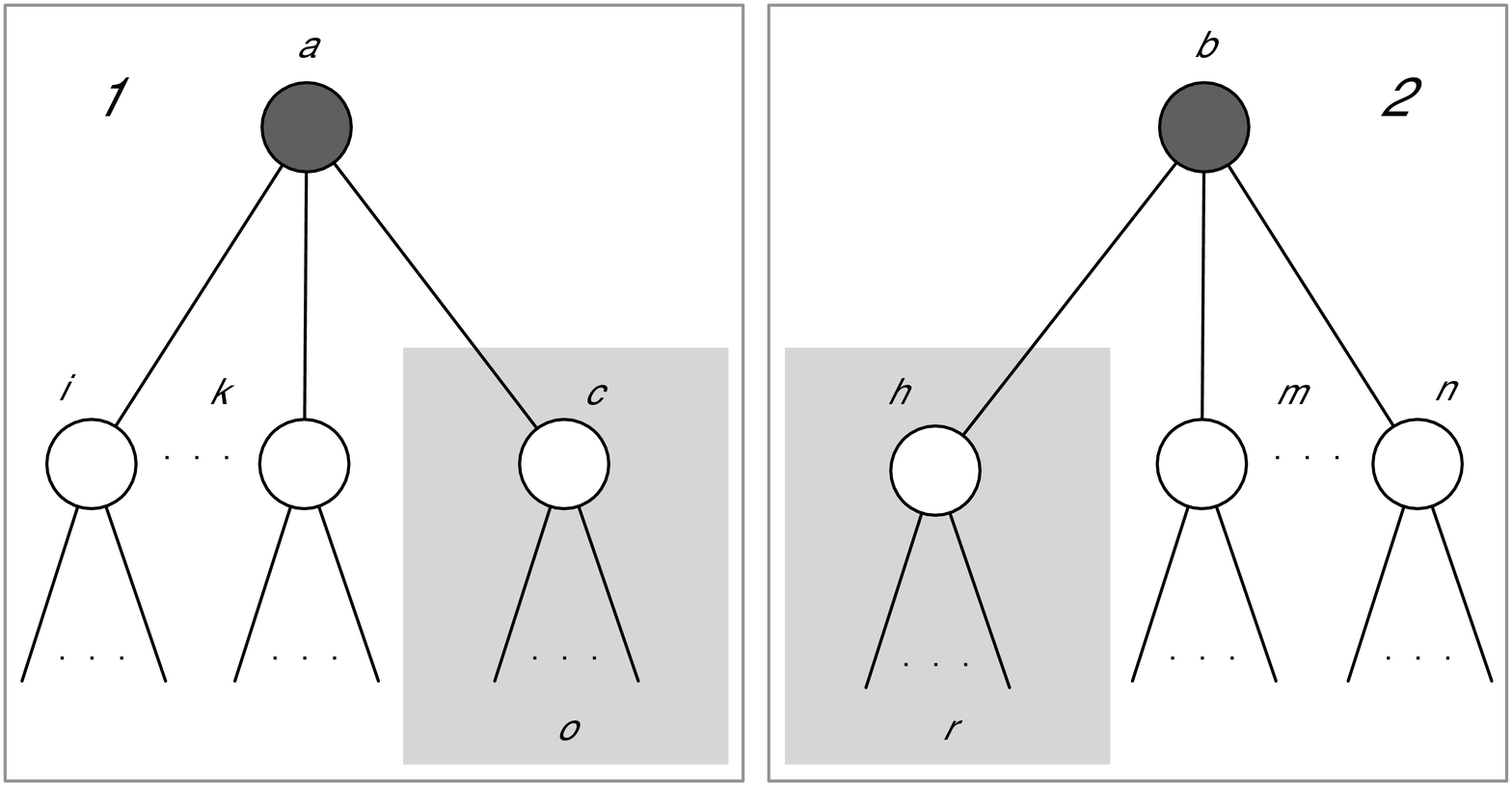}}
  \hspace{0.1in}
  \subfigure[Constructed super node graph $\sng$.]{
    \label{subfig:two_source_network_super}
    \includegraphics[width=0.6\textwidth]{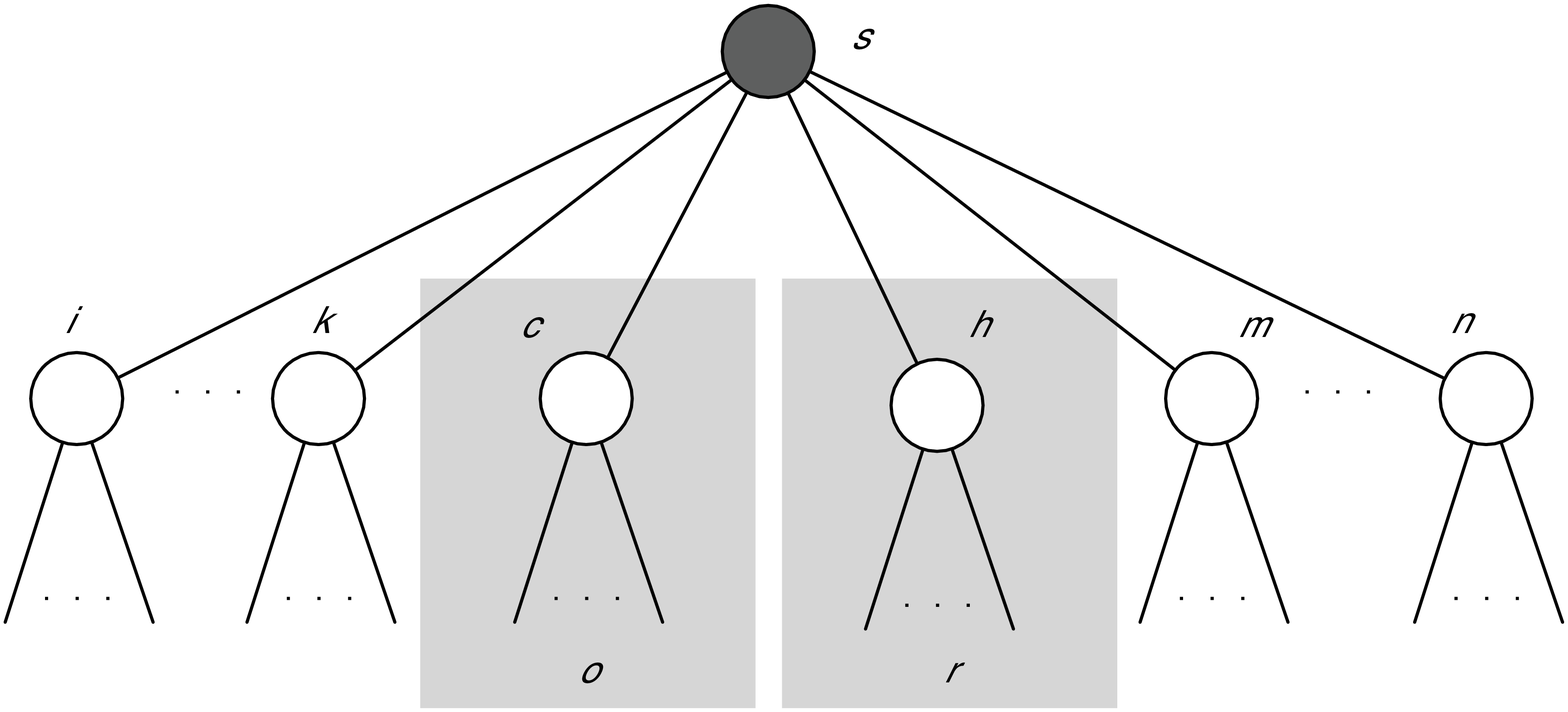}}
\caption{\blue{Illustration of the construction of the super node graph $\sng$ from an infinite tree $G$ with $S=\{s_1,s_2\}$ being the infection sources.}}
\label{fig:super_node_graph}
\end{figure}

In the following, we show how to transform the $k$ sources estimation problem to an equivalent single source estimation problem, then we can use Theorem \ref{theorem:single_source_estimate_Jordan_infection_center} to find the optimal multiple sources estimator. We first introduce the definition of \emph{super node graph}. \blue{See Fig.~\ref{fig:super_node_graph} for an illustration of the super node graph construction}.

\begin{definition}[Super node graph]\label{def:super_node_graph_transformation}
Suppose that $G$ is an infinite tree. Given a set $S=\{s_1,s_2, \cdots, s_{k}\} \subset V$, where $k>1$, and any infection path $X^t$ conditioned on $S$ being the infection sources, the super node graph $\sng$ is constructed using the following procedure for each $\tau = 0, 1, \ldots, t$:
\begin{itemize}
	\item Starting at $\tau = 0$, we initialize $A_i=\{ s_i \}$ for each $i=1,\ldots,k$.
	\item For each $\tau = 1, \ldots, t$, consider every node $v \in V_\Inf$ that becomes susceptible at time $\tau$ in $X^t$ for the first time. Let $N_v$ be the set of neighboring nodes of $v$ that is infected at time $\tau-1$. We choose a node $u \in N_v$ uniformly at random, and include $v$ and the edge $(u,v)$ in the component $A_i$ that $u$ belongs to.
	\item Based on the resulting graph $\cA=\bigcup_{i=1}^{k} A_i$, the \textit{super node graph $\sng$} is constructed by considering all infection sources as a single virtual node, which we call a super node and denote as $\sn{S}$.
\end{itemize}
\end{definition}

\blue{In summary, we trace the infection path $X^t$ and assign each infected node to the tree $A_i$ if its infection comes from $s_i$, with ties broken randomly. This then partitions the infection graph $G$ into disjoint trees rooted at each $s_i \in S$. The trees are connected together to form the super node graph by treating $S$ as a single ``super node''.}

Given any infection path $X^t$ following the SI model, it can be shown that (with probability one) the conditional probability $P_S(X^t)$ is the same for $G$ and any corresponding $\sng$ as defined in Definition \ref{def:super_node_graph_transformation}. Consider any node $v$ with $|N_v|>1$ and assume $v$ becomes susceptible at time slot $t_1$ and becomes infected at time slot $t_2$. Then $P_S(X^t(v,[1,t])) = (1-p_{\Sus})^{t_2-t_1-1} p_{\Sus}$, regardless of the number of infected neighbors $v$ has as long as there is at least one infected neighbor.\footnote{\label{footnote:infection_partition_not_hold}This property does not hold for an infection following the SIR, SIRI or SIS model, where some infected neighbors of $v$ may recover after $t_1$ and $P_S(X^t(v,[1,t]))$ may change if we remove some edges connecting $v$.} We formally present this result in the following lemma.

\begin{lemma}\label{lemma:super_node_graph_equivalency}
Let $S=\{s_1,s_2, \cdots, s_{k}\}\subset V$, where  $k>1$. Given any infection path $X^t$ conditioned on $S$ being the infection sources, $P_S(X^t)$ is the same for both $G$ and any corresponding $\sng$ as defined in Definition \ref{def:super_node_graph_transformation}.
\end{lemma}

Following Lemma \ref{lemma:super_node_graph_equivalency}, instead of searching for a most likely infection path for $S$ in $G$, we can now search for a most likely infection path for $\sn{S}$ in a corresponding super node graph $\sng$. In this way, we transform the $k$ sources estimation problem to an equivalent single source estimation problem. As discussed in Section \ref{sec:single_source_estimation}, Theorem \ref{theorem:single_source_estimate_Jordan_infection_center} shows that a Jordan center of the infected node set is an optimal single source estimator. Therefore, our objective is to find a set of $k$ nodes $S$, where $\sn{S}$ is a Jordan center of the infected node set in $\sng$. We show in the following lemma that $k$-Jordan center set is the solution.

\begin{lemma}\label{lemma:k_jordan_centers_equivalency}
Suppose that $G$ is an infinite tree and the set of infected nodes $V_\Inf$ is non-empty. Given any infection path $X^t$ consistent with $V_\Inf$ under the SI model, if $S=\{s_1,s_2, \cdots, s_{k}\}$ is the $k$-Jordan center set of $V_\Inf$ in $G$, then $\sn{S}$ is a Jordan center of $V_\Inf$ in any corresponding super node graph $\sng$.
\end{lemma}

The proof of Lemma \ref{lemma:k_jordan_centers_equivalency} is provided in Appendix \ref{appendix:lemma:k_jordan_centers_equivalency}. The following theorem follows immediately from Lemma \ref{lemma:k_jordan_centers_equivalency} and Theorem \ref{theorem:single_source_estimate_Jordan_infection_center}.

\begin{theorem}\label{theorem:multiple_source_estimate_Jordan_infection_center}
Suppose that $G$ is an infinite tree and there are $k>1$ infection sources. For an infection in the SI model, a $k$-Jordan center set of $V_\Inf$ is an optimal source set estimator for \eqref{equ:proposed_multiple_sources_estimator}.
\end{theorem}

Theorem \ref{theorem:multiple_source_estimate_Jordan_infection_center} is consistent with Theorem \ref{theorem:single_source_estimate_Jordan_infection_center} for an infection in the SI model. Due to the difficulty described in footnote \ref{footnote:infection_partition_not_hold}, the multiple-sources estimation problem remains an open problem for more complicated infection spreading models including SIR, SIRI and SIS models. To verify the robustness of the proposed estimators, we conduct extensive simulations on both trees and general networks for SI, SIR, SIRI and SIS models in Section \ref{sec:simulation_results}.

\section{Source Estimation for General Graphs} \label{sec:sources_estimation_for_general_graphs}
In this section, we consider the case where the underlying network $G$ is a general graph. Inspired by the robustness of Jordan center estimators in tree networks, we heuristically extend them to general graphs. We first review an algorithm, proposed in \cite{Zhu2012}, that finds the Jordan center for $k=1$. We then propose a heuristic algorithm to find the $k$-Jordan center set for $k>1$.

\subsection{Single Jordan Center Estimation Algorithm}
A simple algorithm was proposed in \cite{Zhu2012} to find the Jordan center of $V_\Inf$ when there is a single source and the underlying network is a general graph. Let any node in $V_\Inf$ broadcast a message containing its own identity. The first node that receives messages from every node in $V_\Inf$ declares itself as a Jordan center and the algorithm terminates. We call this algorithm the Single Jordan Center estimation algorithm (SJC), with a computational complexity of $O(|V||E|)$.

\subsection{Multiple Jordan Center Set Estimation Algorithm}
When $k$ is greater than 1, it is usually impractical to use exhaustive search methods to find the $k$-Jordan center set as the number of possible $k$-Jordan center sets is $\binom{|V|}{k}$. Therefore, we propose a heuristic algorithm to find an approximate $k$-Jordan center set when there are $k>1$ sources and the underlying network is a general graph, which we call the Multiple Jordan Center set estimation  algorithm (MJC). MJC starts with randomly selecting a set of $k$ nodes $\hat{S}^0 = \{s^0_i\}_{i=1}^k$ as the initial guess, and then utilizes an iterative two-step optimization approach. Specifically, in iteration $l$, let $\hat{S}^l = \{s^l_i\}_{i=1}^k$ be the $k$-Jordan center set estimate. We perform the following two steps at each iteration $l$:
\begin{itemize}
	\item \textbf{Partition step}. In this step, MJC partitions $V_\Inf$ into $k$ sets $M_1, M_2, \cdots, M_k$ such that for all $v\in M_i$, $d(s_i^{l-1}, v) \le d(s_j^{l-1}, v)$ if $i \ne j$. We call $M_i$ the \emph{Voronoi set} corresponding to $s_i^{l-1}$. To do this, let each $s_i^{l-1}$ broadcast a message. The broadcasting process terminates when each node $v \in V_\Inf$ receives at least one message from a node in $\hat{S}^{l-1}$. In the broadcasting process, each node $v \in V_\Inf$ learns the distance between itself and the nearest nodes in $\hat{S}^{l-1}$. We choose a nearest node in $\hat{S}^{l-1}$ at random, and add $v$ to the Voronoi set corresponding to this node.
	\item \textbf{Re-optimization step}. In this step, MJC updates each estimate $s_i^{l-1}$ in the Voronoi sets $M_i$. For each Voronoi set $M_i$, run SJC to find the Jordan center of $M_i$ and set it as the new estimate $s_i^{l}$.
\end{itemize}
MJC terminates when $\max_{1 \le i \le k} d(s_i^{l-1}, s_i^{l}) \le \eta$ for some predetermined small positive value $\eta$ or when the number of iterations reach a predetermined positive number MaxIter. For the partition step in each iteration, the computation complexity is dominated by the broadcasting process, with a computational complexity of $O(k|E|)$. For the re-optimization step in each iteration, the computational complexity is $O(|V||E|)$ due to SJC. Therefore, the overall computational complexity for MJC is $O(\text{MaxIter} \cdot |V||E|)$. \blue{We show in the following proposition that the infection range is non-increasing over the iterations of MJC. The proof of Proposition \ref{lemma:MJC_non_increasing} is provided in Appendix \ref{appendix:lemma:MJC_non_increasing}.

\begin{proposition}\label{lemma:MJC_non_increasing}
Suppose that $G$ is a general graph and there are $k>1$ infection sources. The infection range (cf. Definition \ref{def:jordan_infection_centers}) of is non-increasing over the iterations of MJC, i.e.,
    \begin{align*}
      \bar{d}(\hat{S}^l,V_\Inf) \le \bar{d}(\hat{S}^{l-1},V_\Inf).
    \end{align*}
\end{proposition}
}

\section{Simulation Results}\label{sec:simulation_results}
In this section, we present simulation results using both synthetic and real world networks to evaluate the performance of the proposed estimators. We simulate infection spreading under the SI, SIR, SIRI and SIS models in both homogeneous and heterogeneous networks, and for single and multiple infection sources.

\subsection{Single Infection Source}
When there is a single infection source, we use the following six common centrality measures and random guessing as benchmarks to compare with our estimator. The first four definitions are the same as those in \cite{Luo2014}, and are repeated here for the convenience of the reader.
\begin{enumerate}[(i)]
  \item The betweenness center (BC) is defined as
    \begin{align*}
        \text{BC} \triangleq \arg \max_{v \in G} \sum_{i, j \in V_\Inf, i \ne j \ne v} \frac{\sigma_{ij}(v)}{\sigma_{ij}},
    \end{align*}
    where $\sigma_{ij}$ is the number of shortest paths between node $i$ and node $j$, and $\sigma_{ij}(v)$ is the number of those shortest paths that contain $v$.
  \item The closeness center (CC) is defined as
    \begin{align*}
        \text{CC} \triangleq \arg \max_{v \in G} \sum_{i \in V_\Inf, i \ne v} \frac{1}{d(v, i)}.
    \end{align*}
  \item The distance center (DisC) is defined as
    \begin{align*}
        \text{DisC} \triangleq \arg \min_{v \in G} \sum_{i \in V_\Inf} d(v, i).
    \end{align*}
    \blue{For trees, the DisC is the same as the rumor center defined in \cite{Shah2011}, and it is shown in \cite{Shah2011} that the DisC is the ML estimator for regular trees under the SI model with a single source.}
  \item \blue{Let $H$ denote the minimum connected subgraph of $G$ that contains $V_\Inf$. The degree center (DegC) is defined as
       \begin{align*}
        \text{DegC} \triangleq \arg \max_{v \in H} |N_H(v)|,
    \end{align*}
      where $N_H(v)$ is the set of neighbors of $v$ in $H$, and $|N_H(v)|$ is defined to be the degree centrality of $v$.}
  \item \blue{The eigenvector centrality function of $H$ is a function $\text{EC}: H \mapsto \Real$ such that for any node $v$ in $H$, we have
    \begin{align*}
        \text{EC}(v) = \frac{1}{\lambda} \sum_{i \in N_H(v)} \text{EC}(i),
    \end{align*}
    where $N_H(v)$ is the set of neighbors of $v$ in $H$, and $\lambda$ is a constant. Then the eigenvector center (EC) is the node in $H$ with maximum eigenvector centrality.}
  \item \blue{The pagerank centrality of any node $v$ in $H$ is defined as
    \begin{align*}
        \text{PC}(v) = d \sum_{i \in N_H(v)} \frac{\text{PC}(i)}{|N_H(i)|} + \frac{1-d}{|H|},
    \end{align*}
    where $N_H(v)$ is the set of neighbors of $v$ in $H$, and $d$ is a damping factor in $(0,1)$. Then the pagerank center (PC) is the node in $H$ with maximum pagerank centrality.}
  \item \blue{The random guess estimator randomly selects a node in $H$ as the source estimator.}
\end{enumerate}

We evaluate the performance of our proposed estimator on three kinds of networks: random tree networks where the degree of every node is randomly chosen from $[3,5]$, a small part of the Facebook network with 4039 nodes \cite{McAuley2012} and the western states power grid network of the United States \cite{Watts1998}. We consider both homogeneous and heterogeneous networks. In the homogeneous networks, we vary the recovery and relapse probabilities to demonstrate the impact of these spreading parameters on the performance of the proposed estimator. In the heterogeneous networks, we evaluate the robustness of the proposed estimator on a wide range of randomly generated spreading parameters. In the following, we describe the four different simulation experiments.

\subsubsection{SI and SIRI models in homogeneous networks}

For every $v\in V$, we let $\pSus{v}=p_{\Sus}$, $\pInf{v}=p_{\Inf}$ and $\pRec{v}=p_{\Rec}$, where the infection probabilities are set as follows: $p_{\Sus}$ is randomly chosen from $[0,1]$, $p_{\Inf}$ is set to be $0.1, 0.2, \cdots, 1$, respectively, and $p_{\Rec}$ is randomly chosen from $[0, \min\{1, \frac{p_{\Inf}}{1-p_{\Inf}}\}]$. For each kind of network and each value of $p_{\Inf}$, we perform 1000 simulation runs. In each simulation run, we randomly pick a node as the infection source and simulate the infection using the above parameters. The spreading terminates when the number of infected nodes is greater than 100. We then run SJC on the observed infected nodes to estimate the infection source and compare the result with the benchmarks.

\begin{figure}[!ht]
  \centering
  \subfigure[Random trees.]{
    \includegraphics[width=0.45\textwidth]{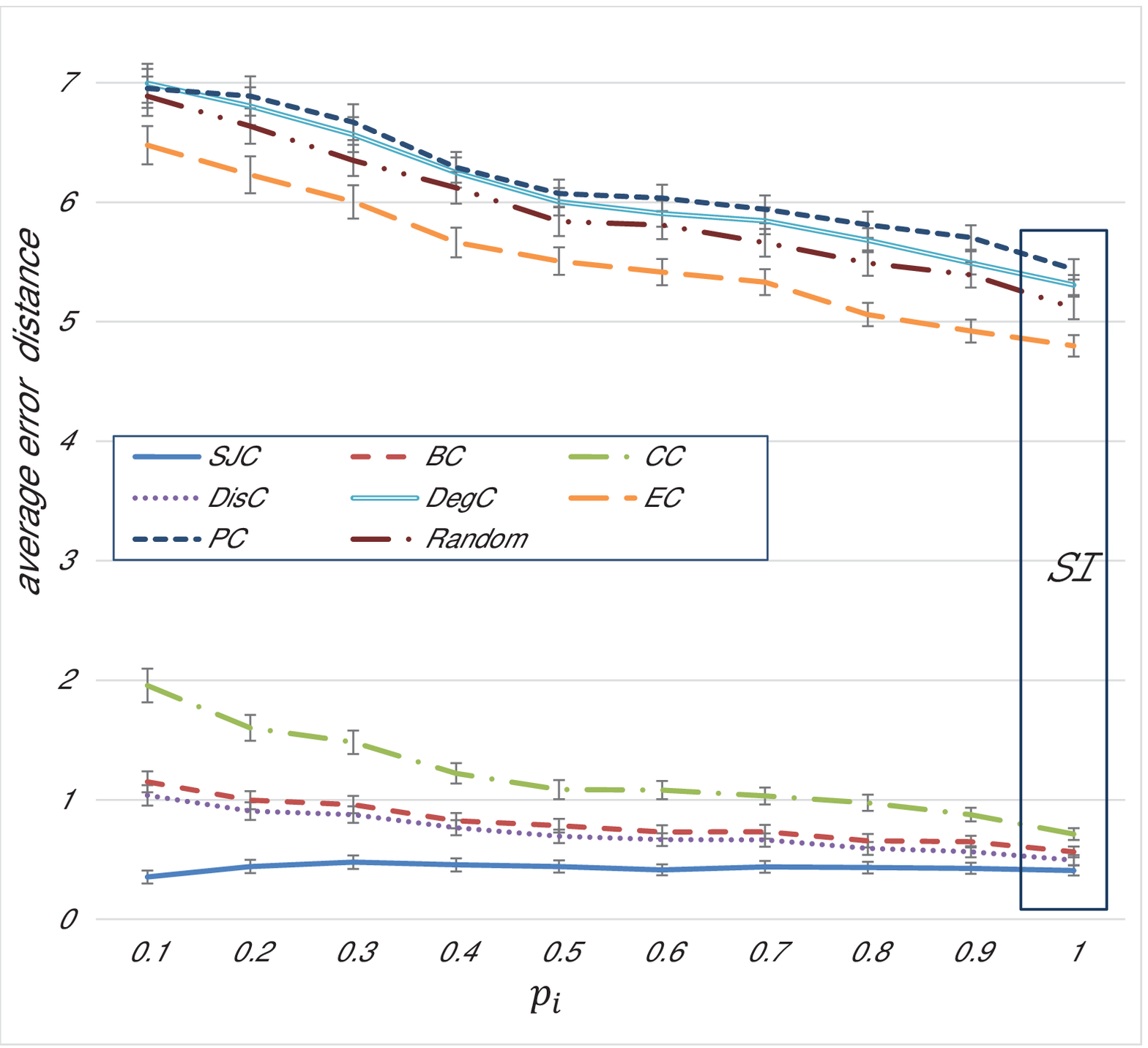}}
  \hspace{0.1in}
  \subfigure[Facebook network.]{
    \includegraphics[width=0.45\textwidth]{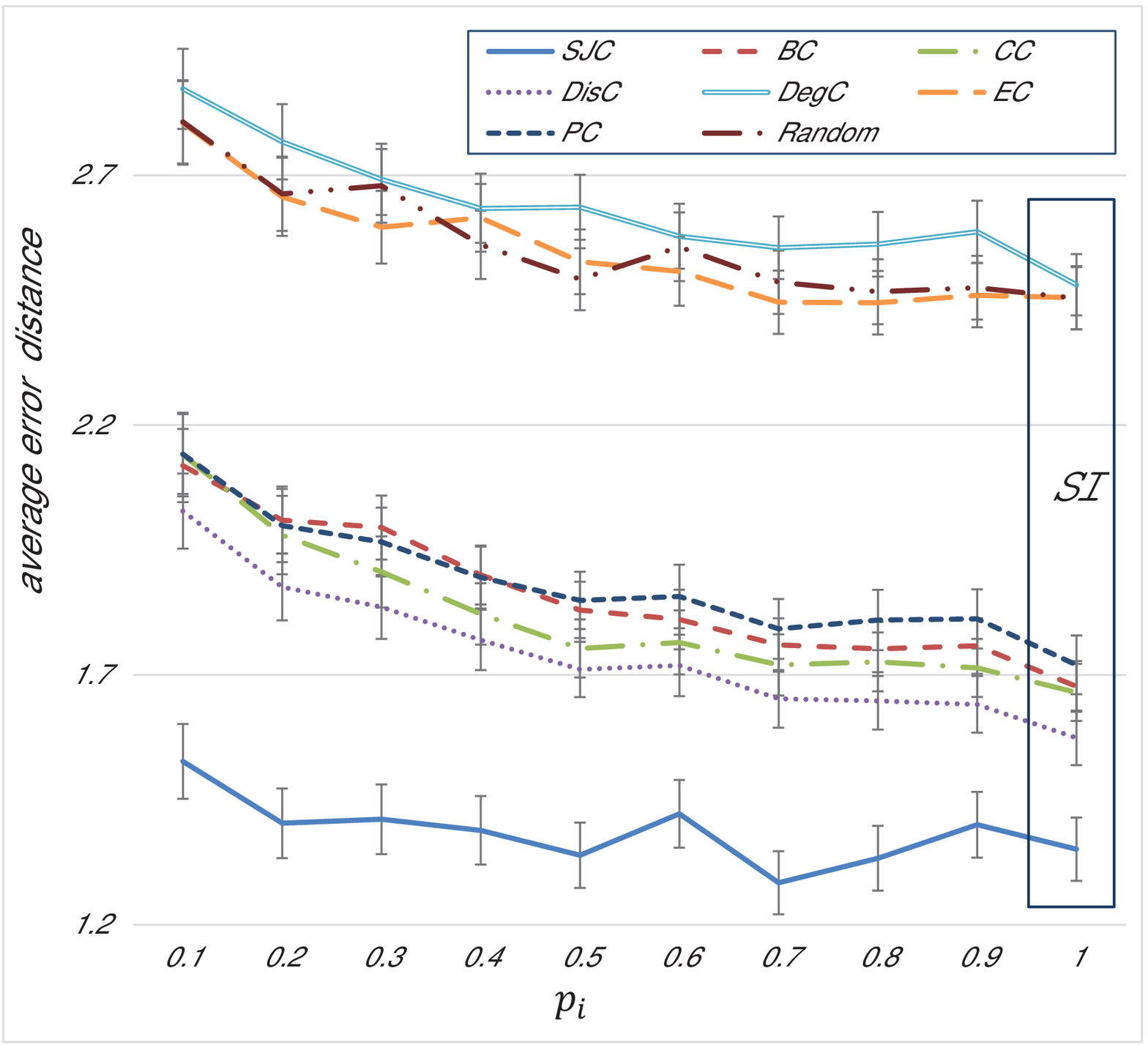}}
  \hspace{0.1in}
  \subfigure[Power grid network.]{
    \includegraphics[width=0.45\textwidth]{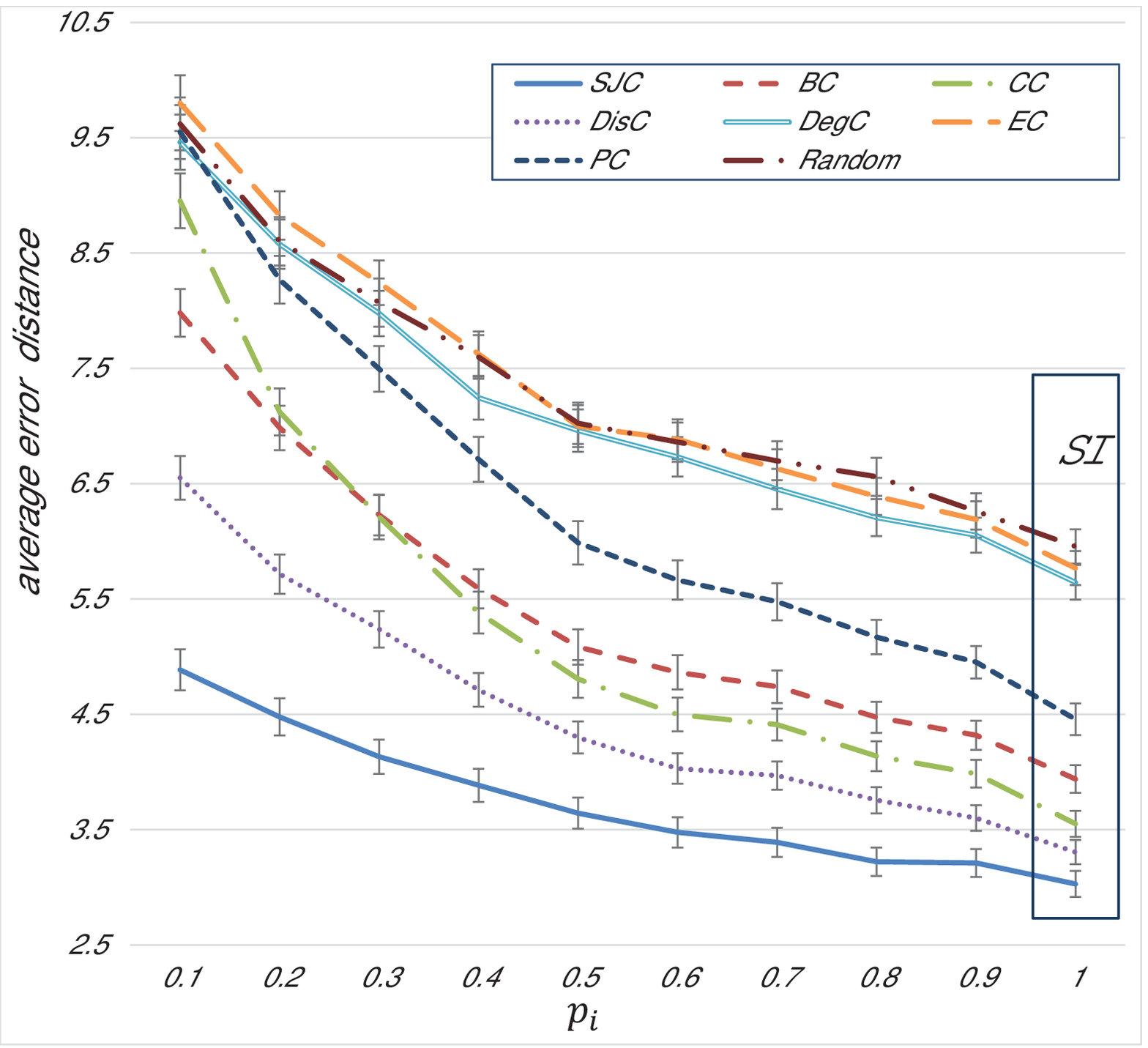}}
\caption{\blue{Average error distances with error bars of 95\% confidence interval for various networks and different values of $p_{\Inf}$ in homogeneous networks. The underlying infection follows the SIRI model for all values of $p_{\Inf}$ and the infection follows the SI model when $p_{\Inf} =1$.}}
\label{fig:error_distance_SIRI_pi}
\end{figure}

The error distance is the number of hops between the estimated and the actual infection source, and is shown in Fig. \ref{fig:error_distance_SIRI_pi}. We see that the proposed estimator performs consistently better than the benchmarks for all considered networks. \blue{The random guess estimator actually performs better or comparable with some estimators like DegC and EC as these estimators only capture the local ``connectivity'' of a node, instead of the topological information inherent in the network.}

\subsubsection{SIR and SIRI models in homogeneous networks}
\begin{figure}[!ht]
  \centering
  \subfigure[Random trees.]{
    \includegraphics[width=0.45\textwidth]{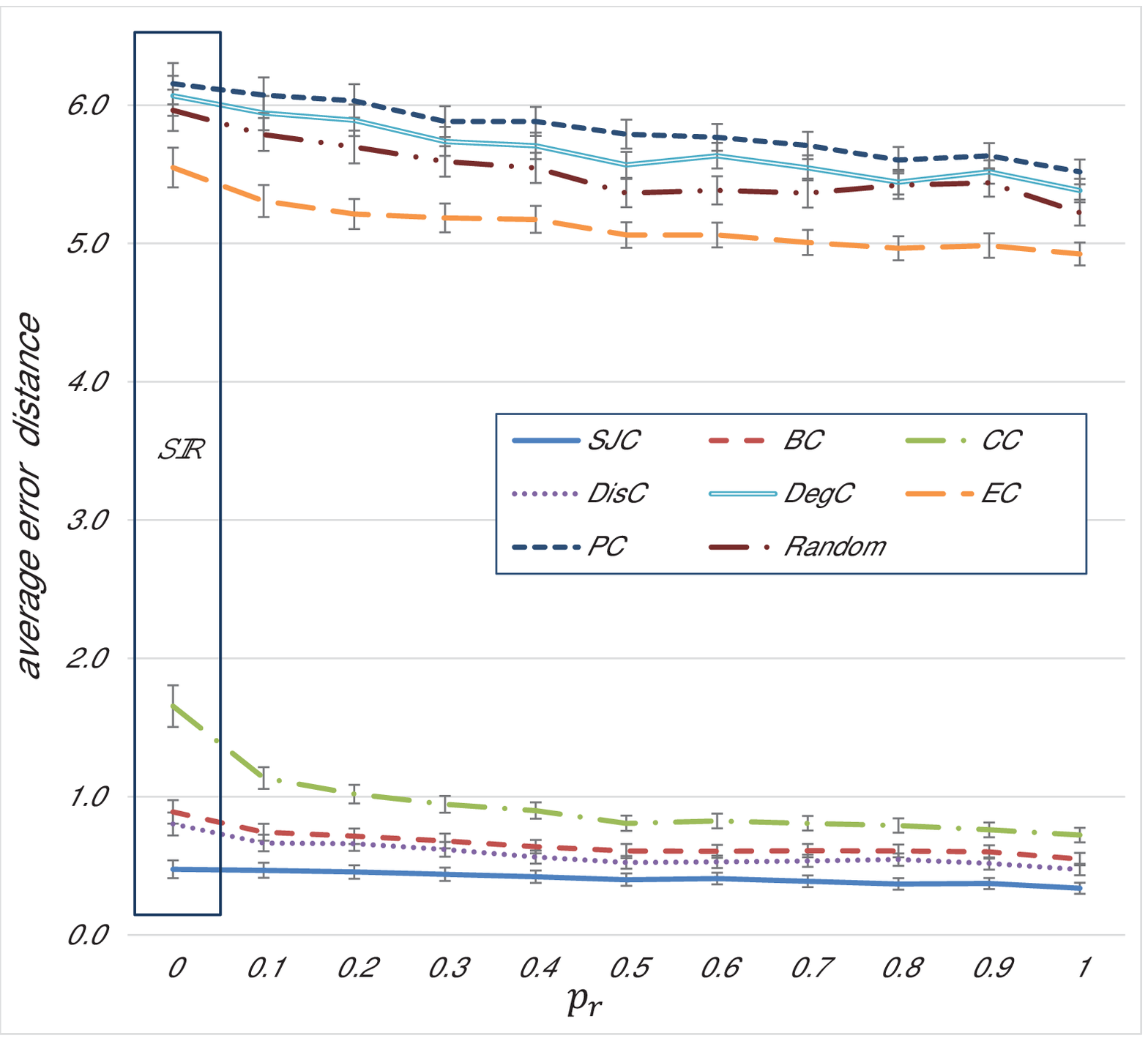}}
  \hspace{0.1in}
  \subfigure[Facebook network.]{
    \includegraphics[width=0.45\textwidth]{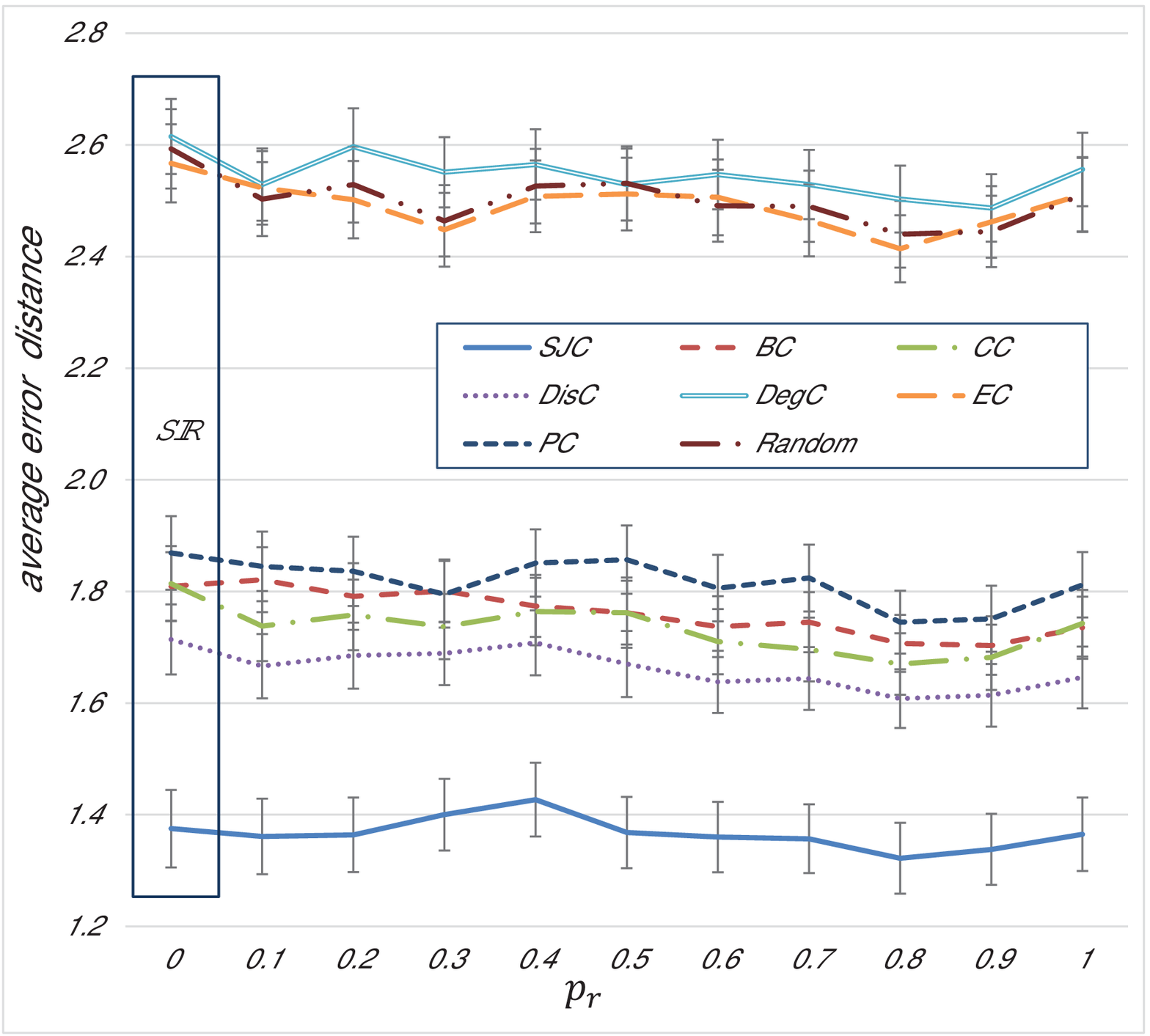}}
  \hspace{0.1in}
  \subfigure[Power grid network.]{
    \includegraphics[width=0.45\textwidth]{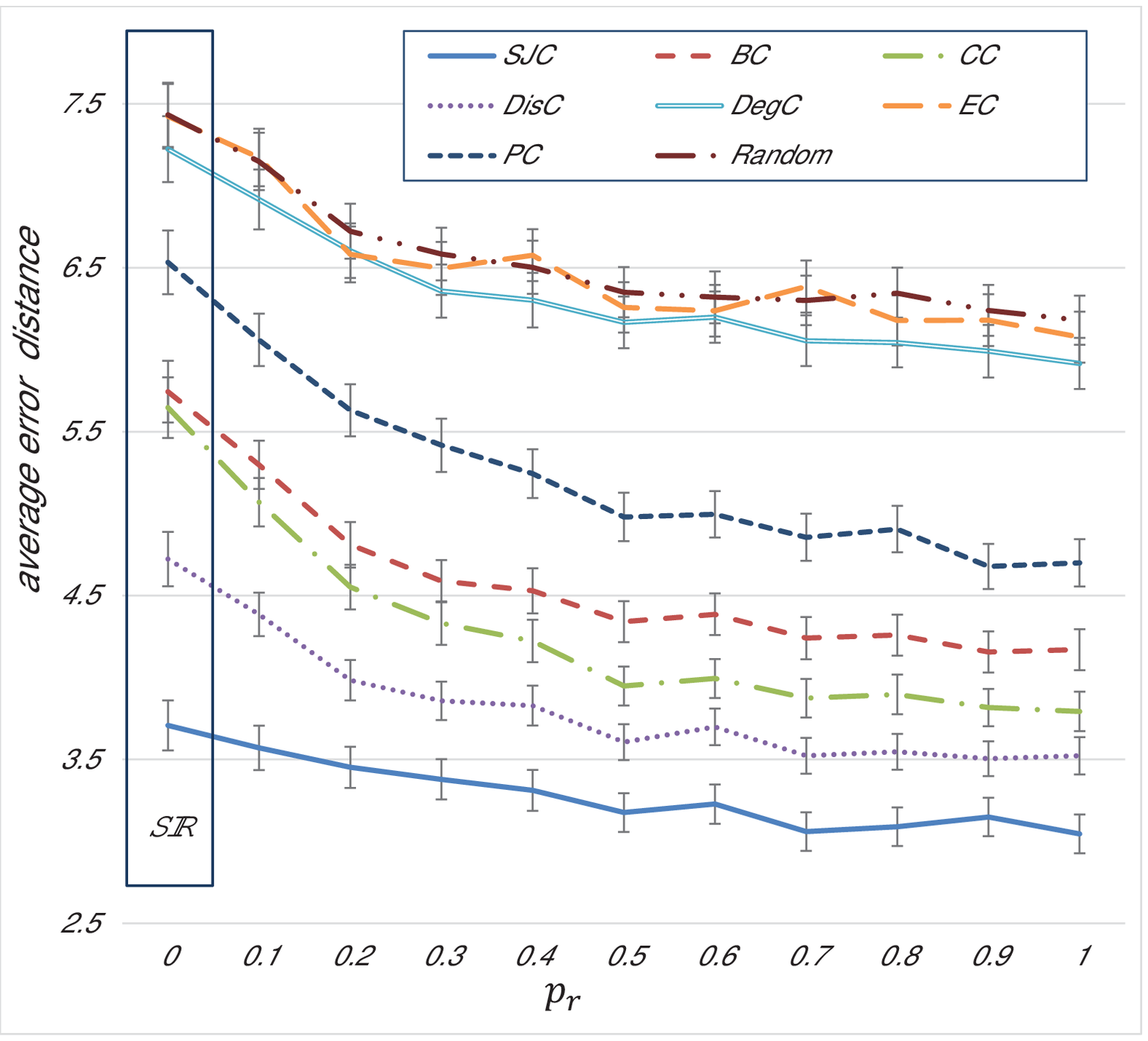}}
\caption{\blue{Average error distances with error bars of 95\% confidence interval for various networks and different values of $p_{\Rec}$ in homogeneous networks. The underlying infection follows the SIRI model for all values of $p_{\Rec}$ and the infection follows the SIR model when $p_{\Rec} =0$.}}
\label{fig:error_distance_SIRI_pr}
\end{figure}

The infection probabilities are set as follows: $p_{\Sus}$ is randomly chosen from $[0,1]$, $p_{\Inf}$ is randomly chosen from $[0.5, 1]$, and $p_{\Rec}$ is set to be $0, 0.1, \cdots, 1$. We compare the performances in Fig. \ref{fig:error_distance_SIRI_pr}. We see that our proposed estimator again performs consistently better than the benchmarks.

\subsubsection{SIS model in homogeneous networks}
\begin{figure}[!ht]
  \centering
  \subfigure[Random trees.]{
    \includegraphics[width=0.45\textwidth]{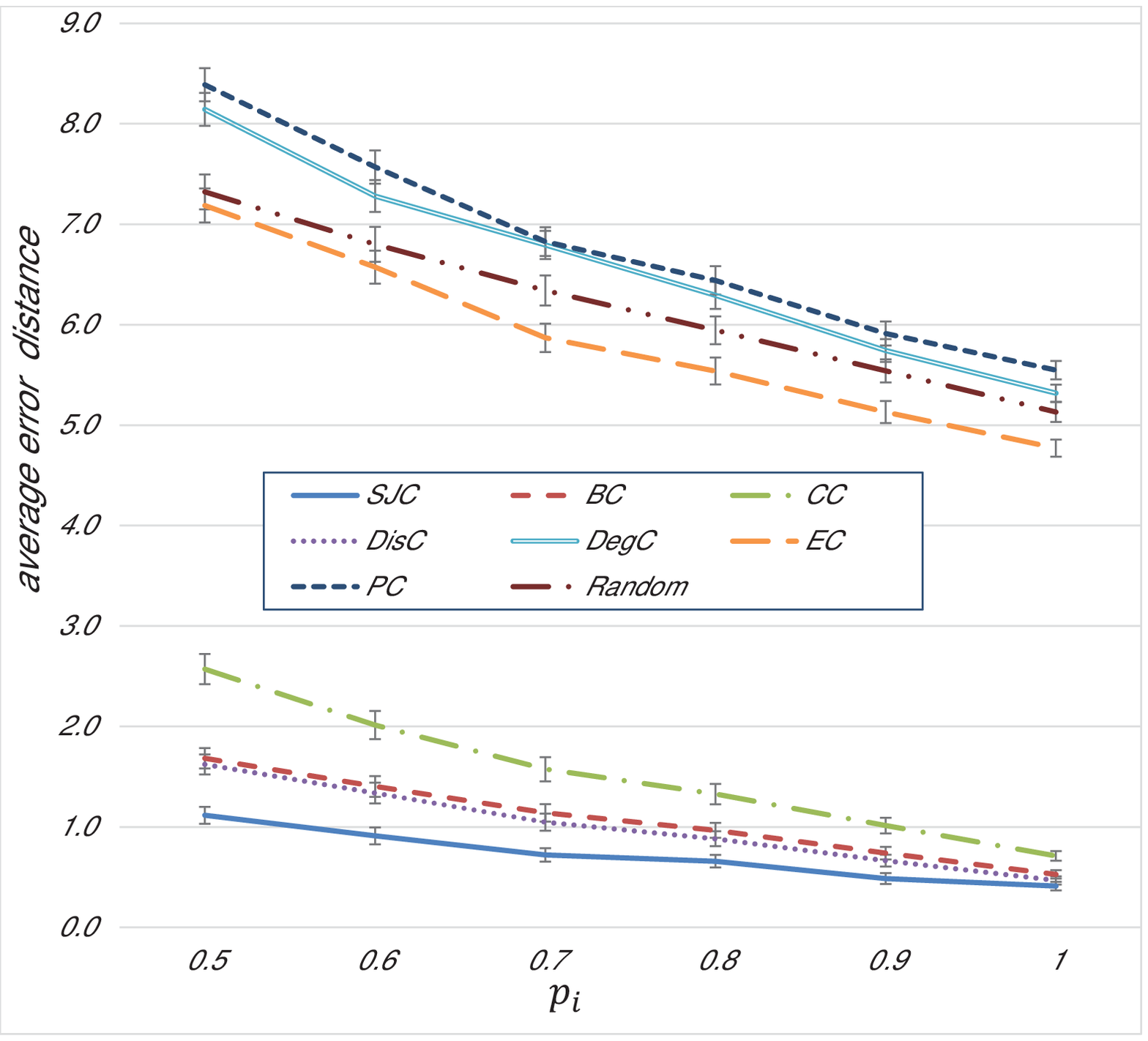}}
  \hspace{0.1in}
  \subfigure[Facebook network.]{
    \includegraphics[width=0.45\textwidth]{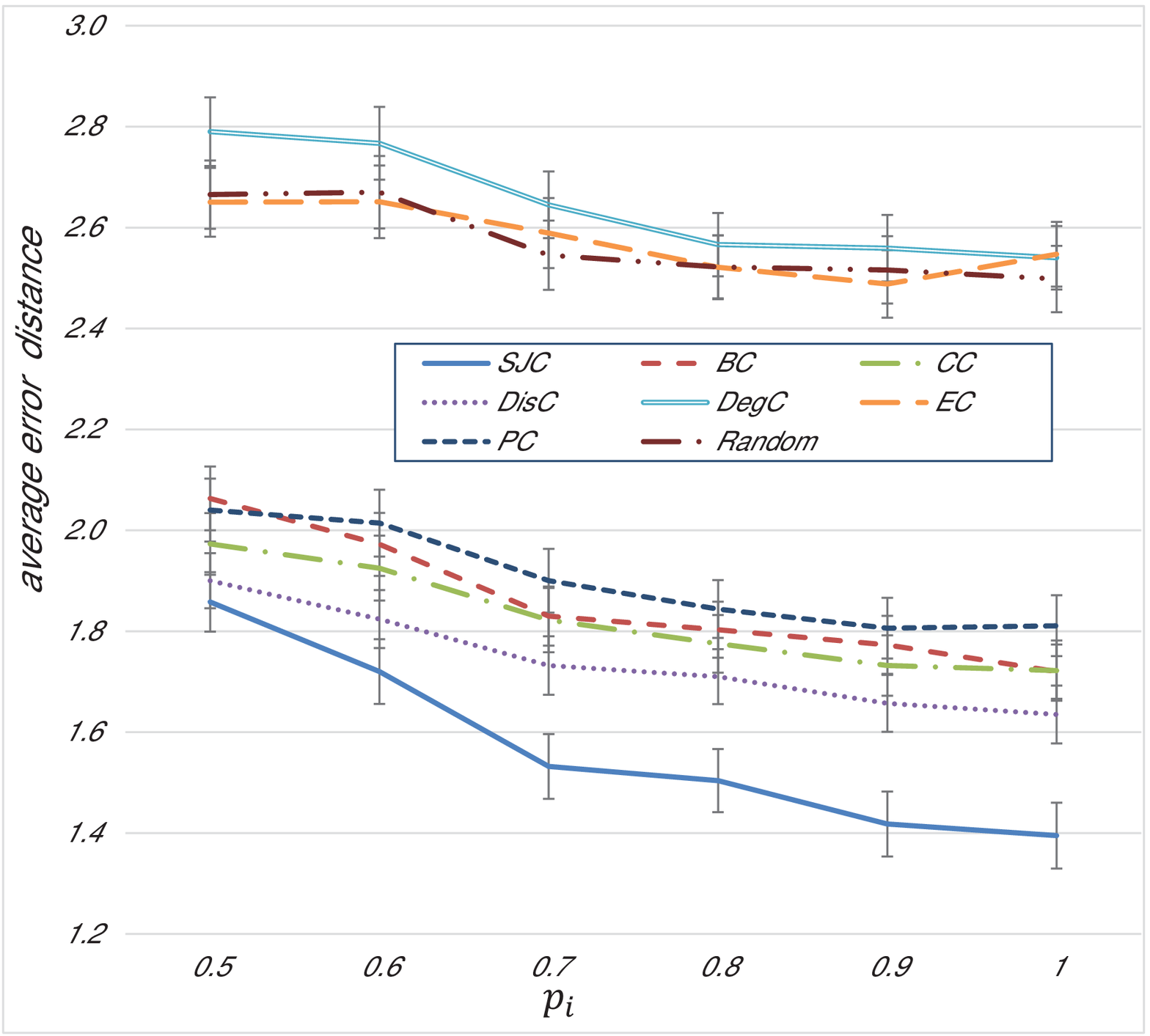}}
  \hspace{0.1in}
  \subfigure[Power grid network.]{
    \includegraphics[width=0.45\textwidth]{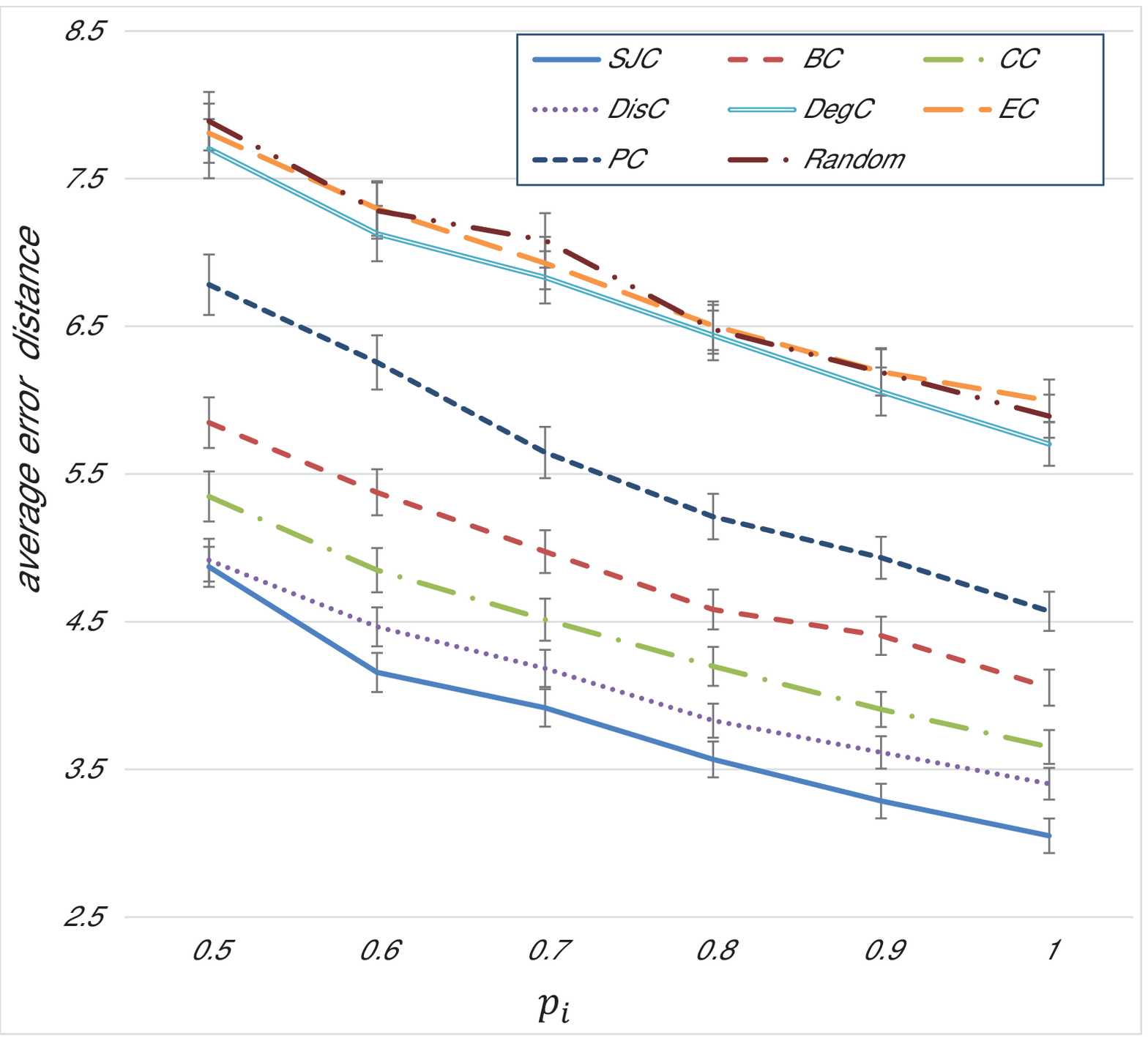}}
\caption{\blue{Average error distances with error bars of 95\% confidence interval for various networks and different values of $p_{\Inf}$ in homogeneous networks under the SIS model.}}
\label{fig:error_distance_SIS_pi}
\end{figure}

We consider the SIS model where $p_{\Inf}$ is set to be $0.5, 0.6, \cdots, 1$, respectively, and $p_{\Sus}$ is randomly chosen from $[0,p_{\Inf}]$. In Fig. \ref{fig:error_distance_SIS_pi}, we observe that our proposed estimator always results in smaller average error distances than the benchmarks for all considered networks.

\subsubsection{SI, SIR, SIRI, and SIS models in heterogeneous networks}

\begin{figure}[!ht]
  \centering
  \subfigure[SI model.]{
    \includegraphics[width=0.45\textwidth]{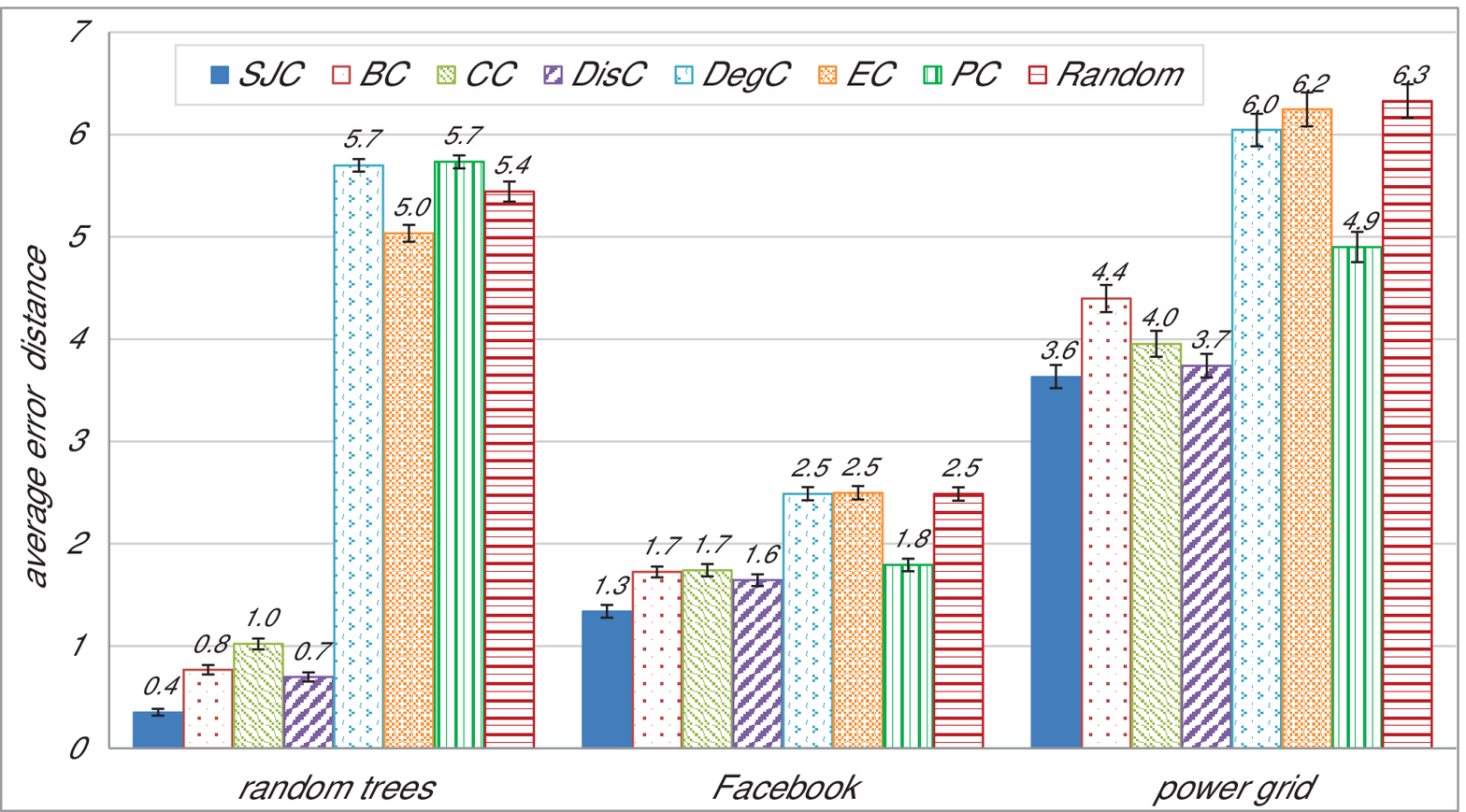}}
  \hspace{0.1in}
  \subfigure[SIR model.]{
    \includegraphics[width=0.45\textwidth]{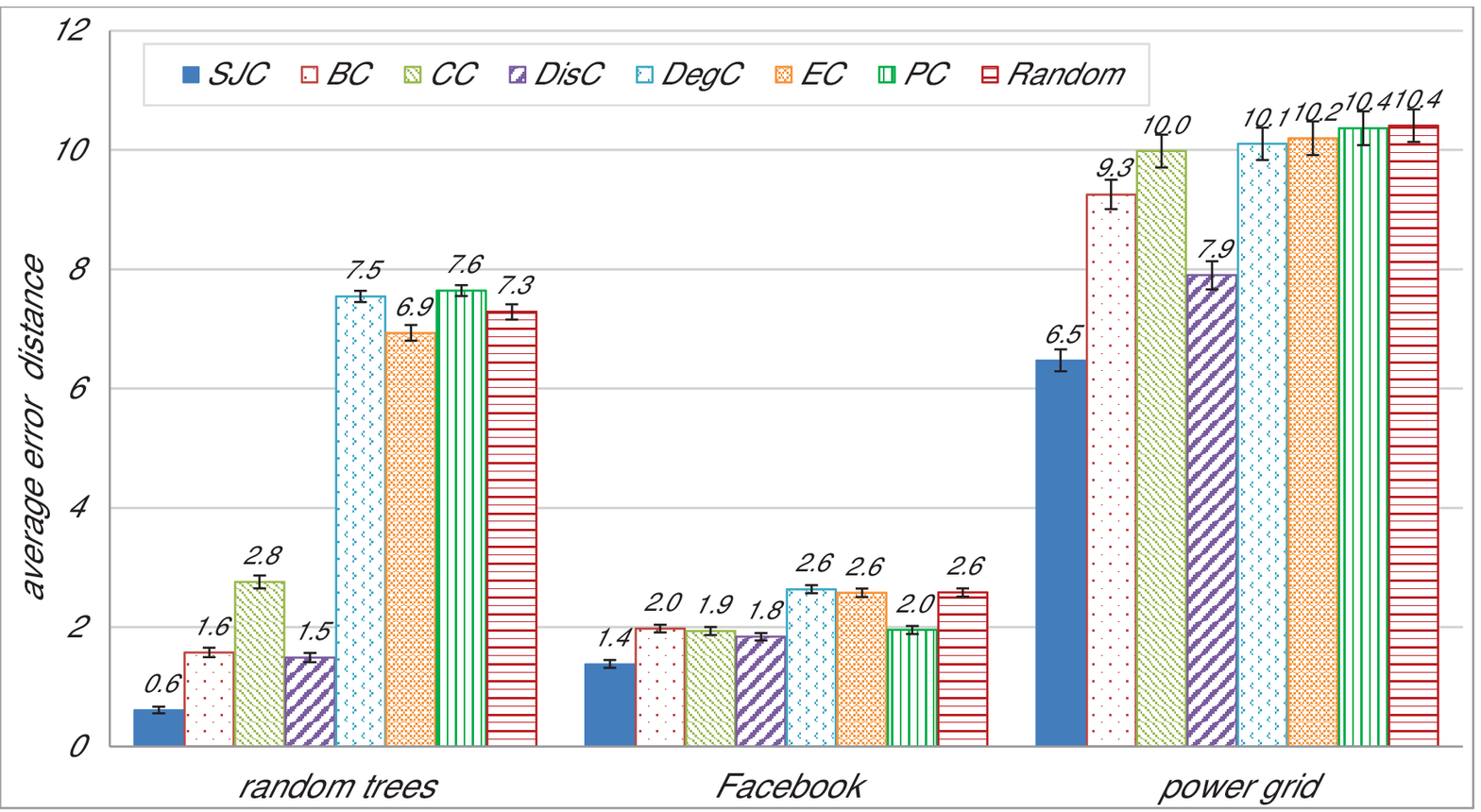}}
  \hspace{0.1in}
  \subfigure[SIRI model.]{
    \includegraphics[width=0.45\textwidth]{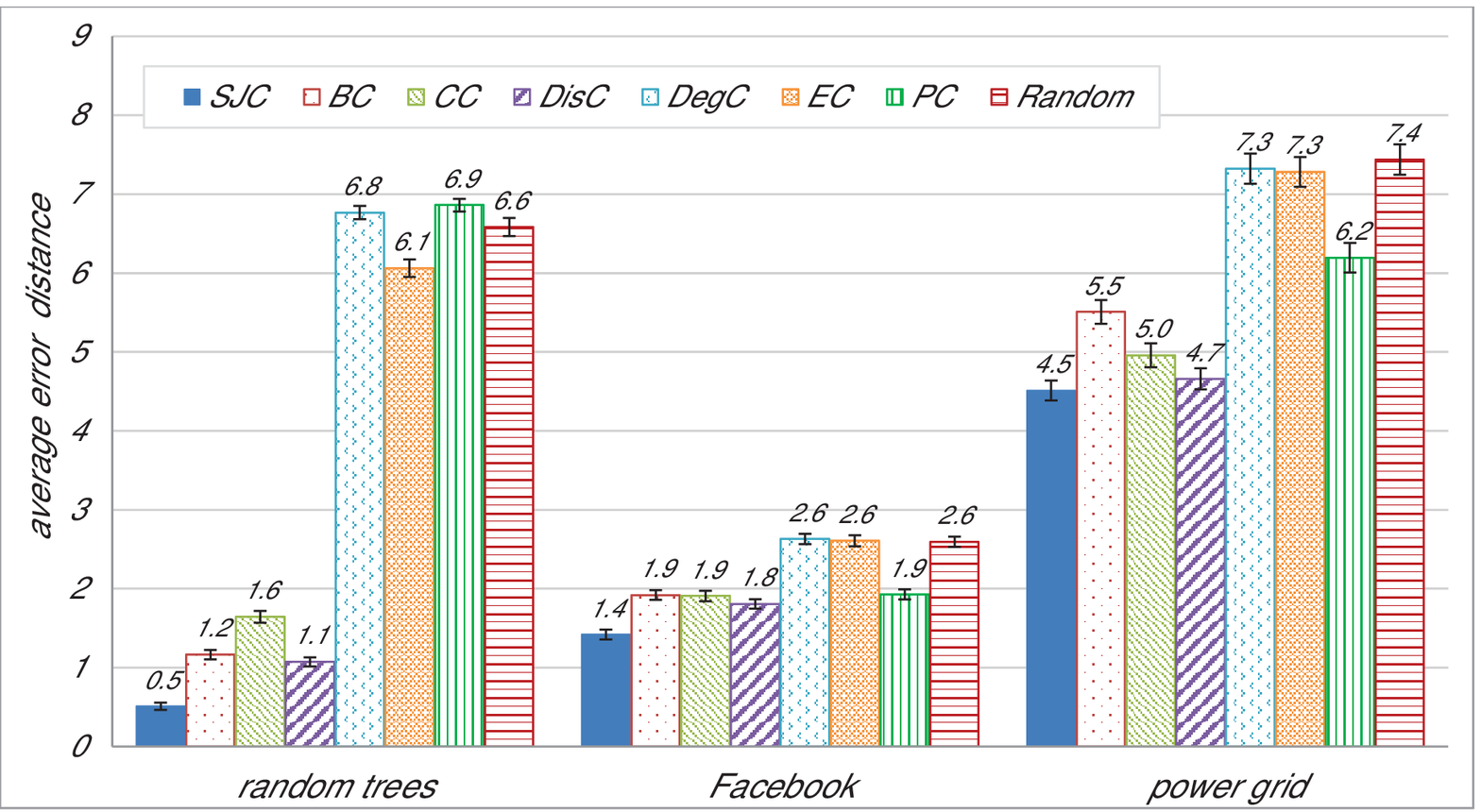}}
  \hspace{0.1in}
  \subfigure[SIS model.]{
    \includegraphics[width=0.45\textwidth]{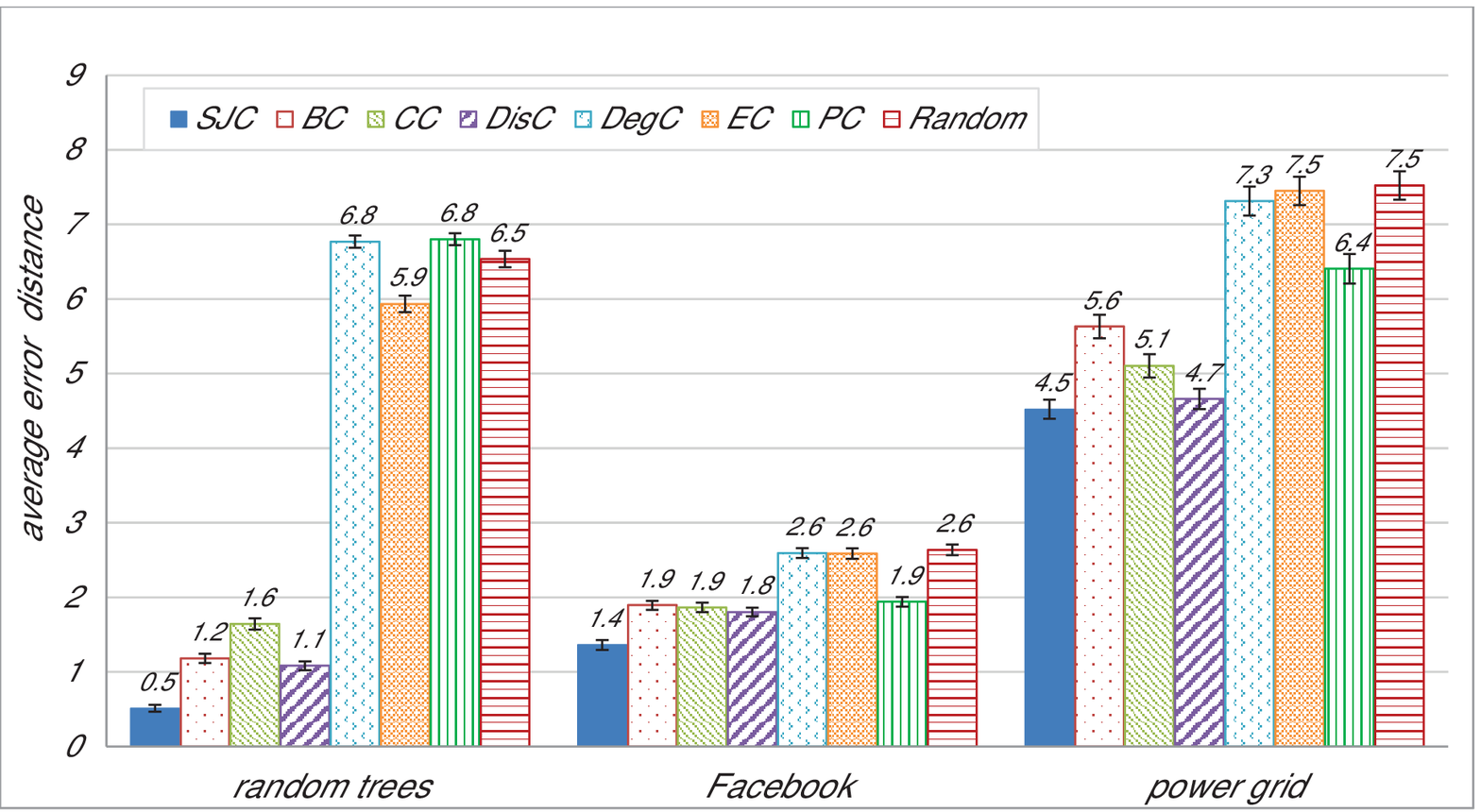}}
\caption{\blue{Average error distances with error bars of 95\% confidence interval for various networks under the SI, SIR, SIRI and SIS models in heterogeneous networks.}}
\label{fig:error_distance_1source_heterogeneous_general}
\end{figure}

In this experiment, we drop the Assumptions \ref{assump:infection_prob_SI}-\ref{assump:infection_prob_SIS} and randomly choose the infection probabilities $\pSus{v}$, $\pInf{v}$, $\pRec{v}$ from $[0,1]$ for any node $v$. We then run simulations under the SI, SIR, SIRI, and SIS models, and compare the performances in Fig. \ref{fig:error_distance_1source_heterogeneous_general}. We see that SJC outperforms all the benchmarks.

%

\subsection{Multiple Infection Sources}
\blue{In this subsection, we consider the cases where $k=2$ or $k=3$ infection sources exist, respectively.} By finding betweenness center, closeness center or distance center of each Voronoi set in the re-optimization step of MJC, we heuristically find multiple betweenness center set (MBC), multiple closeness center set (MCC) or multiple distance center set (MDisC) estimators, respectively. We also extend DegC, EC and PC by finding the $k$ nodes in $H$ with largest degree centralities, eigenvector centralities, pagerank centralities, respectively. Finally, for random guessing, we randomly pick $K$ nodes in $H$ as the estimator. We use MBC, MCC, MDisC, DegC, EC, PC and random guessing as comparison benchmarks.

For the SI, SIR, SIRI and SIS models, we randomly choose the corresponding infection probabilities $\pSus{v}$, $\pInf{v}$, $\pRec{v}$ from $[0,1]$ for any node $v$. For each value of $k$, each kind of network and each infection spreading model, we perform 1000 simulation runs. In each simulation run, we randomly pick $k$ nodes as the infection sources and simulate the infection using the above mentioned spreading model. The spreading terminates when the number of infected nodes is greater than 100. We then run MJC on the observed infected nodes to estimate the infection sources and compare the result with the benchmarks.

\begin{figure}[!ht]
  \centering
  \subfigure[SI model.]{
    \includegraphics[width=0.45\textwidth]{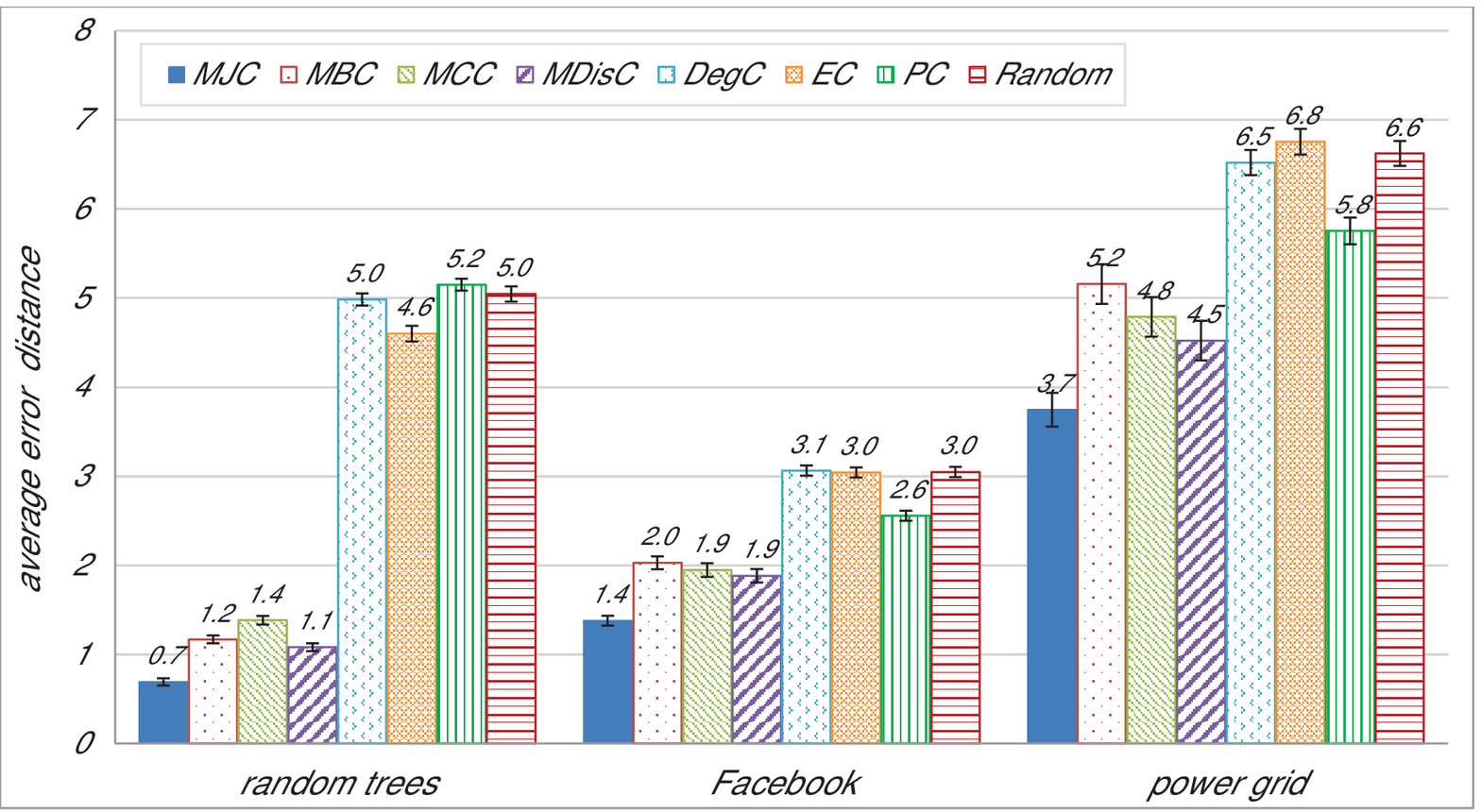}}
  \hspace{0.1in}
  \subfigure[SIR model.]{
    \includegraphics[width=0.45\textwidth]{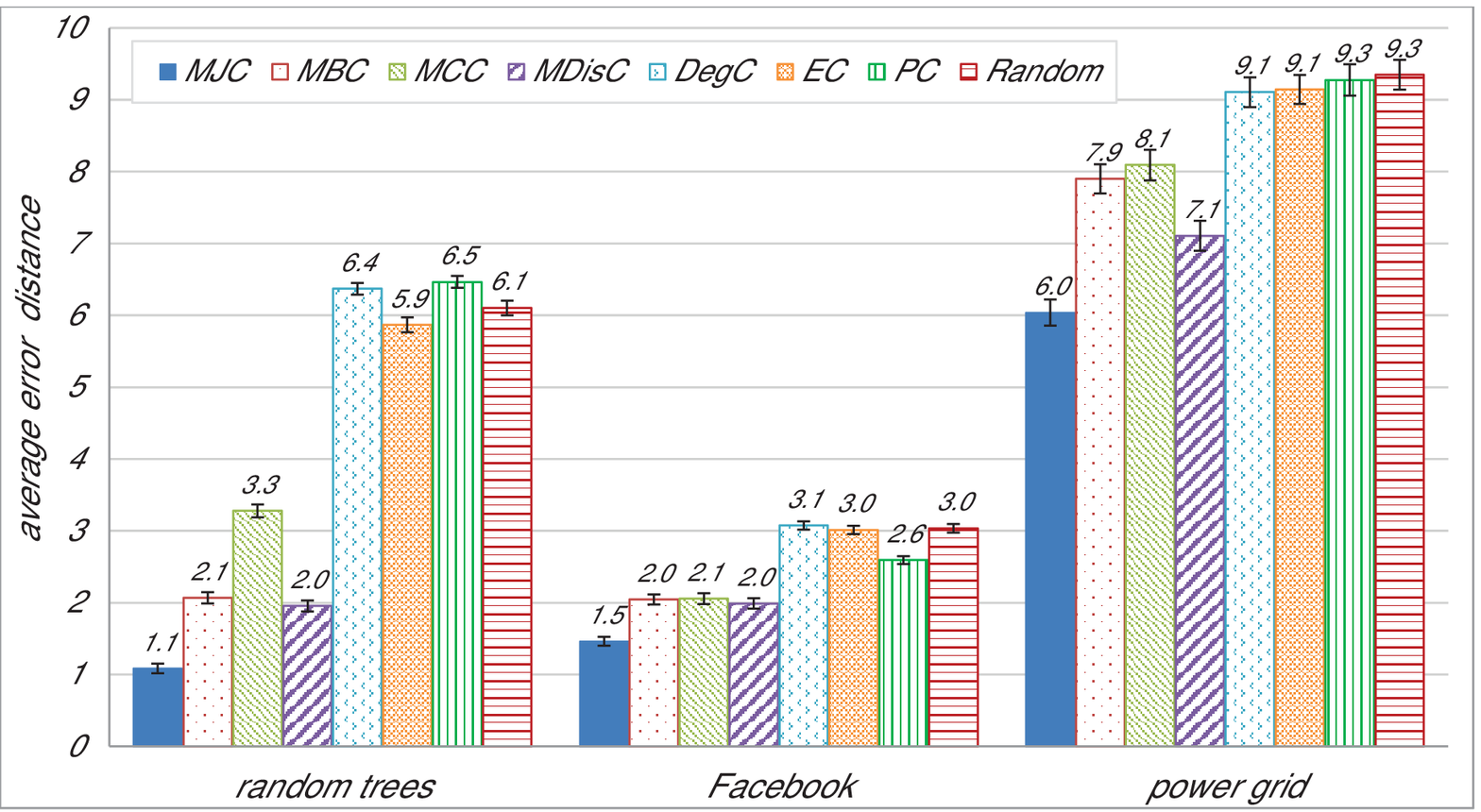}}
  \hspace{0.1in}
  \subfigure[SIRI model.]{
    \includegraphics[width=0.45\textwidth]{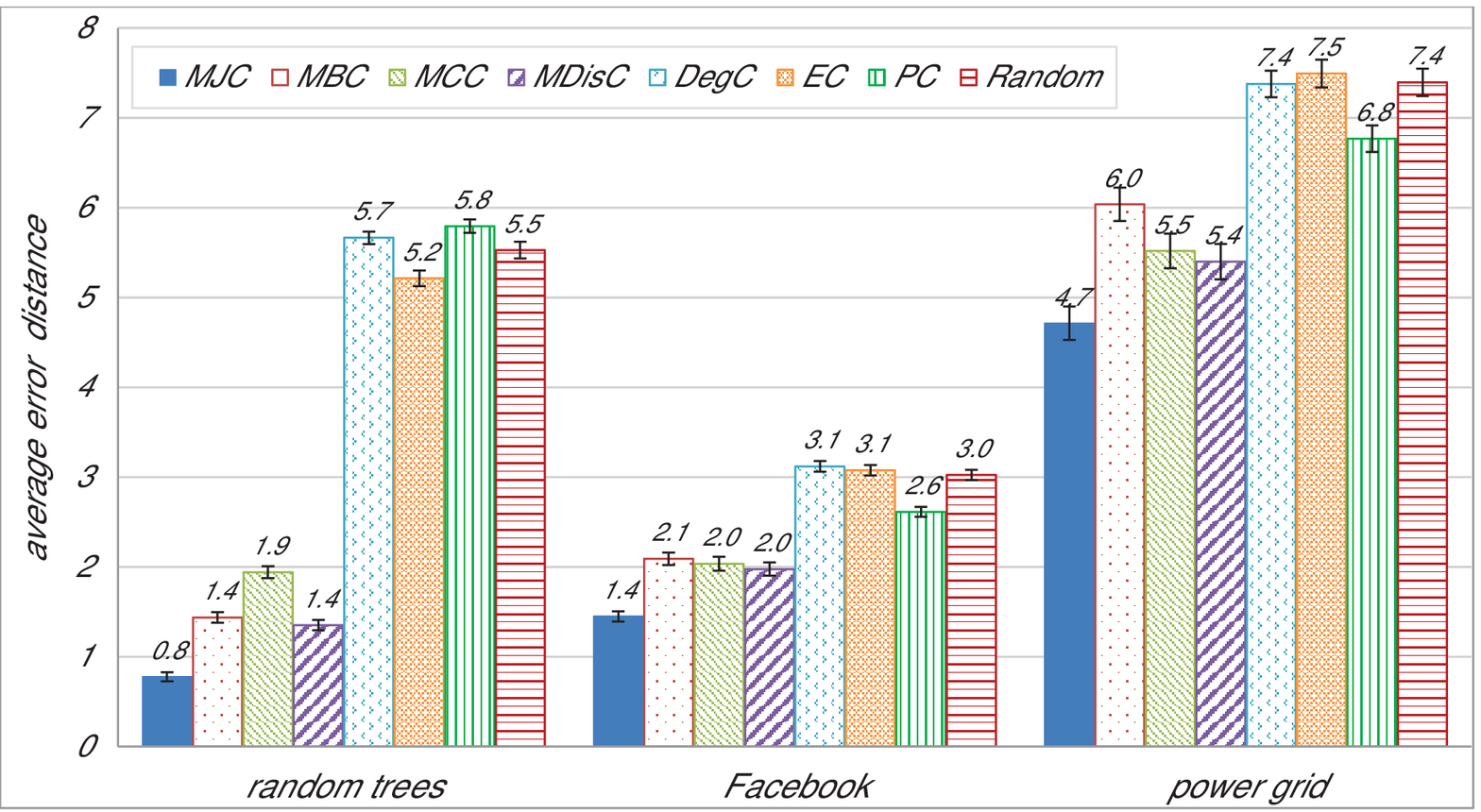}}
  \hspace{0.1in}
  \subfigure[SIS model.]{
    \includegraphics[width=0.45\textwidth]{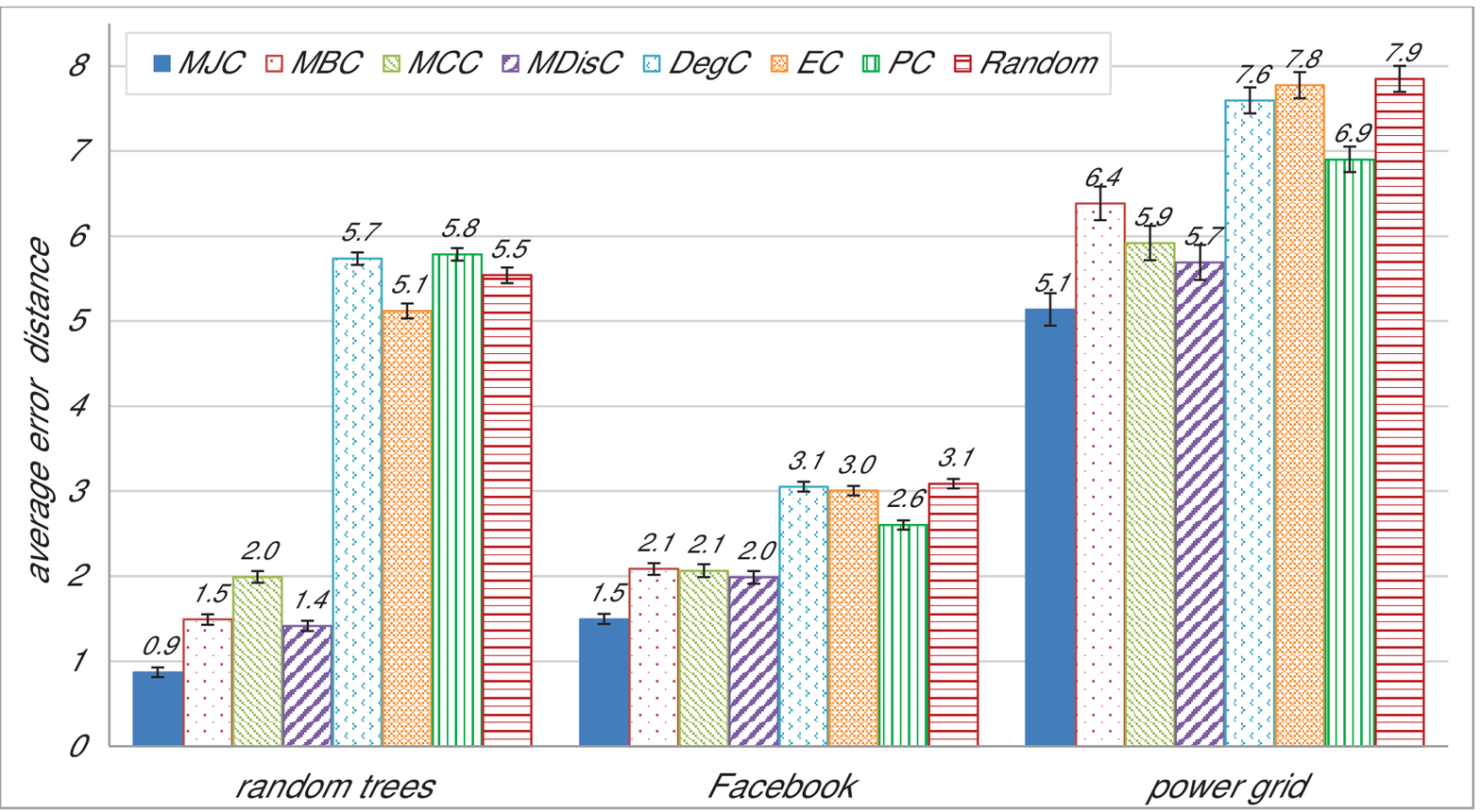}}
\caption{\blue{Average mean error distances with error bars of 95\% confidence interval for various networks under different infection spreading models when there are two infection sources.}}
\label{fig:error_distance_2sources_heterogeneous}
\end{figure}

\begin{figure}[!ht]
  \centering
  \subfigure[SI model.]{
    \includegraphics[width=0.45\textwidth]{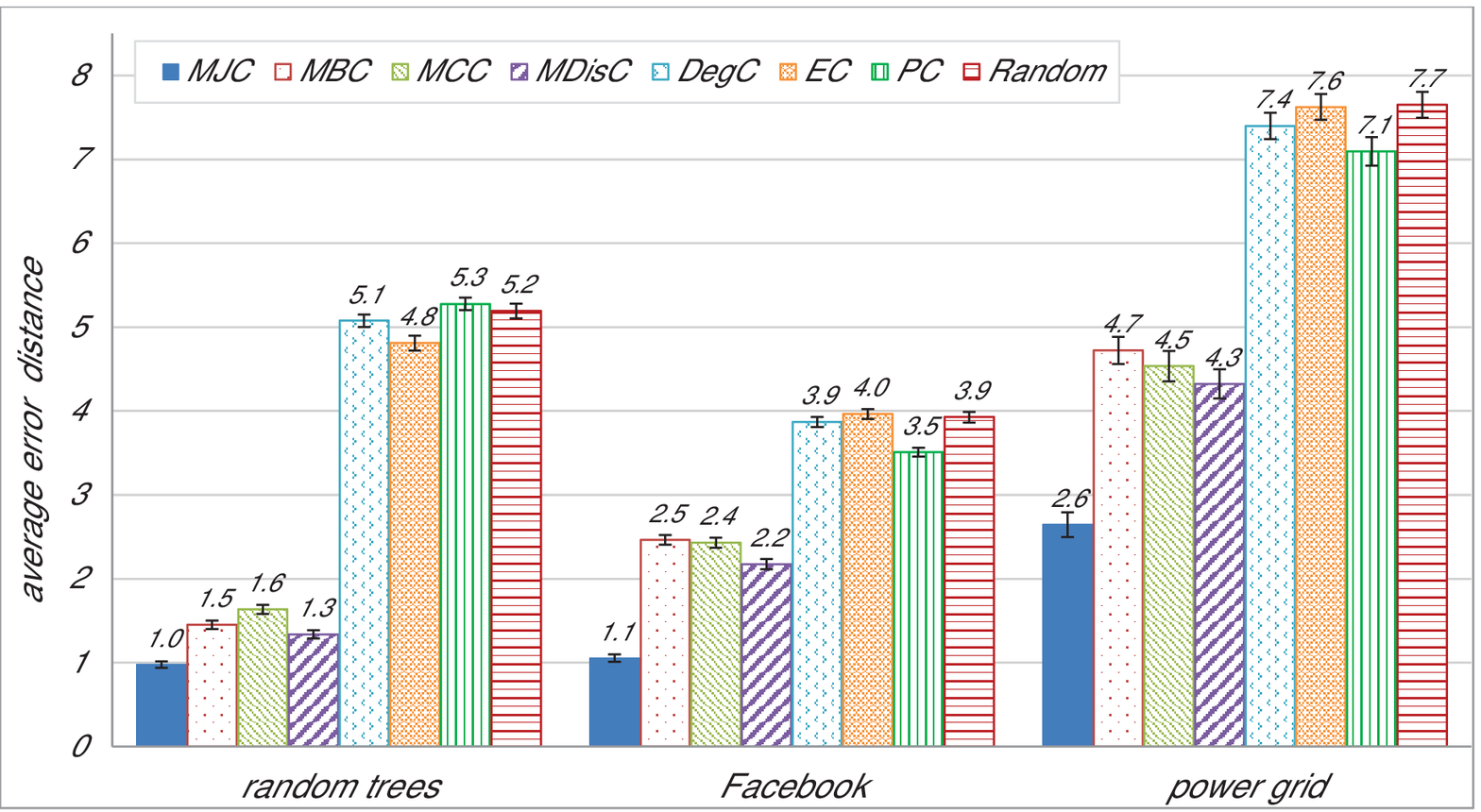}}
  \hspace{0.1in}
  \subfigure[SIR model.]{
    \includegraphics[width=0.45\textwidth]{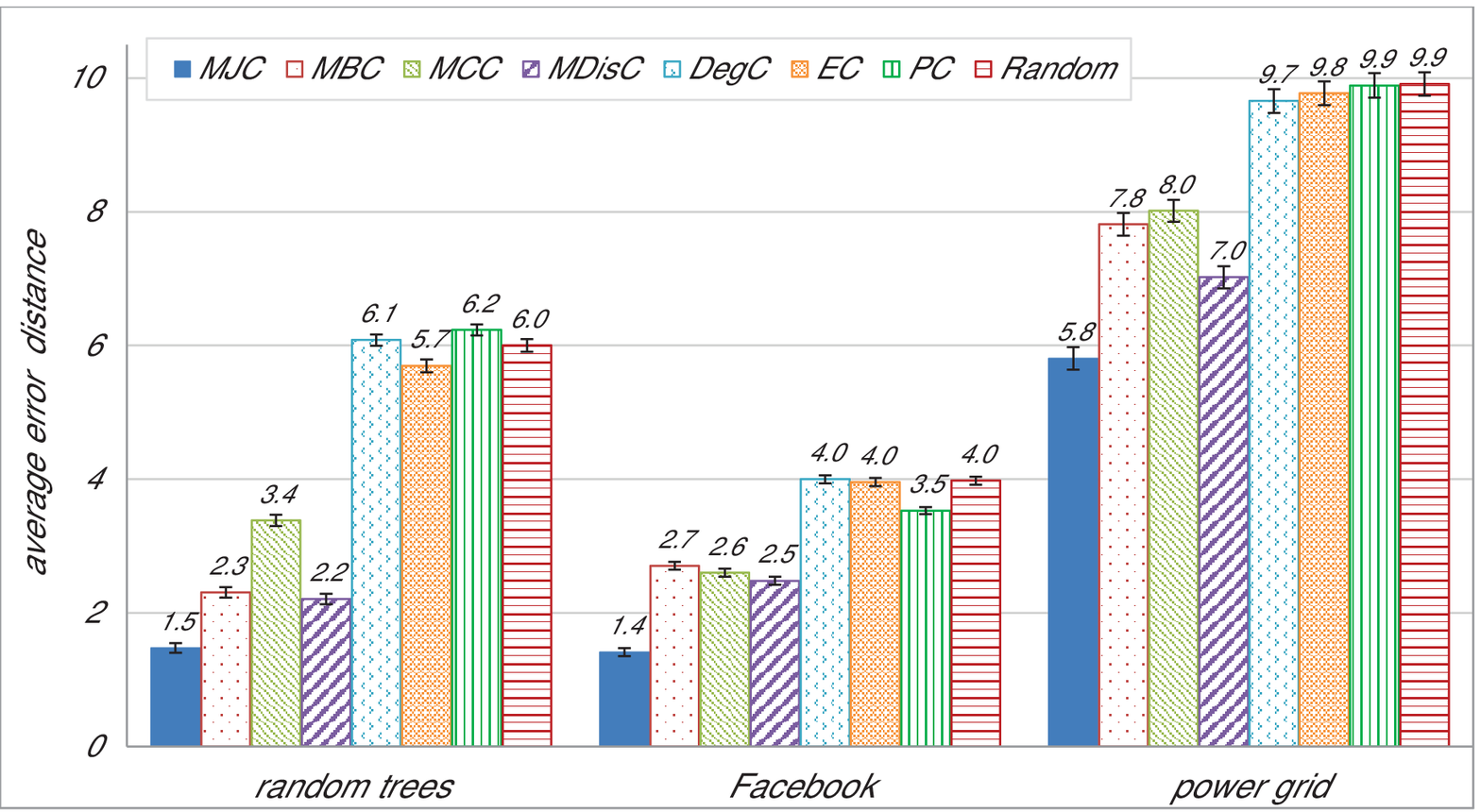}}
  \hspace{0.1in}
  \subfigure[SIRI model.]{
    \includegraphics[width=0.45\textwidth]{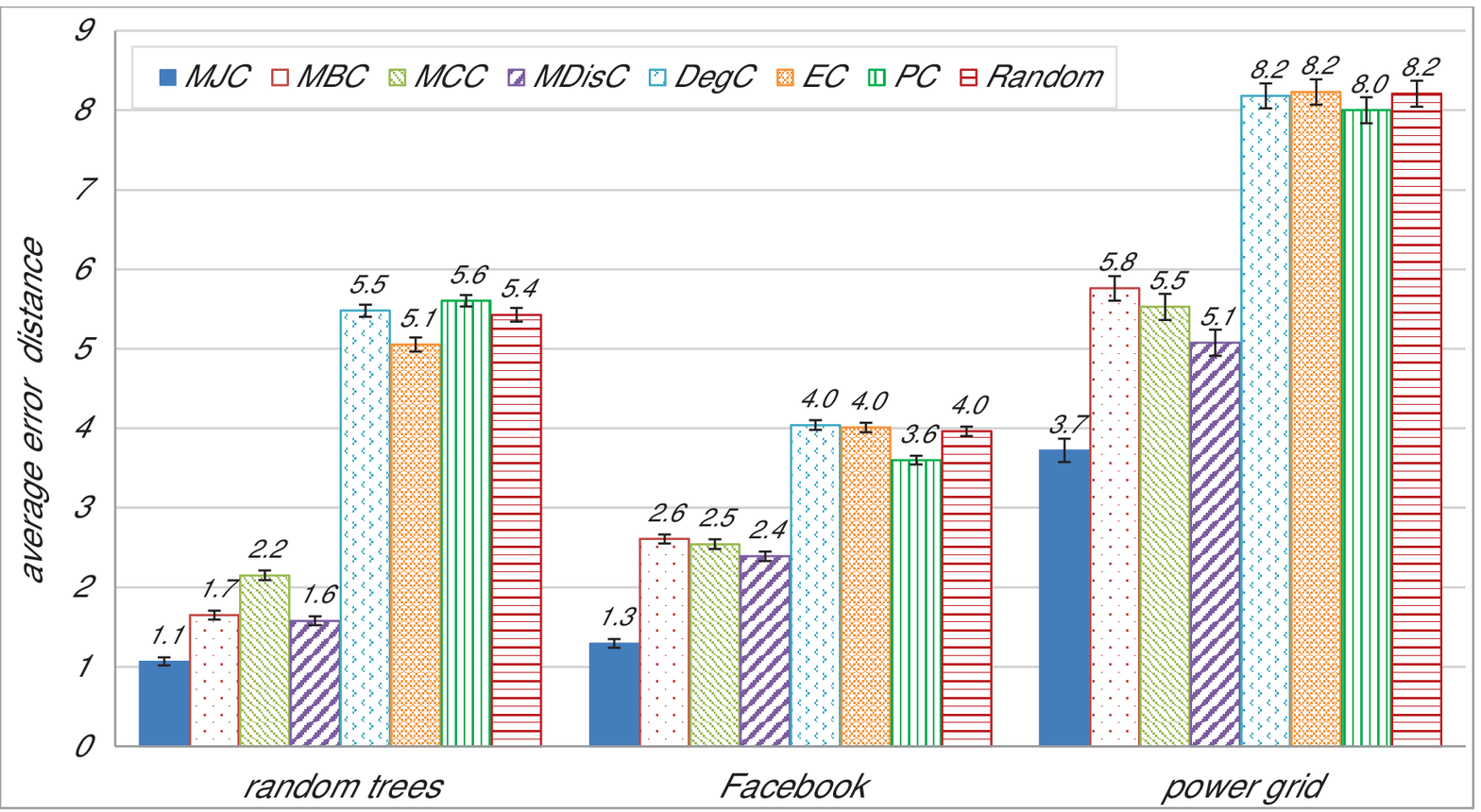}}
  \hspace{0.1in}
  \subfigure[SIS model.]{
    \includegraphics[width=0.45\textwidth]{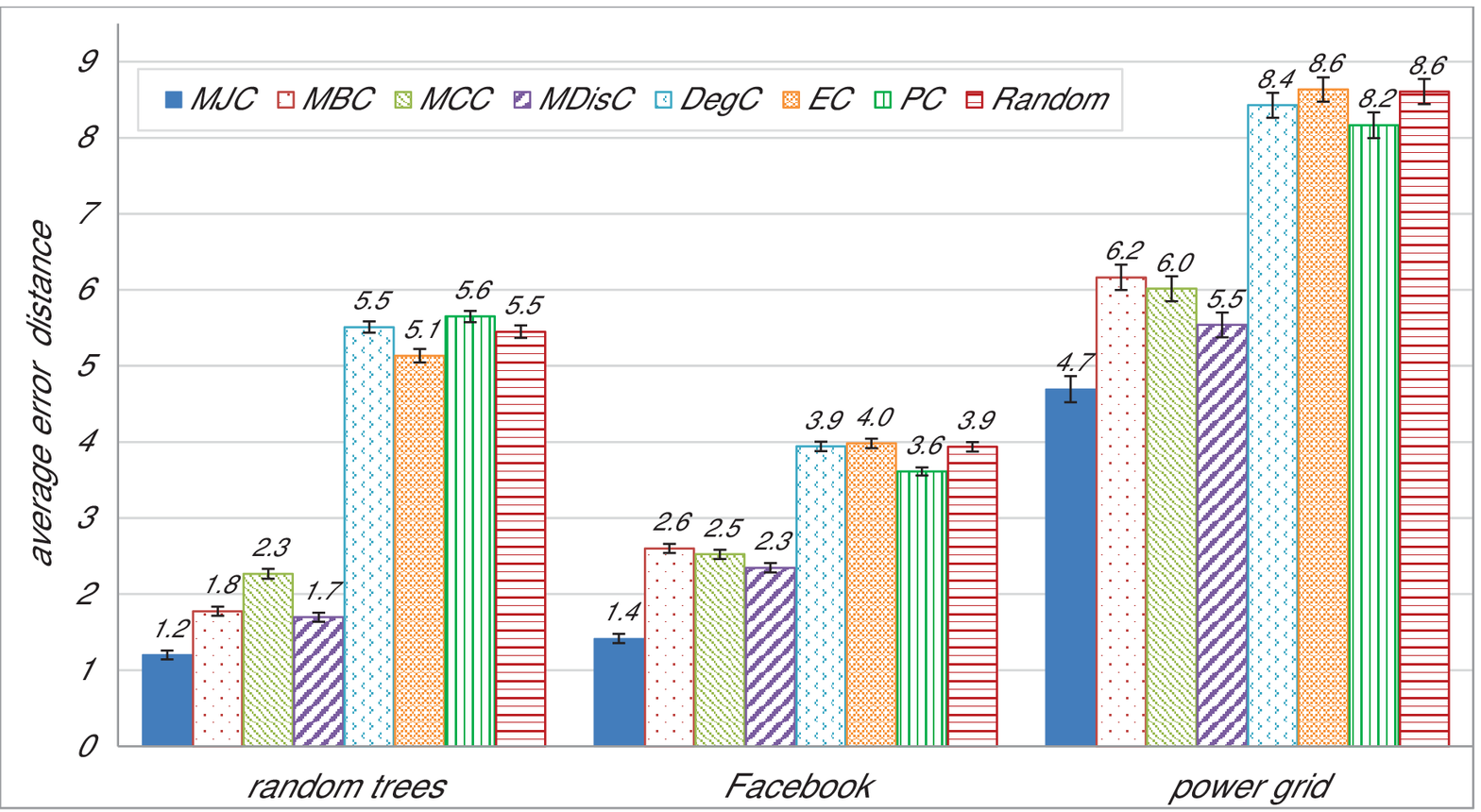}}
\caption{\blue{Average mean error distances with error bars of 95\% confidence interval for various networks under different infection spreading models when there are three infection sources.}}
\label{fig:error_distance_3sources_heterogeneous}
\end{figure}

To quantify the performance of each algorithm, we first match the estimated with the actual sources so that the sum of the error distances between each estimated source and its match is minimized. Then the mean error distance is the average of the error distances for all matched pairs, and is shown in Fig. \ref{fig:error_distance_2sources_heterogeneous} and Fig. \ref{fig:error_distance_3sources_heterogeneous} for $k=2$ and $k=3$, respectively. We see that the proposed estimator performs better than the benchmarks for all considered networks under all considered infection spreading models.

\section{Conclusion}\label{sec:conclusion}
We have investigated the problem of estimating infection sources in a network for the SI, SIR, SIRI and SIS infection spreading models. For the case where a single infection source exists in an infinite tree network and under some technical assumptions, we have shown that the Jordan center of the infected node set is a universal infection source estimator for the SI, SIR, SIRI or SIS model. When there exists more than one infection sources in a tree network, we have shown that the $k$-Jordan center set is an optimal infection source set estimator for the SI model. Simulations have been conducted on random trees, part of the Facebook network and the western states power grid network of the United States. The results suggest that our estimators perform consistently better than the betweenness, closeness, distance, degree, eigenvector, and pagerank centrality based heuristics.

\appendices

\section{Proof of Proposition \ref{lemma:optimal_t}}\label{appendix:lemma:optimal_t}

For any $t \in \cT_v$, consider any most likely infection path $Y^{t+1}$ for $(v,t+1)$. To show claim \eqref{lemma:optimal_t_monotonically_decreasing}, it suffices to construct an infection path $\tilde{X}^t$ for $(v,t)$ such that
\begin{align}
P_v(Y^{t+1}) \le \delta P_v(\tilde{X}^t), \label{equ:optimal_t_decreasing_obj}
\end{align}
since $P_v(\tilde{X}^t) \le P_v(X^t)$.

\subsection{SI model}

We first focus on any neighboring node $u$ of $v$ and consider $T_u(v;G)$. We claim that there exists an infection path $\tilde{X}^t$ such that
\begin{align}
P_v(Y^{t+1}(T_u(v;G), [1,t+1])) \le (1-\psmin) P_v(\tilde{X}^t(T_u(v;G), [1,t])). \label{equ:SI_Tu}
\end{align}
We can see that $T_u(v;G)$ is either an uninfected subtree or infected subtree. In the following, we consider these two cases in order.

Suppose that $T_u(v;G)$ is an uninfected subtree. We have for any $\tilde{X}^t$
\begin{align}
\frac{P_v(Y^{t+1}(T_u(v;G), [1,t+1]))}{P_v(\tilde{X}^t(T_u(v;G), [1,t]))} & = \frac{(1-\pSus{u})^{t+1}}{(1-\pSus{u})^t} \nonumber \\
& \le 1-\psmin. \label{equ:SI_Tu_uninfected}
\end{align}

Suppose that $T_u(v;G)$ is an infected subtree.

If $Y^{t+1}(u,1) = \Sus$, we let $\tilde{X}^t(T_u(v;G), [1,t]) = Y^{t+1}(T_u(v;G), [2,t+1])$, yielding
\begin{align*}
\frac{P_v(Y^{t+1}(T_u(v;G), [1,t+1]))}{P_v(\tilde{X}^t(T_u(v;G), [1,t]))} & = \frac{P_v(Y^{t+1}(u,1)=\Sus)P_v(Y^{t+1}(T_u(v;G), [2,t+1]))}{P_v(\tilde{X}^t(T_u(v;G), [1,t]))} \\
& = 1-\pSus{u} \\
& \le 1-\psmin.
\end{align*}

If $Y^{t+1}(u,1) = \Inf$, we show \eqref{equ:SI_Tu} by mathematical induction on $\bar{d}(v, V_\Inf)$.

\noindent \textbf{Basis step: Suppose that $\bar{d}(v, V_\Inf) = 1$.}

We let $\tilde{X}^t(u,1) = \Inf$. After it gets infected at time slot 1, node $u$ serves as the infection source of the subtree $T_u(v;G)$ with the infection starting at time 1. From the assumption $\bar{d}(v, V_\Inf) = 1$, it follows that $T_w(u;G)$ is an uninfected subtree for any $w \in V(u,1) \bigcap T_u(v;G)$. Then following \eqref{equ:SI_Tu_uninfected}, we have
\begin{align*}
\frac{P_v(Y^{t+1}(T_u(v;G), [1,t+1]))}{P_v(\tilde{X}^t(T_u(v;G), [1,t]))} & = \frac{P_v(Y^{t+1}(u,1)=\Inf)P_v(Y^{t+1}(T_u(v;G), [2,t+1]))}{P_v(\tilde{X}^{t+1}(u,1)=\Inf) P_v(\tilde{X}^t(T_u(v;G), [2,t]))} \\
& = \prod_{w \in V(u,1) \bigcap T_u(u;G)} \frac{P_v(Y^{t+1}(T_w(u;G), [2,t+1]))}{P_v(\tilde{X}^t(T_w(v;G), [2,t]))}\\
& \le (1-\psmin)^{|V(u,1) \bigcap T_u(v;G)|} \\
& \le 1-\psmin,
\end{align*}
where the last inequality follows from Assumption \ref{assump:topology_infinite_regular}. This completes the proof for the basis step.

\noindent \textbf{Inductive step:} Assume \eqref{equ:SI_Tu} holds for $\bar{d}(v, V_\Inf) \le n -1$, where $n \ge 2$. We want to show that \eqref{equ:SI_Tu} also holds for $\bar{d}(v, V_\Inf)=n$.

Assume $\bar{d}(v, V_\Inf)=n$ and let $\tilde{X}^t(u,1) = \Inf$. After it becomes infected at time slot 1, node $u$ serves as the infection source of the subtree $T_u(v;G)$ with the infection starting at time 1. Since $\bar{d}(u, V_\Inf \bigcap T_u(v;G)) \le n-1$, from the induction assumption and for any $w \in V(u,1) \bigcap T_u(v;G)$, we can find a $\tilde{X}^t$ such that
\begin{align*}
P_v(Y^{t+1}(T_w(u;G), [2,t+1])) \le (1-\psmin) P_v(\tilde{X}^t(T_w(u;G), [2,t])).
\end{align*}
We then have,
\begin{align*}
\frac{P_v(Y^{t+1}(T_u(v;G), [1,t+1]))}{P_v(\tilde{X}^t(T_u(v;G), [1,t]))} & = \frac{P_v(Y^{t+1}(u,1)=\Inf)P_v(Y^{t+1}(T_u(v;G), [2,t+1]))}{P_v(\tilde{X}^{t+1}(u,1)=\Inf) P_v(\tilde{X}^t(T_u(v;G), [2,t]))} \\
& = \prod_{w \in V(u,1) \bigcap T_u(u;G)} \frac{P_v(Y^{t+1}(T_w(u;G), [2,t+1]))}{P_v(\tilde{X}^t(T_w(v;G), [2,t]))}\\
& \le (1-\psmin)^{|V(u,1) \bigcap T_u(v;G)|} \\
& \le 1-\psmin,
\end{align*}
where the last inequality follows from Assumption \ref{assump:topology_infinite_regular}. This completes the proof for the inductive step, and the claim is now proved.

By constructing $\tilde{X}^t$ to satisfy \eqref{equ:SI_Tu} for all $u \in V(v,1)$, we have
\begin{align*}
\frac{P_v(Y^{t+1})}{P_v(\tilde{X}^t)} & = \prod_{u \in V(v,1)} \frac{P_v(Y^{t+1}(T_u(v;G), [1,t+1]))}{P_v(\tilde{X}^t(T_u(v;G), [1,t]))} \\
& \le (1-\psmin)^{|V(v,1)|} \\
& \le (1-\psmin)^2,
\end{align*}
where the last inequality follows from Assumption \ref{assump:topology_infinite_regular}. This completes the proof of claim \eqref{lemma:optimal_t_monotonically_decreasing} for the SI model.

\subsection{SIR and SIRI model}

We first present a property of the SIRI model in Lemma \ref{lemma:SIRI}.

\begin{lem}\label{lemma:SIRI}
Suppose that $v \in V$ is the infection source and $v$ has only one neighboring node $u$. Suppose that the set of observed infected nodes $V_\Inf$ is non-empty. Consider an infection under the SIRI model and suppose Assumptions \ref{assump:infection_prob_SIRI} and \ref{assump:topology_infinite_regular} hold. For any $t \in \cT_v$ and any most likely infection path $Y^{t+1}$ for $(v,t+1)$, there exists an infection path $\tilde{X}^t$, such that
\begin{enumerate}[(a)]
      \item \label{lemma:SIRI:v} $P_v(Y^{t+1}(v, [1,t+1])) \le \sqrt{\frac{\psmin}{\psmax}} P_v(\tilde{X}^t(v, [1,t]))$;
      \item \label{lemma:SIRI:T_u}$P_v(Y^{t+1}(T_u(v;G), [1,t+1])) \le P_v(\tilde{X}(T_u(v;G), [1,t]))$; and
      \item \label{lemma:SIRI:V}$P_v(Y^{t+1}) \le \sqrt{\frac{\psmin}{\psmax}} P_v(\tilde{X}^t)$.
\end{enumerate}
\end{lem}

The proof of Lemma \ref{lemma:SIRI} is provided in Appendix \ref{appendix:lemma:SIRI}. Lemma \ref{lemma:SIRI} shows that, in the SIRI model, a most likely elapsed time $t_v$ should be as small as possible when the source has only one neighboring node. We now extend this result to prove Proposition \ref{lemma:optimal_t}\eqref{lemma:optimal_t_monotonically_decreasing} for the SIRI model where $v$ has more than one neighboring node.

In the SIRI model, since $v$ is the source node, $\tilde{X}^t(v, [1,t])$ is independent of the states of other nodes. Furthermore, for any pair of neighboring nodes $u$ and $u'$ of $v$, the states of $T_u(v;G)$ and $T_{u'}(v;G)$ are independent conditioned on the states of node $v$. Therefore, by applying Lemma \ref{lemma:SIRI} to $v$ and each of its neighboring nodes, we have an infection path $\tilde{X}^t$ such that
\begin{align*}
\frac{P_v(Y^{t+1})}{P_v(\tilde{X}^t)}=&\frac{P_v(Y^{t+1}(v,[1,t+1])) \prod_{u\in N_v(1)}P_v(Y^{t+1}(T_u(v;G),[1,t+1]))}{P_v(\tilde{X}^t(v, [1,t])) \prod_{u\in N_v(1)}P_v(\tilde{X}^t(T_u(v;G), [1,t]))} \\
\le & \sqrt{\frac{\psmin}{\psmax}}.
\end{align*}
This completes the proof of claim \eqref{lemma:optimal_t_monotonically_decreasing} for the SIRI model. The proof of claim \eqref{lemma:optimal_t_monotonically_decreasing} for the SIR model is similar to that of the SIRI model, and we omit it here to avoid repetition.

\subsection{SIS model}

For the SIS model, a node can become infected, recover, and then be reinfected again for multiple times by the observation time. We characterize the time when a node is first infected (first infection time) in the following lemma, whose proof is provided in Appendix \ref{appendix:lemma:SIS_better_later}. Recall that $H_v$ is the minimum connected subgraph of $G$ that contains $V_\Inf$ and $v$.

\begin{lem}\label{lemma:SIS_better_later}
Suppose that $v \in V$ is the infection source and a non-empty set of infected nodes $V_\Inf$ is observed. Suppose the infection follows the SIS model and Assumption \ref{assump:infection_prob_SIS} and \ref{assump:topology_infinite_regular} hold. Then, for any $t \in \cT_v$, there exists a most likely infection path $X^t$ for $(v,t)$, such that, for any $u \in H_v \backslash \{v\}$, the first infection time $t_{int}(u)$ of $u$ in $X^t$ is given by
\begin{align}
t_{int}(u)= t-\bar{d}(u, T_u(v;H_v)). \label{equ:SIS_optimal_first_infection_time}
\end{align}
\end{lem}

Lemma \ref{lemma:SIS_better_later} enables us to calculate the first infection time of each node in $H_v$ in a most likely infection path under the SIS model. Moreover, it shows that given the elapsed time, a most likely infection path for a node $v$ is given by a path whose nodes ``resist'' the infection, and each node becomes infected only at the latest possible time. Therefore, intuitively the most likely elapsed time $t_v$ should be as small as possible to minimize the time that nodes ``resist'' the infection spreading, so as to maximize the probability of the infection path.

Since $t \in \cT_v$, we have $t \ge \bar{d}(v, V_\Inf)$. For any $u\in V(v,1)$, from Lemma \ref{lemma:SIS_better_later}, we have that the first infection time $t_{int}(u)$ of $u$ in $Y^{t+1}$ is given by
\begin{align}
t_{int}(u) & = t+1 - \bar{d}(u, T_u(v;H_v)) \nonumber \\
& \ge \bar{d}(v, V_\Inf) + 1 - \bar{d}(u, T_u(v;H_v)) \nonumber \\
& \ge 2. \label{ineq:SIS_better_later:tu_init}
\end{align}

We claim that $Y^{t+1}(v, 1)  = \Inf$. Otherwise, $v$ and all its neighboring nodes are not infected at time 1 because of \eqref{ineq:SIS_better_later:tu_init}. Because the infection can propagate at most 1 hop away from $v$ at time 1, all nodes are uninfected at time 1, and the infection propagation process stops. This contradicts the assumption that the set of observed infected nodes $V_\Inf$ is non-empty. Then, following Lemma \ref{lemma:SIS_better_later}, we can let $\tilde{X}^{t}(V, [1,t]) = Y^{t+1}(V, [2,t+1])$, yielding
\begin{align*}
\frac{P_v(Y^{t+1})}{P_v(\tilde{X}^t)}=&\frac{P_v(Y^{t+1}(v, 1))P_v(Y^{t+1}(V(v,1), 1))P_v(Y^{t+1}(V, [2, t+1]))}{P_v(\tilde{X}^t(V, [1, t]))} \\
=& p_{\Inf} (1-p_{\Sus})^{|V(v,1)|} \\
\le & 1.
\end{align*}

This completes the proof of claim \eqref{lemma:optimal_t_monotonically_decreasing} for the SIS model.

It is easy to see that $\delta \le 1$ for all considered infection spreading models and claim \eqref{lemma:optimal_t_optimal_t} now follows from claim \eqref{lemma:optimal_t_monotonically_decreasing}, and the proof of Proposition \ref{lemma:optimal_t} is complete.

\section{Proof of Proposition \ref{lemma:better_neighbor}}\label{appendix:lemma:better_neighbor}
We first review the following topological property shown in \cite{Luo2014}.
\begin{lem}\label{lemma:tv=tu-1}
Suppose a non-empty set of infected nodes $V_\Inf$ is observed over $G$. For a pair of neighboring nodes $u$ and $v$, if $\bar{d}(v,V_\Inf) < \bar{d}(u,V_\Inf)$, we have
\begin{enumerate}[(a)]
  \item \label{lemma:tv=tu-1_leaf} $l \in T_v(u;H_v \bigcup H_u)$, for all $l \in \arg \max_{x\in V_{\Inf}}d(u,x)$; and
  \item \label{lemma:tv=tu-1_tv=tu-1} $\bar{d}(v,V_\Inf)=\bar{d}(u,V_\Inf)-1$, and there exists $l\in T_v(u;H_v \bigcup H_u)$ such that $d(v,l) = \bar{d}(v,V_\Inf)$.
\end{enumerate}
\end{lem}

To prove Proposition \ref{lemma:better_neighbor}, it suffices to construct an infection path $\tX^{t_v}$ with source node $v$, and show that $P_v(\tX^{t_v}) \ge P_u(Y^{t_u})$. Let $t_{int}(v)$ be the first infection time of node $v$ in the infection path $Y^{t_u}$ with source node $u$. We first show that $t_{int}(v)=1$. Since $u$ is the infection source, the infection can propagate at most $t_u-t_{int}(v)$ hops away from node $v$ within the subtree $T_v(u;H_v \bigcup H_u)$. From Lemma \ref{lemma:tv=tu-1}(\ref{lemma:tv=tu-1_tv=tu-1}), if $t_{int}(v)>1$, we have $t_v=t_u-1>t_u-t_{int}(v)$, a contradiction. Therefore, we must have $t_{int}(v)=1$ in the infection path $Y^{t_u}$.

For the SI, SIR and SIRI models, we let $\tX^{t_v}(T_v(u;G), [1, t_v]) = Y^{t_u}(T_v(u;G), [2, t_u])$, yielding
\begin{align}
\frac{P_u(Y^{t_u}(T_v(u;G), [1,t_u]))}{P_v(\tilde{X}^{t_v}(T_v(u;G),[1,t_v]))} &= \frac{P_u(Y^{t_u}(v,1)=\Inf)P_u(Y^{t_u}(T_v(u;G), [2,t_u]))} {P_v(\tilde{X}^{t_v}(T_v(u;G),[1,t_v]))} \nonumber \\
&= \pSus{v}. \label{equ:better_neighbor_subtree_v}
\end{align}

Let $\tX^{t_v}(u, 1) = \Inf$ and $u$ can be seen as the infection source of the subtree $T_u(v;G)$ with the infection starting at time 1. For the SI model, applying \eqref{equ:SI_Tu} twice, we have
\begin{align}
\frac{P_u(Y^{t_u}(T_u(v;G), [1,t_u]))}{P_v(\tilde{X}^{t_v}(T_u(v;G),[1,t_v]))} &=\frac{P_u(Y^{t_u}(T_u(v;G), [1,t_u]))}{P_v(\tilde{X}^{t_v}(u,1)=\Inf) P_v(\tilde{X}^{t_v}(T_u(v;G),[2,t_v]))} \nonumber \\
&\le \frac{(1-\psmin)^2}{\pSus{u}}.  \label{equ:better_neighbor_subtree_u_SI}
\end{align}

Multiplying \eqref{equ:better_neighbor_subtree_v} by \eqref{equ:better_neighbor_subtree_u_SI}, we obtain
\begin{align*}
\frac{P_u(Y^{t_u})}{P_v(\tX^{t_v})}
& \le \frac{\pSus{v}\cdot (1-\psmin)^2}{\pSus{u}} \\
& \le \frac{\psmax  (1-\psmin)^2}{\psmin} \\
& \le 1,
\end{align*}
where the last inequality follows from \eqref{ineq:SI_ps}.

For the SIR and SIRI models, applying Lemma \ref{lemma:SIRI} twice to $u$ and each of its neighboring nodes in $T_u(v;G)$, we have
\begin{align}
\frac{P_u(Y^{t_u}(T_u(v;G), [1,t_u]))}{P_v(\tilde{X}^{t_v}(T_u(v;G),[1,t_v]))} &=\frac{P_u(Y^{t_u}(T_u(v;G), [1,t_u]))}{P_v(\tilde{X}^{t_v}(u,1)=\Inf) P_v(\tilde{X}^{t_v}(T_u(v;G),[2,t_v]))} \nonumber \\
&\le \frac{\psmin}{\psmax \cdot \pSus{u}}.  \label{equ:better_neighbor_subtree_u_SIR_SIRI}
\end{align}

Multiplying \eqref{equ:better_neighbor_subtree_v} by \eqref{equ:better_neighbor_subtree_u_SIR_SIRI}, we have
\begin{align*}
\frac{P_u(Y^{t_u})}{P_v(\tX^{t_v})} \le \frac{\pSus{v}}{\psmax} \cdot \frac{\psmin}{\pSus{u}} \le 1.
\end{align*}
This completes the proof of Proposition \ref{lemma:better_neighbor} in the SI, SIR and SIRI models.

We next consider the SIS model. Following Lemma \ref{lemma:SIS_better_later}, we have $Y^{t_u}(V(u,1)\backslash \{v\},1) = \Sus$ and we can let $\tilde{X}^{t_v}(V, [1, t_v]) = Y^{t_u}(V, [2, t_u])$. Moreover, following similar arguments as the worst case in \eqref{equ:SIS_better_later_basis_tv_u_uninfected}, we have that $Y^{t_u}(u, 1) \ne \Inf$, yielding
\begin{align*}
\frac{P_u(Y^{t_u})}{P_v(\tX^{t_v})} &= \frac{P_u(Y^{t_u}(v,1))P_u(Y^{t_u}(u,1))P_u(Y^{t_u}(V(u,1)\backslash \{v\},1)) P_u(Y^{t_u}(V, [2, t_u]))}{P_v(\tX^{t_v}(V, [1, t_v]) )} \\
&=p_{\Sus}(1-p_{\Inf})(1-p_{\Sus})^{|V(u,1)\backslash \{v\}|} \\
&\le 1.
\end{align*}
This completes the proof of Proposition \ref{lemma:better_neighbor} in the SIS model. The proof of Proposition \ref{lemma:better_neighbor} is now complete.

\section{Proof of Proposition \ref{lemma:optimal_t_k_sources}} \label{appendix:lemma:optimal_t_k_sources}

We extend the notation of subtree as follows. For any graph $A$, any node $v \in A$ and a set of nodes $S \subset A$, let $T_v(S; A)$ be the subtree of $A$ rooted at node $v$ with the first link in the path from $v$ to each element in $S$ removed. Moreover, for any set of nodes $M \subset A$, let $T_M(S; A) = \bigcup_{v \in M} T_v(S; A)$.

For any $t \in \cT_{S}$, consider any most likely infection path $Y^{t+1}$ for $(S,t+1)$. To show claim \eqref{lemma:optimal_t_k_sources_monotonically_decreasing}, it suffices to construct an infection path $\tilde{X}^t$ for $(S,t)$ such that
\begin{align}
P_{S}(Y^{t+1}) \le P_{S}(\tilde{X}^t). \label{equ:optimal_t_2sources_monotonically_decreasing}
\end{align}
We start with the case where $k=2$ and show \eqref{equ:optimal_t_2sources_monotonically_decreasing} by mathematical induction on $d(s_1,s_2)$.

\noindent \textbf{Basis step (i):} The inequality  \eqref{equ:optimal_t_2sources_monotonically_decreasing} holds for $d(s_1,s_2) = 1$.

The states of $T_{s_1}(S;G)$ and $T_{s_2}(S;G)$ are independent. We can treat $s_1$ and $s_2$ as the infection source of $T_{s_1}(S;G)$ and $T_{s_2}(S;G)$, respectively. Then following Proposition \ref{lemma:optimal_t}, we can find a $\tilde{X}^t$ such that
\begin{dmath*}
\frac{P_{S}(Y^{t+1})}{P_{S}(\tilde{X}^t)} = \frac{P_{S}(Y^{t+1}(T_{s_1}(S;G),[1,t+1]))}{P_{S}(\tilde{X}^t(T_{s_1}(S;G),[1,t]))} \frac{P_{S}(Y^{t+1}(T_{s_2}(S;G),[1,t+1]))}{P_{S}(\tilde{X}^t(T_{s_2}(S;G),[1,t]))}\\
 \le 1.
\end{dmath*}

\noindent \textbf{Basis step (ii):} The inequality \eqref{equ:optimal_t_2sources_monotonically_decreasing} holds for $d(s_1,s_2) = 2$.

Denote the common neighboring node of $s_1$ and $s_2$ to be $u$. Consider the following two possible cases of $Y^{t+1}$.

\noindent \emph{Case 1:} $Y^{t+1}(u,1) = \Inf$.

We let $\tilde{X}^t(u,1) = \Inf$, conditioning on which the states of $T_{s_1}(S;G)$, $T_{s_2}(S;G)$ and $T_{u}(S;G)$ are independent. Moreover, $u$ can be seen as the infection source of $T_{u}(S;G)$ with the infection starting at time 1. Then following Proposition \ref{lemma:optimal_t}, we can find a $\tilde{X}^t$ such that
\begin{dmath*}
\frac{P_{S}(Y^{t+1})}{P_{S}(\tilde{X}^t)} = \frac{P_{S}(Y^{t+1}(T_{s_1}(S;G),[1,t+1]))}{P_{S}(\tilde{X}^t(T_{s_1}(S;G),[1,t]))} \frac{P_{S}(Y^{t+1}(T_{s_2}(S;G),[1,t+1]))}{P_{S}(\tilde{X}^t(T_{s_2}(S;G),[1,t]))}\\
\frac{P_{S}(Y^{t+1}(u,1))}{P_{S}(\tilde{X}^t(u,1))} \frac{P_{S}(Y^{t+1}(T_{u}(S;G),[2,t+1]))}{P_{S}(\tilde{X}^t(T_{u}(S;G),[2,t]))}
\le 1.
\end{dmath*}

\noindent \emph{Case 2:} $Y^{t+1}(u,1) = \Sus$.

We let $\tilde{X}^t(T_{u}(S;G),[1,t]) = Y^{t+1}(T_{u}(S;G),[2,t+1])$. Then following Proposition \ref{lemma:optimal_t}, we can find a $\tilde{X}^t$ such that
\begin{dmath*}
\frac{P_{S}(Y^{t+1})}{P_{S}(\tilde{X}^t)} = \frac{P_{S}(Y^{t+1}(T_{s_1}(S;G),[1,t+1]))}{P_{S}(\tilde{X}^t(T_{s_1}(S;G),[1,t]))} \frac{P_{S}(Y^{t+1}(T_{s_2}(S;G),[1,t+1]))}{P_{S}(\tilde{X}^t(T_{s_2}(S;G),[1,t]))}\\
\frac{P_{S}(Y^{t+1}(u,1)) P_{S}(Y^{t+1}(T_{u}(S;G),[2,t+1]))}{P_{S}(\tilde{X}^t(T_{u}(S;G),[1,t]))} \\
\le 1-\pSus{u}
\le 1.
\end{dmath*}

\noindent \textbf{Inductive step:} If \eqref{equ:optimal_t_2sources_monotonically_decreasing} holds for $d(s_1,s_2) \le n$, then \eqref{equ:optimal_t_2sources_monotonically_decreasing} also holds for $d(s_1,s_2) = n+1$, where $n \ge 2$.

Let $\rho(v,u)$ be the path between two nodes $v$ and $u$. Denote the neighboring node of $s_1$ and $s_2$ in $\rho(s_1, s_2)$ to be $u_1$ and $u_2$, respectively. Consider the following four possible cases of $Y^{t+1}$.

\noindent \emph{Case 1:} $Y^{t+1}(u_1,1) = \Inf$ and $Y^{t+1}(u_2,1) = \Inf$.

We let $\tilde{X}^t(u_1,1) = \Inf$ and $\tilde{X}^t(u_2,1) = \Inf$. Then $u_1$ and $u_2$ can be seen as the pair of infection sources of $T_{\rho(u_1,u_2)}(S;G)$ with the infection starting at time 1. Moreover, we have $d(u_1, u_2) = d(s_1, s_2) - 2 = n-1$. Then by induction assumption, we can find a $\tilde{X}^t$ such that
\begin{dmath*}
P_{S}(Y^{t+1}(T_{\rho(u_1, u_2)}(S;G),[2,t+1])) \le P_{S}(\tilde{X}^t(T_{\rho(u_1, u_2)}(S;G),[2,t])).
\end{dmath*}

Then by Proposition \ref{lemma:optimal_t}, we can find a $\tilde{X}^t$ such that \eqref{equ:optimal_t_2sources_monotonically_decreasing} holds.

\noindent \emph{Case 2:} $Y^{t+1}(u_1,1) = \Inf$ and $Y^{t+1}(u_2,1) = \Sus$.

We let $\tilde{X}^t(u_1,1) = \Inf$ and $\tilde{X}^t(u_2,1) = \Sus$. Then $u_1$ and $s_2$ can be seen as the pair of infection sources of $T_{\rho(u_1,u_2)} \bigcup \{s_2\}$ with the infection starting at time 1. Moreover, we have $d(u_1, u_2) = d(s_1, s_2) - 2 = n-1$. Then by induction assumption, we can find a $\tilde{X}^t$ such that
\begin{dmath*}
P_{S}(Y^{t+1}(T_{\rho(u_1, u_2)}(S;G),[2,t+1])) \le P_{S}(\tilde{X}^t(T_{\rho(u_1, u_2)}(S;G),[2,t])).
\end{dmath*}

Then by Proposition \ref{lemma:optimal_t}, we can find a $\tilde{X}^t$ such that \eqref{equ:optimal_t_2sources_monotonically_decreasing} holds.

\noindent \emph{Case 3:} $Y^{t+1}(u_1,1) = \Sus$ and $Y^{t+1}(u_2,1) = \Inf$.

Following similar arguments as that in Case 2, we can find a $\tilde{X}^t$ such that \eqref{equ:optimal_t_2sources_monotonically_decreasing} holds.

\noindent \emph{Case 4:} $Y^{t+1}(u_1,1) = \Sus$ and $Y^{t+1}(u_2,1) = \Sus$.

We let $\tilde{X}^t(T_{\rho(u_1, u_2)}(S;G),[1,t]) = Y^{t+1}(T_{\rho(u_1, u_2)}(S;G),[2,t+1])$. Then following Proposition \ref{lemma:optimal_t}, we can find a $\tilde{X}^t$ such that
\begin{dmath*}
\frac{P_{S}(Y^{t+1})}{P_{S}(\tilde{X}^t)} = \frac{P_{S}(Y^{t+1}(T_{s_1}(S;G),[1,t+1]))}{P_{S}(\tilde{X}^t(T_{s_1}(S;G),[1,t]))} \frac{P_{S}(Y^{t+1}(T_{s_2}(S;G),[1,t+1]))}{P_{S}(\tilde{X}^t(T_{s_2}(S;G),[1,t]))}\\
\frac{P_{S}(Y^{t+1}(u_1,1)) P_{S}(Y^{t+1}(u_2,1)) P_{S}(Y^{t+1}(T_{\rho(u_1, u_2)}(S;G),[2,t+1]))}{P_{S}(\tilde{X}^t(T_{\rho(u_1, u_2)}(S;G),[1,t]))} \\
\le (1- \pSus{u_1})(1-\pSus{u_2}) \\
\le 1.
\end{dmath*}

This completes the proof for the inductive step. By the spirit of mathematical induction, we have shown that \eqref{equ:optimal_t_2sources_monotonically_decreasing} holds for $k=2$. When $k>2$, similar arguments can be applied to each pair of source nodes, and this completes the proof of claim \eqref{lemma:optimal_t_k_sources_monotonically_decreasing}.

We show that $\cT_{S}=[\bar{d}(S,V_\Inf), +\infty)$. Consider any node $l \in V_\Inf$ such that $d(S,l) = \bar{d}(S,V_\Inf)$. The infection can propagate at most one hop further from any source node in one time slot. If $t<\bar{d}(S,V_\Inf)$, the infection can not reach node $l$. Claim \eqref{lemma:optimal_t_k_sources_optimal_t} now follows from claim \eqref{lemma:optimal_t_k_sources_monotonically_decreasing}, and the proof of Proposition \ref{lemma:optimal_t_k_sources} is complete.

\section{Proof of Lemma \ref{lemma:k_jordan_centers_equivalency}} \label{appendix:lemma:k_jordan_centers_equivalency}
We first show that the value of the minimum infection range in $\sng$ can not be less than $\bar{d}(S,V_\Inf)$. Assume there is a super node $\sn{S'}$ in $\sng$ that is associated with a set of $k$ nodes $S' \subset V$ such that, $\bar{d}(\sn{S'},V_\Inf) < \bar{d}(S,V_\Inf)$. Then it is implied that $\bar{d}(S',V_\Inf) < \bar{d}(S,V_\Inf)$, which contradicts with the assumption that $S$ is a $k$-Jordan center set.

We then show that $\sn{S}$ is a Jordan center of $V_\Inf$ in the transformed super node graph $\sng$, i.e., $\sn{S}$ has the minimum infection range in $\sng$. In other words, we want to show that $d(\sn{S},v) \le \bar{d}(S,V_\Inf)$ for any node $v \in V_\Inf$. From Definition \ref{def:super_node_graph_transformation}, it suffices to show that $d(s_i,v) \le \bar{d}(S,V_\Inf)$ for any node $v \in A_i$, where $i  \in \{1, 2, \cdots, k\}$. Suppose that there exists a node $v \in A_i$ such that $d(s_i, v) \ge \bar{d}(S,V_\Inf)+1$. Then the first infection time $t_{int}(v)$ of $v$ in $X^{t_S}$ is
\begin{align*}
t_{int}(v) &\ge  d(s_i, v) \\
&\ge \bar{d}(S,V_\Inf)+1,
\end{align*}
because the infection can spread at most one hop further from $s_i$ in one time slot. Following Proposition \ref{lemma:optimal_t_k_sources}\eqref{lemma:optimal_t_k_sources_optimal_t}, we have that $t_S = \bar{d}(S,V_\Inf) < t_{int}(v)$, a contradiction. Therefore we have $d(s_i,v) \le \bar{d}(S,V_\Inf)$ for any $v \in A_i$, where $i  \in \{1, 2, \cdots, k\}$. This competes the proof of Lemma \ref{lemma:k_jordan_centers_equivalency}.

\section{Proof of Proposition \ref{lemma:MJC_non_increasing}} \label{appendix:lemma:MJC_non_increasing}
For a node $u \in V_\Inf$, suppose that $\hat{s}_j^{l-1}$ is a nearest node in $\hat{S}^{l-1}$ to $u$, and $M_j$ be the Voronoi set corresponding to $\hat{s}_j^{l-1}$. Following Definition \ref{def:jordan_infection_centers}, it suffices to show that
\begin{align*}
d(\hat{s}_j^l, u) \le \bar{d}(\hat{S}^{l-1},V_\Inf),
\end{align*}
where $\hat{s}_j^l$ is the Jordan center of $M_j$. We have $d(\hat{s}_j^l, u) \le \max_{z \in V_\Inf \cap M_j} d(\hat{s}_j^l, z) \le  \max_{z \in V_\Inf \cap M_j} d(\hat{s}_j^{l-1}, z) \le \bar{d}(\hat{S}^{l-1},V_\Inf)$. The proof of Proposition \ref{lemma:MJC_non_increasing} is now complete.

\section{Proof of Lemma \ref{lemma:SIRI}}\label{appendix:lemma:SIRI}
We first show the following property of the SIRI model in a network with only one node.
\begin{lem}\label{lemma:SIRI_single_node}
Suppose that $G$ has only one node $v$. For any $t \in [1, +\infty)$, consider any two most likely infection paths $X^t$ for $(v,t)$ and $Y^{t+1}$ for $(v,t+1)$ under the SIRI model. Assume Assumption \ref{assump:infection_prob_SIRI} holds. We have
\begin{align}
\frac{P_v(Y^{t+1})}{P_v(X^t)} \le \sqrt{\frac{\psmin}{\psmax}}.
\end{align}
\end{lem}
\begin{IEEEproof}
Given any most likely infection path $Y^{t+1}$ for $(v,t+1)$ with $t \in [1, +\infty)$, it suffices to construct another infection path $\tX^t$ for $(v,t)$ such that
\begin{align}
\frac{P_v(Y^{t+1})}{P_v(\tX^t)} \le \sqrt{\frac{\psmin}{\psmax}}.  \label{equ:SIRI_signle_node}
\end{align}

Let $\tX^t(v,[1,t]) = Y^{t+1}(v, [2,t+1])$, we have three cases for $Y^{t+1}$ which are discussed in the following.

\noindent \emph{Case 1:} If $Y^{t+1}(v,1) = \Inf$, we have
\begin{align*}
\frac{P_v(Y^{t+1})}{P_v(\tX^t)} & = \frac{P_v(Y^{t+1}(v ,1)) P_v(Y^{t+1}(v ,[2,t+1]))}{P_v(\tX^{t}(v ,[1, t]))} \\
&=\pInf{v} \\
&\le \sqrt{\frac{\psmin}{\psmax}},
\end{align*}
where the last inequality holds from \eqref{ineq:SIRI_pi}.

\noindent \emph{Case 2:} If $Y^{t+1}(v,1) = \Rec$ and $Y^{t+1}(v,2) = \Inf$, we have
\begin{align*}
\frac{P_v(Y^{t+1})}{P_v(\tX^t)} & = \frac{P_v(Y^{t+1}(v ,[1,2])) P_v(Y^{t+1}(v ,[3,t+1]))}{P_v(\tX^{t}(v ,1)) P_v(\tX^{t}(v ,[2, t]))} \\
&=\frac{(1-\pInf{v}) \pRec{v}}{\pInf{v}} \\
&\le \sqrt{\frac{\psmin}{\psmax}},
\end{align*}
where the last inequality holds from \eqref{ineq:SIRI_pr}.

\noindent \emph{Case 3:} If $Y^{t+1}(v,1) = \Rec$ and $Y^{t+1}(v,2) = \Rec$, we have
\begin{align*}
\frac{P_v(Y^{t+1})}{P_v(\tX^t)} & = \frac{P_v(Y^{t+1}(v ,[1,2])) P_v(Y^{t+1}(v ,[3,t+1]))}{P_v(\tX^{t}(v ,1)) P_v(\tX^{t}(v ,[2, t]))} \\
& = \frac{(1-\pInf{v}) (1-\pRec{v})}{1 - \pInf{v}} \\
&\le \sqrt{\frac{\psmin}{\psmax}},
\end{align*}
where the last inequality holds from \eqref{ineq:SIRI_pr}.

We see that \eqref{equ:SIRI_signle_node} holds for all three possible cases. The proof for Lemma \ref{lemma:SIRI_single_node} is now complete.
\end{IEEEproof}

We note that $T_u(v;G)$ is either an uninfected subtree or infected subtree. In the following, we prove these two cases separately.

\subsection{Proof of Lemma \ref{lemma:SIRI} for Uninfected Subtree} \label{appendix:lemma:SIRI_uninfected_graph}
If $T_u(v;G)$ is an uninfected subtree, we can easily see that $\cT_v = [1,+\infty)$. It is clear that claim \eqref{lemma:SIRI:V} follows from claim \eqref{lemma:SIRI:v} and \eqref{lemma:SIRI:T_u}. In the following, we prove claim \eqref{lemma:SIRI:v} and \eqref{lemma:SIRI:T_u} by mathematical induction on the elapsed time $t$.

\begin{figure}[!ht] 
  \centering
  \psfrag{a}[][][0.8][0]{$\Inf$}
  \psfrag{c}[][][0.8][0]{$\Sus$}
  \psfrag{e}[][][0.8][0]{$\Rec$}
  \psfrag{n}[][][0.8][0]{$\Non$}
  \psfrag{v}[r][][0.8][0]{$v$}
  \psfrag{u}[r][][0.8][0]{$u$}
  \psfrag{x}[r][][0.8][0]{$\children{u}$}
  \includegraphics[width=0.6\textwidth]{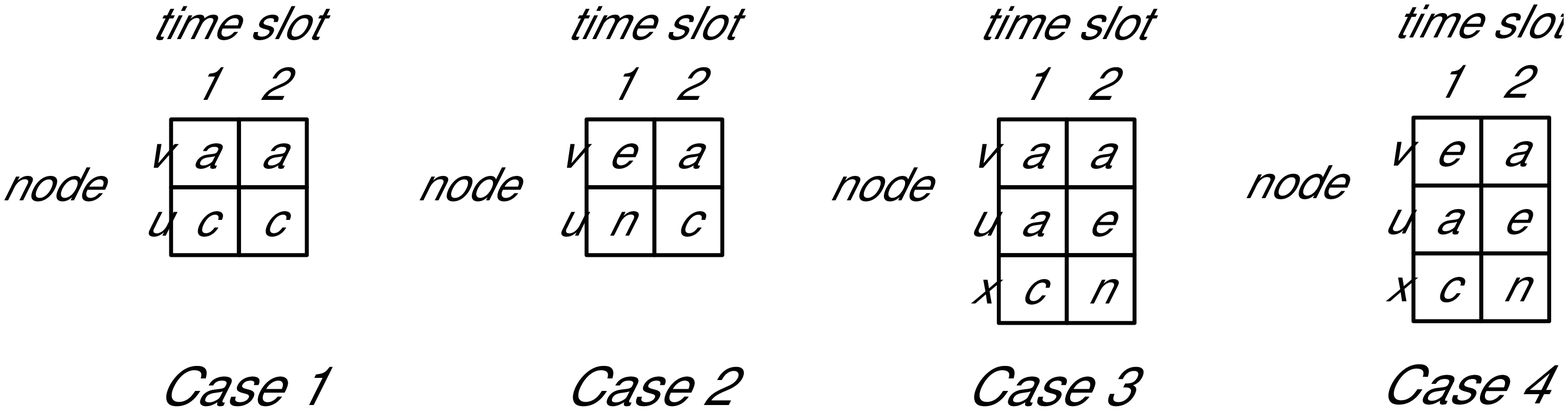}
  \caption{Illustration of four possible cases for $Y^{2}$, where we omit the states for any node that only have non-susceptible state. We have $Y^{2}(v, [1,2])=\pInf{v}^2$, $(1-\pInf{v})\pRec{v}$, $\pInf{v}^2$, or $(1-\pInf{v})\pRec{v}$ for four cases respectively. Moreover, we have $X^{2}(T_u(v;G), [1,2])=(1-\pSus{u})^2$, $1-\pSus{u}$, $\pSus{u}(1-\pInf{u})\prod_{w \in \children{u}}(1-\pSus{w})$, or $\pSus{u}(1-\pInf{u})\prod_{w \in \children{u}}(1-\pSus{w})$ for four cases respectively.}
  \label{fig:SIRI_uninfected_basis_step}
\end{figure}

\noindent \textbf{Basis step: $t=1$.}

If $v \in V_\Inf$, we let $\tilde{X}^1(v,1) = \Inf$ and $\tilde{X}^1(u,1) = \Sus$, then $P_v(\tilde{X}^1(v, 1)) = \pInf{v}$ and $P_v(\tilde{X}^1(T_u(v;G), 1)) = P_v(\tilde{X}^1(u,1)) = 1-\pSus{u}$. As shown in Figure \ref{fig:SIRI_uninfected_basis_step}, there are four possible cases for $Y^{2}$. Following Assumption \ref{assump:topology_infinite_regular}, we have
\begin{align}
\frac{\pSus{u}(1-\pInf{u})\prod_{w \in \children{u}}(1-\pSus{w})}{1-\pSus{u}} & \le \frac{(1-\pInf{u})(1-\psmin)}{1-\psmax} \nonumber \\
&\le 1, \label{ineq:SIRI_uninfected_basis_step}
\end{align}
where the last inequity holds from \eqref{ineq:SIRI_pi}. Then following \eqref{ineq:SIRI_pi}, \eqref{ineq:SIRI_pr} and \eqref{ineq:SIRI_uninfected_basis_step}, we have
\begin{align*}
\frac{P_v(Y^{2}(v,[1,2]))}{P_v(\tilde{X}^1(v,1))} &=  \frac{\max \left\{ \pInf{v}^2, (1-\pInf{v})\pRec{v} \right\}}{\pInf{v}} \\
&\le \sqrt{\frac{\psmin}{\psmax}}, \\
\frac{P_v(Y^{2}(T_u(v;G),[1,2]))}{P_v(\tilde{X}^1(T_u(v;G),1))} &= \frac{\maximum{(1-\pSus{u})^2, 1-\pSus{u},  \pSus{u}(1-\pInf{u})\prod_{w \in \children{u}}(1-\pSus{w})}}{1-\pSus{u}} \\
& = 1.
\end{align*}

If $v \notin V_\Inf$, we have $\tilde{X}^1(v,1) = \Rec$ and $\tilde{X}^1(u,1) = \Non$, then $P_v(\tilde{X}^1(v, 1)) = 1-\pInf{v}$ and $P_v(\tilde{X}^1(T_u(v;G), 1)) = P_v(\tilde{X}^1(u,1)) = 1-\pSus{u}$. Change the states of node $v$ at time slot 2 for all four cases in Figure \ref{fig:SIRI_uninfected_basis_step} from infected to recovered. Then following \eqref{ineq:SIRI_pi}, \eqref{ineq:SIRI_pr} and \eqref{ineq:SIRI_uninfected_basis_step}, we have
\begin{align*}
\frac{P_v(Y^{2}(v,[1,2]))}{P_v(\tilde{X}^1(v,1))} &= \frac{\maximum{ \pInf{v}(1-\pInf{v}), (1-\pInf{v})(1-\pRec{v})}}{1-\pInf{v}}, \\
& \le \sqrt{\frac{\psmin}{\psmax}}, \\
\frac{P_v(Y^{2}(T_u(v;G),[1,2]))}{P_v(\tilde{X}^1(T_u(v;G),1))} &= \frac{\maximum{ 1-\pSus{u},  \pSus{u}(1-\pInf{u})\prod_{w \in \children{u}}(1-\pSus{w})}}{1-\pSus{u}} \\
&= 1.
\end{align*}

This completes the proof for the basis step.

\noindent \textbf{Inductive step:} assume claim \eqref{lemma:SIRI:v} and \eqref{lemma:SIRI:T_u} hold for $t=\tau -1$, where $\tau \ge 2$. We want to show that claim \eqref{lemma:SIRI:v} and \eqref{lemma:SIRI:T_u} also hold for $t=\tau$.

Assume $t=\tau$ and consider the following six possible cases for $Y^{t+1}$.

\noindent \emph{Case 1:} $Y^{t+1}(v,1) = \Inf$ and $Y^{t+1}(u,1) = \Sus$.

Let $\tilde{X}^t(V, [1,t]) = Y^{t+1}(V, [2,t+1])$. Then following \eqref{ineq:SIRI_pi}, we have
\begin{align*}
\frac{P_v(Y^{t+1}(v, [1,t+1]))}{P_v(\tilde{X}^t(v, [1,t]))} &= \frac{P_v(Y^{t+1}(v,1)=\Inf)P_v(Y^{t+1}(v, [2,t+1]))}{P_v(\tilde{X}^t(v, [1,t]))} \\
&=\pInf{v}\\
&\le \sqrt{\frac{\psmin}{\psmax}},\\
\frac{P_v(Y^{t+1}(T_u(v;G),[1,t+1]))}{P_v(\tilde{X}^t(T_u(v;G), [1,t]))} &= \frac{P_v(Y^{t+1}(u,1)=\Sus)P_v(Y^{t+1}(T_u(v;G), [2,t+1]))}{P_v(\tilde{X}^t(T_u(v;G), [1,t]))} \\
&=1 - \pSus{u} \\
&\le 1.
\end{align*}

\noindent \emph{Case 2:} $Y^{t+1}(v,1) = \Inf$ and $Y^{t+1}(u,1) = \Inf$.

Let $\tilde{X}^t(v,[1,t]) = Y^{t+1}(v, [2,t+1])$ and $\tilde{X}^t(u,1) = \Inf$. In this case, the states of $v$ do not depend on the states of any other nodes, therefore, it can be seen as the infection source of a graph containing only itself with the infection starting at time 1. Then by Lemma \ref{lemma:SIRI_single_node}, we can find a $\tilde{X}^t$ such that
\begin{align*}
\frac{P_v(Y^{t+1}(v,[2,t+1]))}{P_v(\tilde{X}^t(v,[2,t]))} \le  \sqrt{\frac{\psmin}{\psmax}} .
\end{align*}
We then have
\begin{align*}
\frac{P_v(Y^{t+1}(v, [1,t+1]))}{P_v(\tilde{X}^t(v, [1,t]))} &= \frac{P_v(Y^{t+1}(v,1)=\Inf) P_v(Y^{t+1}(v,[2,t+1]))} {P_v(\tilde{X}^t(v,1)=\Inf) P_v(\tilde{X}^t(v,[2,t]))}\\
&\le \sqrt{\frac{\psmin}{\psmax}}.
\end{align*}

After it gets infected at time $1$, node $u$ serves as the infection source of $T_u(v;G)$ with the infection starting at time 1. By the induction assumption, we can find a $\tilde{X}^t$ such that
\begin{align*}
\frac{P_v(Y^{t+1}(T_u(v;G), [2,t+1]))}{P_v(\tilde{X}^t(T_u(v;G),[2,t]))} & \le \sqrt{\frac{\psmin}{\psmax}} \le 1.
\end{align*}
The following inequality then holds,
\begin{align*}
\frac{P_v(Y^{t+1}(T_u(v;G),[1,t+1]))}{P_v(\tilde{X}^t(T_u(v;G), [1,t]))} &= \frac{P_v(Y^{t+1}(u,1)=\Inf) P_v(Y^{t+1}(T_u(v;G), [2,t+1]))}{P_v(\tilde{X}^t(u,1)=\Inf) P_v(\tilde{X}^t(T_u(v;G),[2,t])} \\
&\le 1.
\end{align*}

\noindent \emph{Case 3:} $Y^{t+1}(v,1) = \Rec$, $Y^{t+1}(v,2) = \Inf$ and $Y^{t+1}(u,1) = \Non$.

Let $\tilde{X}^t(V, [1,t]) = Y^{t+1}(V, [2,t+1])$. Following \eqref{ineq:SIRI_pr}, we have
\begin{align*}
\frac{P_v(Y^{t+1}(v, [1,t+1]))}{P_v(\tilde{X}^t(v, [1,t]))} &= \frac{P_v(Y^{t+1}(v,1)=\Rec) P_v(Y^{t+1}(v,2)=\Inf) P_v(Y^{t+1}(v, [3,t+1]))}{P_v(\tilde{X}^t(v,1)=\Inf) P_v(\tilde{X}^t(v, [2, t])} \\
&=\frac{(1-\pInf{v})\pRec{v}}{\pInf{v}} \\
&\le \sqrt{\frac{\psmin}{\psmax}},\\
\frac{P_v(Y^{t+1}(T_u(v;G),[1,t+1]))}{P_v(\tilde{X}^t(T_u(v;G), [1,t]))} &= \frac{P_v(Y^{t+1}(u,1)=\Non)P_v(Y^{t+1}(u,2)=\Sus) P_v(Y^{t+1}(T_u(v;G), [3,t+1]))} {P_v(\tilde{X}^t(u,1)=\Sus) P_v(\tilde{X}^t(T_u(v;G), [2, t])} \\
&=1.
\end{align*}

\noindent \emph{Case 4:} $Y^{t+1}(v,1) = \Rec$, $Y^{t+1}(v,2) = \Inf$ and $Y^{t+1}(u,1) = \Inf$.

Let $\tilde{X}^t(v, [1,t]) = Y^{t+1}(v, [2,t+1])$ and $\tilde{X}^t(u, 1) = \Inf$. Following \eqref{ineq:SIRI_pr}, we have
\begin{align*}
\frac{P_v(Y^{t+1}(v, [1,t+1]))}{P_v(\tilde{X}^t(v, [1,t]))} &= \frac{P_v(Y^{t+1}(v,1)=\Rec) P_v(Y^{t+1}(v,2)=\Inf) P_v(Y^{t+1}(v, [3,t+1]))}{P_v(\tilde{X}^t(v,1)=\Inf) P_v(\tilde{X}^t(v,[2,t]))} \\
&=\frac{(1-\pInf{v})\pRec{v}}{\pInf{v}} \\
&\le \sqrt{\frac{\psmin}{\psmax}}.\\
\end{align*}

Following the same arguments as that in case 2, we can find a $\tilde{X}^t$ such that
\begin{align*}
\frac{P_v(Y^{t+1}(T_u(v;G),[1,t+1]))}{P_v(\tilde{X}^t(T_u(v;G), [1,t]))} &= \frac{P_v(Y^{t+1}(u,1)=\Inf) P_v(Y^{t+1}(T_u(v;G), [2,t+1]))} {P_v(\tilde{X}^t(u,1) = \Inf) P_v(\tilde{X}^t(T_u(v;G), [2,t]))} \\
&\le 1.
\end{align*}

\noindent \emph{Case 5:} $Y^{t+1}(v,1) = \Rec$, $Y^{t+1}(v,2) = \Rec$ and $Y^{t+1}(u,1) = \Non$.

Let $\tilde{X}^t(V, [1,t]) = Y^{t+1}(V, [2,t+1])$. Following \eqref{ineq:SIRI_pr}, we have
\begin{align*}
\frac{P_v(Y^{t+1}(v, [1,t+1]))}{P_v(\tilde{X}^t(v, [1,t]))}&= \frac{P_v(Y^{t+1}(v,1)=\Rec) P_v(Y^{t+1}(v,2)=\Rec) P_v(Y^{t+1}(v, [3,t+1])) }{P_v(\tilde{X}^t(v,1)=\Rec) P_v(\tilde{X}^t(v, [2, t]))} \\
&=\frac{(1-\pInf{v})(1-\pRec{v})}{1-\pInf{v}} \\
&\le \sqrt{\frac{\psmin}{\psmax}}, \\
\frac{P_v(Y^{t+1}(T_u(v;G),[1,t+1]))}{P_v(\tilde{X}^t(T_u(v;G), [1,t]))}&= \frac{P_v(Y^{t+1}(u,1)=\Non) P_v(Y^{t+1}(u,2)=\Non) P_v(Y^{t+1}(T_u(v;G), [3,t+1]))} {P_v(\tilde{X}^t(u,1)=\Non) P_v(\tilde{X}^t(T_u(v;G), [2, t]))} \\
&=1.
\end{align*}

\noindent \emph{Case 6:} $Y^{t+1}(v,1) = \Rec$, $Y^{t+1}(v,2) = \Rec$ and $Y^{t+1}(u,1) = \Inf$.

Let $\tilde{X}^t(v, [1,t]) = Y^{t+1}(v, [2,t+1])$ and $\tilde{X}^t(u, 1) = \Inf$. Following the same arguments as that in case 2, we can find a $\tilde{X}^t$ such that
\begin{align*}
\frac{P_v(Y^{t+1}(T_u(v;G), [2,t+1]))}{P_v(\tilde{X}^t(T_u(v;G),[2,t]))} \le 1.
\end{align*}
Then following \eqref{ineq:SIRI_pr}, we have
\begin{align*}
\frac{P_v(Y^{t+1}(v, [1,t+1]))}{P_v(\tilde{X}^t(v, [1,t]))}&= \frac{P_v(Y^{t+1}(v,1)=\Rec) P_v(Y^{t+1}(v,2)=\Rec) P_v(Y^{t+1}(v, [3,t+1]))}{P_v(\tilde{X}^t(v,1)=\Rec)P_v(\tilde{X}^t(v,[2,t]))} \\
&=\frac{(1-\pInf{v})(1-\pRec{v})}{1-\pInf{v}} \\
&\le \sqrt{\frac{\psmin}{\psmax}}, \\
\frac{P_v(Y^{t+1}(T_u(v;G),[1,t+1]))}{P_v(\tilde{X}^t(T_u(v;G), [1,t]))}&= \frac{P_v(Y^{t+1}(u,1)=\Inf) P_v(Y^{t+1}(T_u(v;G), [2,t+1]))}{ P_v(\tilde{X}^t(u,1)=\Inf) P_v(\tilde{X}^t(T_u(v;G), [2,t]))}\\
&\le 1.
\end{align*}

Therefore, we have shown that claim \eqref{lemma:SIRI:v} and \eqref{lemma:SIRI:T_u} hold for all six possible cases. This completes the proof for the inductive step. The proof of Lemma \ref{lemma:SIRI} for uninfected subtree is now complete.

\subsection{Proof of Lemma \ref{lemma:SIRI} for Infected Subtree}\label{appendix:lemma:SIRI_infected_graph}
If $T_u(v;G)$ is an infected subtree, we can see that $\cT_v = [\bar{d}(v, V_\Inf),+\infty)$. We prove claim \eqref{lemma:SIRI:v} and \eqref{lemma:SIRI:T_u} for infected subtree by mathematical induction on $\bar{d}(v, V_\Inf)$.

\noindent \textbf{Basis step: $\bar{d}(v, V_\Inf) = 1$.}

For any $t \ge 1$, we consider any most likely infection path $Y^{t+1}$ for $(v,t+1)$. In the following, six possible cases for $Y^{t+1}$ are discussed in order.

\noindent \emph{Case 1:} $Y^{t+1}(v,1) = \Inf$ and $Y^{t+1}(u,1) = \Sus$.

Let $\tilde{X}^t(V, [1,t]) = Y^{t+1}(V, [2,t+1])$. Then following \eqref{ineq:SIRI_pi}, we have
\begin{align*}
\frac{P_v(Y^{t+1}(v,[1,t+1]))}{P_v(\tilde{X}^t(v,[1,t]))} &= \frac{P_v(Y^{t+1}(v,1)=\Inf)P_v(Y^{t+1}(v, [2,t+1]))}{P_v(\tilde{X}^t(v, [1,t]))} \\
&=\pInf{v} \\
&\le \sqrt{\frac{\psmin}{\psmax}}, \\
\frac{P_v(Y^{t+1}(T_u(v;G), [1,t+1]))}{P_v(\tilde{X}^t(T_u(v;G),[1,t]))} &= \frac{P_v(Y^{t+1}(u,1)=\Sus)P_v(Y^{t+1}(T_u(v;G), [2,t+1]))}{P_v(\tilde{X}^t(T_u(v;G), [1,t]))} \\
&=1-\pSus{u}\\
&\le 1.
\end{align*}

\noindent \emph{Case 2:} $Y^{t+1}(v,1) = \Inf$ and $Y^{t+1}(u,1) = \Inf$.

Let $\tilde{X}^t(v,[1,t]) = Y^{t+1}(v, [2,t+1])$ and $\tilde{X}^t(u,1) = \Inf$, following \eqref{ineq:SIRI_pi}, we have
\begin{align*}
\frac{P_v(Y^{t+1}(v,[1,t+1]))}{P_v(\tilde{X}^t(v,[1,t]))}&= \frac{P_v(Y^{t+1}(v,1)=\Inf) P_v(Y^{t+1}(v, [2,t+1])) }{P_v(\tilde{X}^t(v, [1,t]))} \\
&=\pInf{v} \\
&\le \sqrt{\frac{\psmin}{\psmax}}.
\end{align*}

After it gets infected at time slot 1, node $u$ serves as the infection source of the subtree $T_u(v;G)$ with the infection starting at time 1. From the assumption $\bar{d}(v, V_\Inf) = 1$, it follows that $T_w(u;G)$ is an uninfected subtree for any $w \in V(u,1) \bigcap T_u(v;G)$. Then by Lemma \ref{lemma:SIRI}(\ref{lemma:SIRI:V}) for uninfected subtree, we can find a $\tilde{X}^t$ such that
\begin{align}
\frac{P_v(Y^{t+1}(T_u(v;G),[2,t+1]))}{P_v(\tilde{X}^t(T_u(v;G),[2,t]))} &\le \sqrt{\frac{\psmin}{\psmax}} \nonumber \\
&\le 1. \label{equ:SIRI_infected_graph_basis}
\end{align}
We then have,
\begin{align*}
\frac{P_v(Y^{t+1}(T_u(v;G),[1,t+1]))}{P_v(\tilde{X}^t(T_u(v;G),[1,t]))}&= \frac{P_v(Y^{t+1}(u,1)=\Inf) P_v(Y^{t+1}(T_u(v;G), [2,t+1]))}{P_v(\tilde{X}^t(u,1)=\Inf) P_v(\tilde{X}^t(T_u(v;G), [2,t]))} \\
&\le 1.
\end{align*}

\noindent \emph{Case 3:} $Y^{t+1}(v,1) = \Rec$, $Y^{t+1}(v,2) = \Inf$ and $Y^{t+1}(u,1) = \Non$.

Let $\tilde{X}^t(V, [1,t]) = Y^{t+1}(V, [2,t+1])$. Following \eqref{ineq:SIRI_pr}, we have
\begin{align*}
\frac{P_v(Y^{t+1}(v,[1,t+1]))}{P_v(\tilde{X}^t(v,[1,t]))} &= \frac{P_v(Y^{t+1}(v,1)=\Rec) P_v(Y^{t+1}(v,2)=\Inf) P_v(Y^{t+1}(v, [3,t+1]))}{P_v(\tilde{X}^t(v,1)=\Inf) P_v(\tilde{X}^t(v, [2, t])} \\
&=\frac{(1-\pInf{v})\pRec{v}}{\pInf{v}} \\
&\le \sqrt{\frac{\psmin}{\psmax}}, \\
\frac{P_v(Y^{t+1}(T_u(v;G),[1,t+1]))}{P_v(\tilde{X}^t(T_u(v;G),[1,t]))} &= \frac{P_v(Y^{t+1}(u,1)=\Non) P_v(Y^{t+1}(u,2)=\Sus) P_v(Y^{t+1}(T_u(v;G), [3,t+1]))}{P_v(\tilde{X}^t(u,1)=\Sus) P_v(\tilde{X}^t(T_u(v;G), [2, t])}\\
&=1.
\end{align*}

\noindent \emph{Case 4:} $Y^{t+1}(v,1) = \Rec$, $Y^{t+1}(v,2) = \Inf$ and $Y^{t+1}(u,1) = \Inf$.

Let $\tilde{X}^t(v, [1,t]) = Y^{t+1}(v, [2,t+1])$ and $\tilde{X}^t(u, 1) = \Inf$. Following the same arguments as that in case 2, we can find a $\tilde{X}^t$ such that \eqref{equ:SIRI_infected_graph_basis} holds. Then following \eqref{ineq:SIRI_pr}, we have
\begin{align*}
\frac{P_v(Y^{t+1}(v,[1,t+1]))}{P_v(\tilde{X}^t(v,[1,t]))}&= \frac{P_v(Y^{t+1}(v,1)=\Rec) P_v(Y^{t+1}(v,2)=\Inf) P_v(Y^{t+1}(v, [3,t+1])) }{P_v(\tilde{X}^t(v,1)=\Inf) P_v(\tilde{X}^t(v,[2,t]))} \\
&=\frac{(1-\pInf{v})\pRec{v}}{\pInf{v}} \\
&\le \sqrt{\frac{\psmin}{\psmax}}, \\
\frac{P_v(Y^{t+1}(T_u(v;G),[1,t+1]))}{P_v(\tilde{X}^t(T_u(v;G),[1,t]))}&=\frac{P_v(Y^{t+1}(u,1)=\Inf) P_v(Y^{t+1}(T_u(v;G), [2,t+1]))}{P_v(\tilde{X}^t(u,1)=\Inf) P_v(\tilde{X}^t(T_u(v;G),[2,t]))} \\
&\le 1.
\end{align*}

\noindent \emph{Case 5:} $Y^{t+1}(v,1) = \Rec$, $Y^{t+1}(v,2) = \Rec$ and $Y^{t+1}(u,1) = \Non$.

Let $\tilde{X}^t(V, [1,t]) = Y^{t+1}(V, [2,t+1])$. Then following \eqref{ineq:SIRI_pr}, we have
\begin{align*}
\frac{P_v(Y^{t+1}(v,[1,t+1]))}{P_v(\tilde{X}^t(v,[1,t]))} &= \frac{P_v(Y^{t+1}(v,1)=\Rec) P_v(Y^{t+1}(v,2)=\Rec) P_v(Y^{t+1}(v, [3,t+1]))}{P_v(\tilde{X}^t(v,1)=\Rec) P_v(\tilde{X}^t(v, [2, t]))} \\
&=\frac{(1-\pInf{v})(1-\pRec{v})}{1-\pInf{v}} \\
&\le \sqrt{\frac{\psmin}{\psmax}}, \\
\frac{P_v(Y^{t+1}(T_u(v;G),[1,t+1]))}{P_v(\tilde{X}^t(T_u(v;G),[1,t]))}&=\frac{P_v(Y^{t+1}(u,1)=\Non) P_v(Y^{t+1}(u,2)=\Non) P_v(Y^{t+1}(T_u(v;G), [3,t+1]))}{P_v(\tilde{X}^t(u,1)=\Non) P_v(\tilde{X}^t(T_u(v;G), [2, t]))} \\
&=1.
\end{align*}

\noindent \emph{Case 6:} $Y^{t+1}(v,1) = \Rec$, $Y^{t+1}(v,2) = \Rec$ and $Y^{t+1}(u,1) = \Inf$.

Let $\tilde{X}^t(v, [1,t]) = Y^{t+1}(v, [2,t+1])$ and $\tilde{X}^t(u, 1) = \Inf$. Following the same arguments as that in case 2, we can find a $\tilde{X}^t$ such that \eqref{equ:SIRI_infected_graph_basis} holds. Then following \eqref{ineq:SIRI_pr}, we have
\begin{align*}
\frac{P_v(Y^{t+1}(v,[1,t+1]))}{P_v(\tilde{X}^t(v,[1,t]))} &= \frac{P_v(Y^{t+1}(v,1)=\Rec) P_v(Y^{t+1}(v,2)=\Rec) P_v(Y^{t+1}(v, [3,t+1]))}{P_v(\tilde{X}^t(v,1)=\Rec)P_v(\tilde{X}^t(v,[2,t]))}  \\
&=\frac{(1-\pInf{v})(1-\pRec{v})}{1-\pInf{v}} \\
&\le \sqrt{\frac{\psmin}{\psmax}}, \\
\frac{P_v(Y^{t+1}(T_u(v;G),[1,t+1]))}{P_v(\tilde{X}^t(T_u(v;G),[1,t]))}&=\frac{P_v(Y^{t+1}(u,1)=\Inf) P_v(Y^{t+1}(T_u(v;G), [2,t+1]))}{P_v(\tilde{X}^t(u,1)=\Inf) P_v(\tilde{X}^t(T_u(v;G),[2,t]))} \\
&\le 1.
\end{align*}

We have shown that claim \eqref{lemma:SIRI:v} and \eqref{lemma:SIRI:T_u} hold for all six possible cases. This completes the proof for the basis step.

\noindent \textbf{Inductive step:} assume claim \eqref{lemma:SIRI:v} and \eqref{lemma:SIRI:T_u} hold for $\bar{d}(v, V_\Inf) \le n -1$, where $n \ge 2$. We want to show that claim \eqref{lemma:SIRI:v} and \eqref{lemma:SIRI:T_u} also hold for $\bar{d}(v, V_\Inf)=n$.

Assume $\bar{d}(v, V_\Inf)=n$ and consider any most likely infection path $Y^{t+1}$ for $(v,t+1)$, where $t \ge n$. We first show that \eqref{equ:SIRI_infected_graph_basis} in case 2 also holds in the inductive step. For case 2, we have $Y^{t+1}(v,1) = \Inf$ and $Y^{t+1}(u,1) = \Inf$. Let $\tilde{X}^t(v,[1,t]) = Y^{t+1}(v, [2,t+1])$ and $\tilde{X}^t(u,1) = \Inf$, after it gets infected at time slot 1, node $u$ will serve as the infection source of the subtree $T_u(v;G)$ with the infection starting at time 1. Since $\bar{d}(u, V_\Inf \bigcap T_u(v;G)) \le n-1$, from the induction assumption, we can find a $\tilde{X}^t$ such that \eqref{equ:SIRI_infected_graph_basis} holds. From the same arguments as that in the basis step, it follows that claim \eqref{lemma:SIRI:v} and \eqref{lemma:SIRI:T_u} hold for all six possible cases. This completes the proof for the inductive step. By the spirit of mathematical induction, the proof of Lemma \ref{lemma:SIRI} for infected subtree is now complete. This completes the proof of Lemma \ref{lemma:SIRI}.

\section{Proof of Lemma \ref{lemma:SIS_better_later}}\label{appendix:lemma:SIS_better_later}
Let $d$ be the degree of any node in $G$. Fix the elapsed time to be $t$ and consider any most likely infection path $X^t$ for $(v,t)$. Given any $u \in H_v \backslash \{v\}$, we first show that
\begin{align}
t_{int}(u) \in [d(v,u), t-\bar{d}(u, T_u(v;H_v))]. \label{equ:SIS_first_infection_time_bound}
\end{align}

Firstly, it is easy to see that any node in $H_v$ has been infected at least once due to the assumption that the underlying network $G$ is a tree, otherwise, the infection can not reach the leaf nodes of $H_v$. We now consider the lower bound of $t_{int}(u)$ in \eqref{equ:SIS_first_infection_time_bound}. Since the infection can spread at most one hop away from $v$ in each time slot, the earliest time for $u$ to get the infection is $d(v,u)$. Then we consider the upper bound of $t_{int}(u)$ in \eqref{equ:SIS_first_infection_time_bound}. After node $u$ gets infected for the first time at $t_{int}(u)$, the infection can spread at most $t-t_{int}(u)$ hops away from $u$. Consider a node $w$ such that $d(u,w) = \bar{d}(u, T_u(v;H_v))$. In order for the infection to reach node $w$, $t-t_{int}(u) \geq d(u,w)$. So $t_{int}(u) \leq t - d(u,w) = t - \bar{d}(u, T_u(v;H_v))$. The proof for \eqref{equ:SIS_first_infection_time_bound} is now complete.

Suppose that there exists a node $u \in H_v \backslash \{v\}$ such that the first infection time $t_{int}(u)$ of $u$ in $X^t$ is less than $t-\bar{d}(u, T_u(v;H_v))$. To prove Lemma \ref{lemma:SIS_better_later}, following \eqref{equ:SIS_first_infection_time_bound}, it suffices to show that we can construct another infection path $\tilde{X}^t$ for $(v,t)$ that occurs with at least the same probability as $X^t$, where the first infection time of $u$ in $\tilde{X}^t$ is $\tilde{t}(u) = t-\bar{d}(u, T_u(v;H_v))$.

We let the states of $G\backslash (T_u(v;G)\bigcup\{\parent{u}\})$ in $\tilde{X}^t$ to be the same as those in $X^t$, i.e.
\begin{align*}
\tilde{X}^t(G\backslash (T_u(v;G)\bigcup\{\parent{u}\}), [1,t]) = X^t(G\backslash (T_u(v;G)\bigcup\{\parent{u}\}), [1,t]).
\end{align*}
Let $\tilde{X}^t(\parent{u}, [1,t_{int}(u)-1]) = X^t(\parent{u}, [1,t_{int}(u)-1])$ and  $A = T_u(v;G)\bigcup \{\parent{u}\} \bigcup V(\parent{u},1)$. It suffices to show that
\begin{dmath}
P_v(\tilde{X}^t(A,[t_{int}(u),t])) \ge P_v(X^t(A,[t_{int}(u),t])). \label{equ:SIS_better_later_obj}
\end{dmath}

We show \eqref{equ:SIS_better_later_obj} by mathematical induction on $\bar{d}(u, T_u(v;H_v))$.

\noindent \textbf{Basis step:} show \eqref{equ:SIS_better_later_obj} holds for $\bar{d}(u, T_u(v;H_v)) = 0$.

Let $B$ denote the set of nodes $V(\parent{u},1) \backslash \{u\}$. Consider a time slot $\tau < \tilde{t}(u)$ where $\tilde{X}^t(\parent{u}, \tau) = \Inf$. We show the worst case for $\tilde{X}^t$ at time $\tau+1$.

If $X^t(\parent{u}, \tau) = \Inf$, we have
\begin{align}
P_v(\tilde{X}^t(B, \tau+1)) = P_v(X^t(B, \tau+1).  \label{equ:SIS_better_later_basis_tv_u_infected}
\end{align}

If $X^t(\parent{u}, \tau) \ne \Inf$, we have
\begin{align}
\frac{P_v(\tilde{X}^t(B, \tau+1))}{P_v(X^t(B, \tau+1)} \ge (1-p_\Sus)^{d-1},  \label{equ:SIS_better_later_basis_tv_u_uninfected}
\end{align}
where the equality holds when every node in $B$ is susceptible in $\tilde{X}^t$ and non-susceptible in $X^t$ at time $\tau$. By \eqref{equ:SIS_better_later_basis_tv_u_infected} and \eqref{equ:SIS_better_later_basis_tv_u_uninfected}, we can see that the worst case for $\tilde{X}^t$ at time $\tau+1$ is that $X^t(\parent{u}, \tau) \ne \Inf$ and $X^t(B, \tau) = \Non$.

We divide the time interval $[t_{int}(u), t]$ into three parts: $t_{int}(u)$, $[t_{int}(u)+1, \tilde{t}(u)-1]$ and $\tilde{t}(u)$, where $\tilde{t}(u) = t-\bar{d}(u, T_u(v;H_v)) = t$.

\noindent \emph{Part 1:} time $\tau = t_{int}(u)$.

Since node $u$ is infected for the first time at time slot $t_{int}(u)$ in $X^t$, we know that node $\parent{u}$ must be infected at time $t_{int}(u)-1$, which in turn suggests that $\tilde{X}^t(\parent{u}, \tau-1)  = X^t(\parent{u}, \tau-1) = \Inf$, yielding
\begin{align}
P_v(\tilde{X}^t(B, \tau))  = P_v(X^t(B, \tau)). \label{equ:SIS_better_later_basis_tv_B}
\end{align}

We let $\tilde{X}^t(\parent{u}, \tau) = \Inf$ and consider the worst case in \eqref{equ:SIS_better_later_basis_tv_u_uninfected}. Following \eqref{ineq:SIS_ps} and \eqref{equ:SIS_better_later_basis_tv_B}, we have
\begin{align}
\frac{P_v(\tilde{X}^t(A, t_{int}(u)))}{P_v(X^t(A, t_{int}(u)))} & = \frac{P_v(\tilde{X}^t(\parent{u}, t_{int}(u)))P_v(\tilde{X}^t(u, t_{int}(u)))P_v(\tilde{X}^t(B, t_{int}(u)))}{P_v({X}^t(\parent{u}, t_{int}(u)))P_v({X}^t(u, t_{int}(u)))P_v({X}^t(B, t_{int}(u)))} \nonumber \\
& \ge \frac{p_{\Inf}(1-p_{\Sus})}{(1-p_{\Inf})p_{\Sus}} \label{equ:SIS_better_later_basis_tv_equation} \\
& \ge 1. \label{equ:SIS_better_later_basis_tv}
\end{align}

\noindent \emph{Part 2:} time $\tau \in [t_{int}(u)+1, t-1]$.

We first consider the case that at least one node in $T_u(v;G)$ is infected at time $\tau$. We let $\tilde{X}^t(\parent{u}, \tau) = \Inf$ and consider the worst case, i.e., $X^t(\parent{u}, \tau-1) \ne \Inf$ and $X^t(B, \tau-1) = \Non$. We then have
\begin{dmath*}
P_v(\tilde{X}^t(A, \tau)) \ge p_{\Inf} (1-p_{\Sus})^d.
\end{dmath*}

Since $X^t(\parent{u}, \tau-1) \ne \Inf$ and at least one node in $T_u(v;G)$ is infected at time $\tau$, there must exist a node $w \in T_u(v;G)$, s.t., $X^t(w, \tau -1) =\Inf$. Consider any neighboring node $z$ of $w$. If $X^t(z, \tau -1) = \Inf$, due to the fact that $X^t(\parent{u}, \tau-1) \ne \Inf$ and the assumption that $G$ is an infinite tree, we can always find a node $y \in T_z(w; G) \bigcap \left( T_u(v;G) \bigcup \{\parent{u}\}\right)$, s.t., $X^t(y, \tau -1) = \Sus$. If $X^t(z, \tau -1) = \Sus$, following similar arguments as the worst case in \eqref{equ:SIS_better_later_basis_tv_u_uninfected}, we can see that $X^t(z, \tau) \ne \Inf$. We then have
\begin{dmath*}
P_v(X^t(T_z(w; G) \bigcap (T_u(v;G) \bigcup \{\parent{u}\}), \tau)) \le 1-p_{\Sus},
\end{dmath*}
for any neighboring node $z$ of $w$. Moreover, we have at least one node in $T_u(v;G)$ being infected at time $\tau$, yielding
\begin{align}
P_v(X^t(A, \tau)) &\le \max\{p_{\Inf}, p_{\Sus}\} (1-p_{\Sus})^d \nonumber \\
&= p_{\Inf} (1-p_{\Sus})^d \nonumber \\
&\le P_v(\tilde{X}^t(A, \tau)). \label{equ:SIS_better_later_basis_tau1}
\end{align}

We then consider the case that no node in $T_u(v;G)$ is infected at time $\tau$. Without loss of generality, we assume $\tau$ is the earliest time after $t_{int}(u)$ that no node in $T_u(v;G)$ is infected. We let $\tilde{X}(\parent{u},\tau) = X^t(\parent{u}, \tau)$ and consider the worst case for $\tilde{X}^t(\parent{u},\tau)$. If $X^t(\parent{u}, \tau) = \Inf$, we have
\begin{align}
\frac{P_v(\tilde{X}^t(A, \tau))}{P_v(X^t(A, \tau))} & \ge \frac{p_{\Inf}(1-p_{\Sus})^d}{p_{\Sus} (1-p_{\Inf})(1-p_{\Sus})^{d-1}} \nonumber \\
& \ge 1. \label{equ:SIS_better_later_basis_tau2}
\end{align}

If $X^t(\parent{u}, \tau) \ne \Inf$, we have
\begin{align}
\frac{P_v(\tilde{X}^t(A, \tau))}{P_v(X^t(A, \tau))} & \ge \frac{(1-p_{\Inf})(1-p_{\Sus})^d}{(1-p_{\Sus})(1-p_{\Inf})(1-p_{\Sus})^{d-1}} \nonumber \\
& = 1. \label{equ:SIS_better_later_basis_tau3}
\end{align}

From \eqref{equ:SIS_better_later_basis_tau1}-\eqref{equ:SIS_better_later_basis_tau3}, we have
\begin{align}
P_v(\tilde{X}^t(A, [t_{int}(u)+1, \tau])) & \ge P_v(X^t(A, [t_{int}(u)+1, \tau])). \label{equ:SIS_better_later_basis_tau4}
\end{align}

We have now $\tilde{X}^t(V, \tau) = X^t(V, \tau)$. If there are other time slots after $\tau$ that  no node in $T_u(v;G)$ is infected, we can apply the same arguments again. Then by \eqref{equ:SIS_better_later_basis_tau1} and \eqref{equ:SIS_better_later_basis_tau4}, we have
\begin{align}
P_v(\tilde{X}^t(A, [t_{int}(u)+1, t-1]))& \ge P_v(X^t(A, [t_{int}(u)+1, t-1])). \label{equ:SIS_better_later_basis_tau}
\end{align}

\noindent \emph{Part 3:} time $\tau = t$.

If $\parent{u} \in V_\Inf$, we have
\begin{align}
\frac{P_v(\tilde{X}^t(A, t))}{P_v(X^t(A, t))} & \ge \frac{p_{\Inf} p_{\Sus} (1-p_{\Sus})^{d-1}}{p_{\Sus} p_{\Inf} (1-p_{\Sus})^{d-1}} \nonumber \\
& = 1. \label{equ:SIS_better_later_basis_t1}
\end{align}
Then \eqref{equ:SIS_better_later_obj} holds from \eqref{equ:SIS_better_later_basis_tv}, \eqref{equ:SIS_better_later_basis_tau} and \eqref{equ:SIS_better_later_basis_t1}.

If $\parent{u} \notin V_\Inf$, we have
\begin{align}
\frac{P_v(\tilde{X}^t(A, t))}{P_v(X^t(A, t))} & \ge \frac{(1-p_{\Inf}) p_{\Sus} (1-p_{\Sus})^{d-1}}{(1-p_{\Sus}) p_{\Inf} (1-p_{\Sus})^{d-1}} \nonumber \\
& = \frac{(1-p_{\Inf})p_{\Sus}}{p_{\Inf}(1-p_{\Sus})}. \label{equ:SIS_better_later_basis_t2}
\end{align}
Then \eqref{equ:SIS_better_later_obj} holds from \eqref{equ:SIS_better_later_basis_tv_equation}, \eqref{equ:SIS_better_later_basis_tau} and \eqref{equ:SIS_better_later_basis_t2}. This competes the proof of the basis step.

\noindent \textbf{Inductive step:} assume \eqref{equ:SIS_better_later_obj} holds for $\bar{d}(u, T_u(v;H_v)) \le n$, where $0 \le n \le \bar{d}(v, H_v) -1$. Show \eqref{equ:SIS_better_later_obj} also holds for $\bar{d}(u, T_u(v;H_v)) = n+1$.

We divide the time interval $[t_{int}(u), t]$ into four parts: $t_{int}(u)$, $[t_{int}(u)+1, \tilde{t}(u)-1]$, $\tilde{t}(u)$ and $[\tilde{t}(u)+1,t]$, where $\tilde{t}(u)=t-\bar{d}(u, T_u(v;H_v))$. For any node $w \in \children{u}$, we have $\bar{d}(w, T_w(v;H_v)) \le \bar{d}(u, T_u(v;H_v))-1 = n$. By induction assumption, we have that node $w$ get infected for the first time at $t(w) = t-\bar{d}(w, T_w(v;H_v))$ in $X^t$, which in turn suggests that $X^t(u, \tilde{t}(u)) = \Inf$. For the time range $[\tilde{t}(u)+1,t]$, we let $\tilde{X}^t(A, [\tilde{t}(u)+1,t]) = X^t(A, [\tilde{t}(u)+1,t])$, yielding
\begin{align*}
P_v(\tilde{X}^t(A, [\tilde{t}(u)+1,t]))& = P_v(X^t(A, [\tilde{t}(u)+1,t])).
\end{align*}

For the first three parts, following similar arguments as in the basis step, we have
\begin{align*}
P_v(\tilde{X}^t(A, [t_{int}(u)+1, \tilde{t}(u)]))& \ge P_v(X^t(A, [t_{int}(u)+1, \tilde{t}(u)])).
\end{align*}

We can now conclude that \eqref{equ:SIS_better_later_obj} holds for the inductive step. By the spirit of mathematical induction, \eqref{equ:SIS_better_later_obj} holds and the proof of Lemma \ref{lemma:SIS_better_later} is now complete.

\bibliography{IEEEabrv,SIS}{}
\bibliographystyle{IEEEtran}

\end{document}